\pdfoutput=1
\documentclass[12pt,english]{article}
\usepackage[hyphens]{url} %
\usepackage{afterpage}
\usepackage{float}
\usepackage{changes}
\usepackage{etoolbox}
\AtBeginEnvironment{tabular}{\normalsize}
\usepackage[full]{textcomp}
\usepackage[skip=0.1\baselineskip]{caption}
\usepackage[a4paper,
            left=1in,
            right=1in,
            top=1in,
            bottom=1in]{geometry}
\usepackage[english]{babel}
\usepackage[utf8x]{inputenc}
\usepackage{amsthm}  %
\newtheorem{proposition}{Proposition}
\newtheorem{lemma}{Lemma}

\usepackage{amssymb} 
\usepackage[retainorgcmds]{IEEEtrantools}
\usepackage[flushleft]{threeparttable}
\usepackage{rotating,booktabs}
\usepackage{graphicx}
\usepackage{tabularx}
\usepackage{kpfonts}    %
\usepackage{microtype} 
\usepackage{booktabs}   %
\usepackage{bm}         %
\usepackage{listings}   %
\usepackage{verbatim}   %
\usepackage{color}  
\usepackage{hyperref}
\hypersetup{
	colorlinks = true,
	linkcolor = black,
	citecolor = black,
	runcolor = black,
	urlcolor = black
}
\usepackage{url}
\usepackage[toc,page,header]{appendix}
    \usepackage[]{minitoc}
    \noptcrule %

\usepackage{lscape}
\usepackage{subcaption}
\usepackage{mwe}
\usepackage{natbib}
\usepackage{multibib}
\newcites{secvthree}{References in the Online Appendix}
\usepackage{setspace,caption}
\usepackage{lipsum}
\usepackage{amsthm}
\theoremstyle{plain}

\newtheorem{assumption}{Assumption}

\newtheorem*{proposition*}{Proposition}
\newtheorem*{proof*}{Proof}
\setcitestyle{round,citesep={,}}
\usepackage{subcaption}
\usepackage{adjustbox}
\usepackage{pdflscape}

\usepackage{dsfont}

\makeatletter
\let\oldsection\section
\renewcommand{\section}{\vspace{-\parskip}\oldsection}

\let\oldsubsection\subsection
\renewcommand{\subsection}{\vspace{-\parskip}\oldsubsection}

\let\oldparagraph\paragraph
\renewcommand{\paragraph}{\vspace{-\parskip}\oldparagraph}
\makeatother

\usepackage{longtable}
\DeclareRobustCommand{\auditrev}[1]{#1}
\begin{document}

\title{\Huge{NGO Activism:\\ Exposure vs. Influence}\thanks{We are grateful to Martino Banchio, Jacob Conway, Agathe Denis, Nickolay Gantchev, Julian Koelbel, Filippo Lancieri, Juliane Proelss, Federica Salve, Carlo Schwarz, Catherine Thomas, seminar participants at ETH Zurich (2026) and participants at the EFA Annual Meetings (2026), FIBS Workshop (2026), Sciences Po Econ Alumni Conference (2026), and ESSEC Research Day (2025) for their constructive comments. The views expressed herein are those of the authors and do not necessarily reflect the views or policies of the Board of Governors of the Federal Reserve System. All errors and omissions are ours.}}

\author{\textbf{Michele Fioretti}\thanks{Bocconi, Department of Economics, IGIER, and CEPR. e-mail: \href{fioretti.m@unibocconi.it}{fioretti.m@unibocconi.it}} \and \textbf{Victor Saint-Jean}\thanks{ESSEC Business School, Department of Finance. e-mail: \href{saintjean@essec.edu}{saintjean@essec.edu}} \and \textbf{Simon C. Smith}\thanks{Federal Reserve Board. e-mail: \href{simon.c.smith@frb.gov}{simon.c.smith@frb.gov}}}

\date{October 14, 2026}

\maketitle

\begin{abstract}\singlespacing
This paper studies how the timing of NGO activism shapes its effectiveness in influencing corporate behavior. Using data on 2,350 campaigns targeting U.S. firms, we show that campaigns timed at annual general meetings (AGMs) generate large visibility gains but little contemporaneous influence, while campaigns launched before the AGM significantly increase shareholder proposal success and improve firms’ environmental and social performance. We develop a dynamic model in which NGOs trade off awareness building and credibility formation, generating a \auditrev{life cycle} in activism from visibility-seeking to influence-oriented engagement. Therefore, NGOs' objectives evolve endogenously to coordinate shareholder pressure and shape corporate behavior.
\\

\end{abstract}

\newpage\onehalfspacing
\section{Introduction}

Non-governmental organizations (NGOs) increasingly shape firms’ environmental and social policies \citep[e.g.,][]{baron2001private, yaziji2009ngos}. Yet, because NGOs rarely disclose their internal decision-making, their strategic behavior remains largely unobserved \citep{maxwell2012economic}. This opacity has divided the literature into two competing views: NGOs are seen either as principled social advocates \citep{besley2005competition,benabou2006incentives,daubanes2019rise} or as agents of strategic capture \citep{grossman1994protection,bertrand2020tax,akey2021hacking}. We provide an alternative framework and empirical evidence showing that NGOs respond systematically to stakeholder incentives, coordinate their social demands and translate them into corporate action.

 \begin{figure}[!htb]
\centering
\caption{Timing of NGO campaigns and media attention around AGMs \label{fig:intro_fig}}
\begin{subfigure}{0.49 \linewidth}
    \centering
    \includegraphics[width=\linewidth]{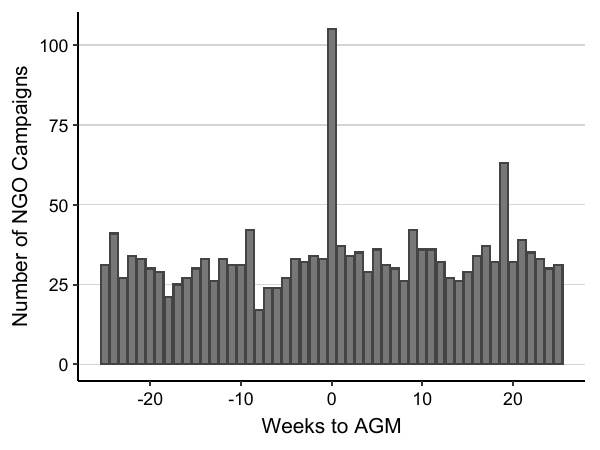}
    \subcaption{Campaigns peak during AGMs \label{fig:NGO_time}}
    \end{subfigure}
    \begin{subfigure}{0.49 \linewidth}
        \centering
    \includegraphics[width=\linewidth]{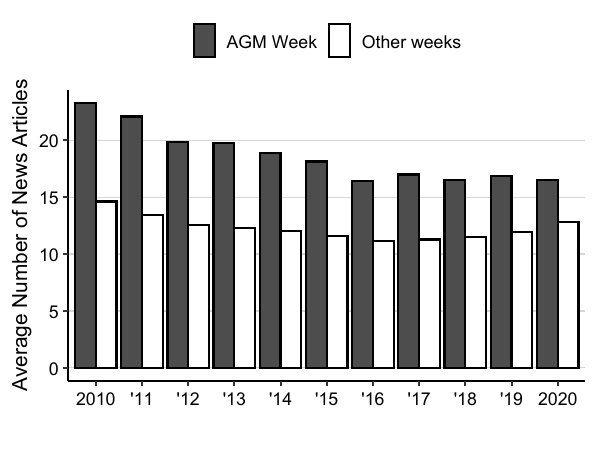}
    \subcaption{Media coverage rises during AGMs \label{fig:AGM_coverage}}
    \end{subfigure}
     \begin{minipage}{1 \textwidth} %
 {\footnotesize Note: Panel (a) plots the weekly frequency of NGO campaign launches relative to the target firm's nearest AGM (Week 0), \auditrev{for the 1,705 campaign records matched to an AGM within the plotted window of $-25$ to $+25$ rounded weeks.} Panel (b) compares average weekly firm news counts in AGM weeks (dark gray) and other weeks (white with black borders) by year, using RavenPack observations matched to ISS AGM dates. Data: Sigwatch, ISS, and RavenPack. \par}
 \end{minipage}
\end{figure}

A rare opportunity to observe these strategic responses arises from campaign timing. NGO campaigns must be public to be effective, and their timing provides one of the few visible choices NGOs make. We focus on how NGOs align campaigns with a moment when firms are uniquely exposed to outside scrutiny: the Annual General Meeting (AGM). AGMs are predictable focal events at which executives report results, shareholders vote, and media attention peaks \citep{fioretti2024shared}. Panel~(a) of Figure~\ref{fig:intro_fig} shows that NGO campaigns are about 3.3 times more likely to occur in AGM weeks than at other times, and Panel~(b) shows that media coverage of targeted firms peaks at the same moment.  These patterns raise a central question: do NGOs target AGMs primarily to maximize \emph{exposure}, or do they use campaign timing to exert \emph{influence} on firms? Is there a tradeoff between exposure and influence? More broadly, what incentives shape NGO strategy, and how do their choices affect the firms they target?

Leveraging the AGM as a predictable focal point, we address these questions and uncover a \auditrev{life cycle} in NGO activism. Younger NGOs concentrate on AGM-day campaigns that build visibility, while more established NGOs shift to pre-AGM campaigns that influence shareholder decisions and firms’ environmental and social performance. A simple dynamic model, in which NGOs trade off awareness building against credibility formation, rationalizes this timing pattern. Taken together, our results show that NGO behavior is governed by incentives that determine which stakeholders are mobilized and when, and thus whether campaigns generate symbolic exposure or substantive corporate change. Empirically, the effectiveness of influence-oriented campaigns depends on the ability to align dispersed shareholder responses, pointing to coordination frictions in corporate governance. 

Imagine an NGO uncovers troubling information about a firm and seeks to change its behavior. Because AGMs occur at predictable times, the NGO can choose whether to campaign before the AGM (when shareholder votes and negotiations are still open) or at the AGM itself, when public and media attention peak but most proxy votes have already been cast. If NGOs simply campaigned as soon as concerns arose, we would not expect the sharp clustering of campaigns around AGMs shown in Figure~\ref{fig:intro_fig}. The pronounced spike instead suggests that campaign timing is a strategic choice.

To study this choice, we assemble a novel dataset of 2{,}350 campaigns by international NGOs targeting U.S.\ firms, linked to media coverage, Google search activity, NGO financial statements, shareholder proposals, and environmental and social (E\&S) outcomes. SEC rules tie proxy filings and AGM dates to the fiscal year, so each firm’s AGM recurs within a narrow window each year, creating a predictable surge in visibility alongside a fixed deadline for shareholder voting. This institutional structure generates sharp variation in when campaigns occur relative to when shareholder decisions are made, which we exploit to distinguish exposure-oriented from influence-oriented activism and their drivers.

\begingroup\color{black}%
Our first set of results shows that campaign timing is tightly linked to how NGOs affect firms. When NGOs target the AGM date, they raise visibility for themselves and the issues they advocate (raising media coverage, online search activity, younger NGOs' operating margins, and the likelihood that related shareholder proposals are filed at the next AGM) but have little effect on contemporaneous proposal success or firms’ E\&S performance. This pattern suggests that shaming campaigns alone may have limited influence on corporate decisions. 
\endgroup%

\auditrev{NGOs' first few campaigns are disproportionately timed at the AGM.} As NGOs gain experience and reputation, campaigns shift into the months before the AGM, when proxy voting and negotiations are still open. These earlier actions significantly increase the withdrawal and vote-passage probabilities of E\&S proposals and are followed by improvements in firms’ E\&S scores, consistent with genuine changes in corporate policy rather than pure reputational damage.

To understand how NGO credibility affects firm performance, we distinguish between two primary channels. If NGOs operate through information provision, their influence should act symmetrically across firms, affecting all investors regardless of ownership structure. In contrast, if NGOs act as coordinators, their effectiveness should depend on a firm’s shareholder composition: coordination is less likely to emerge in highly concentrated firms where large owners act unilaterally, or in extremely fragmented firms where free-riding persists. \auditrev{Indeed, the filing response is strongest at firms with intermediate levels of ownership concentration, while the association between campaigning and proposal success weakens as concentration rises.} These patterns suggest that NGOs do not merely transmit information; they solve collective action problems to translate social pressure into substantive corporate change.

Our results are not explained by NGOs gradually moving from easier to harder targets: observable characteristics of target firms do not change systematically as NGOs accumulate campaigns. To address selection on unobservables, we exploit quasi-exogenous variation from the publication date of the \textit{Fashion Transparency Index}. The Index is released each year on a fixed date unrelated to firms' AGMs, which fixes the publication date and limits discretion over the timing of exposure. Using the distance between index publication and firms’ AGM dates as a source of exogenous timing variation, we find effects similar in sign and magnitude to those estimated in the full dataset.

We then show that NGOs’ use of AGM visibility versus pre-AGM influence evolves systematically over their organizational life cycle. NGOs with little prior AGM exposure or public visibility concentrate campaigns on the AGM date, where media attention peaks and visibility gains are largest. As NGOs accumulate visibility (proxied by past AGM campaigns, search intensity, and organizational size), they progressively shift campaigns earlier in the AGM cycle, when shareholder votes have not yet been cast and managerial concessions are more likely.

\begingroup\color{black}%
NGO income statement data point to different returns from campaigning early or late. For younger NGOs, AGM actions are associated with a sizable rise in operating margin, mostly driven by increased revenue. This is consistent with a financial benefit from visibility relative to its cost. For more established NGOs, the marginal fundraising benefit of AGM campaigns is small, and non-AGM campaigns do not produce comparable financial gains.
\endgroup%

The empirical patterns point to a sharp distinction between campaigns launched before the AGM and those timed on the meeting date. Campaigns before the AGM are associated with greater influence on shareholder votes and firm outcomes, while AGM-day campaigns primarily generate visibility. We interpret these findings through a dynamic model in which an NGO, shareholders, and the firm interact repeatedly over time, and in which campaign timing is the NGO’s central strategic choice.

In each period, the public state is summarized by two variables. The first is an awareness stock, which captures how salient the issue is among dispersed shareholders and the broader public. Awareness increases with highly visible actions (most notably campaigns timed on the AGM) and decays over time in the absence of renewed attention. The second is a credibility belief, which reflects stakeholders’ assessment of the NGO’s resolve and effectiveness in sustaining pressure on firms. This belief evolves endogenously as shareholders and firms observe the NGO’s actions and their outcomes.

Given the current state, the NGO chooses whether to campaign before shareholder votes are cast or at the AGM. AGM-day campaigns maximize visibility and generate warm-glow benefits for the NGO, but they occur after most proxy votes are locked in and therefore have little contemporaneous influence on firm decisions. Early campaigns, by contrast, are less visible and forgo some warm-glow benefits, but they can affect shareholder voting and negotiations with the firm. Importantly, early campaigns are also more informative: they require sustained engagement and, when successful, reveal persistence and competence, strengthening the NGO’s credibility. AGM-day campaigns are comparatively cheap and visible, but once an NGO is established they convey little new information about its type.

Shareholders observe campaign timing and, when relevant, campaign outcomes. Based on the current levels of awareness and credibility, they decide whether to file a proposal and how much support to give it. Higher awareness increases the likelihood that shareholders notice and coordinate around an issue, while higher credibility raises the expected effectiveness of activism and the probability of success. The firm, anticipating these responses, adjusts its behavior accordingly.

This dynamic interaction yields a simple equilibrium structure. When credibility and awareness are low, the NGO optimally campaigns at the AGM: visibility and warm-glow benefits dominate, and early engagement is unlikely to be persuasive. As awareness accumulates and credibility improves, the NGO crosses a threshold at which the influence and reputational gains from early campaigning outweigh the loss in visibility. At that point, it switches to pre-AGM campaigns to exert direct pressure on shareholder decisions and firm policies. This cutoff behavior reflects evolving stakeholder expectations: actions that initially signal resolve later appear routine, leading established NGOs to rely less on public shaming and more on early, substantive engagement \citep{baur2011moral,mitchell2017reputations}.

A purely visibility-driven model would predict persistent AGM-day campaigning. Matching the observed shift from late, high-visibility actions to earlier, influence-oriented engagement instead requires a dynamic reputation-building mechanism in the spirit of intertemporal signaling models \citep{kreps1982reputation,milgrom1982limit,cabral2010dynamics}, combined with a deterministic awareness stock that separates visibility-driven exposure from credibility-driven influence. This framework provides a unified explanation for why NGOs initially seek attention and later convert reputation into corporate impact.

Our analysis yields three main contributions. First, we offer a new quantitative framework to examine NGO strategy and to infer how NGOs respond to incentives, which stakeholders they prioritize at different stages, and how their objectives evolve. While \citet{lo2025risk} focuses on NGO financing and financial management, our focus is on how stakeholder interpretation and credibility concerns shape the evolution of NGO objectives. Classic corporate-finance approaches tie organizational objectives to ownership structure \citep[e.g.,][]{grossman1977value,grossman1979theory}, while stakeholder theory emphasizes salience based on power, legitimacy, and urgency \citep{freeman1984strategic,mitchell1997toward}. Our results show that NGO objectives shift with how donors and investors interpret NGOs’ actions and with credibility concerns about timing. In this sense, we provide a quantitative setting to study how organizational objectives emerge from interactions with multiple stakeholders rather than being fixed by a single principal, for example, an entrepreneur defining the firm’s goals \citep[e.g.,][]{scott2002love,romer2006firms,duarte2025competitive,fioretti2024shared}.\footnote{Our approach also relates to work on the strategic timing of actions around news cycles and focal dates \citep{eisensee2007news,madestam2013do,durante2018attack,djourelova2022media}, but shifts the focus to NGOs in the corporate domain and to the management of reputational capital over time.}

Second, while firms often respond to ESG pressure, recent work shows that such responses can be largely symbolic or involve asset reallocation rather than changes in underlying firm behavior \citep{duchin2025sustainability}. This raises the question of when external pressure leads to substantive corporate change rather than mere exposure or reputational adjustment. In this respect, we contribute to the corporate-finance and stakeholder-governance literature by identifying mechanisms through which firms internalize social and environmental externalities \citep{fioretti2022caring,allcott2022economic}. Our setting complements work on shareholder activism and gadflies \citep[e.g.,][]{brav2008hedge,gantchev2021costs,malenko2024information}: NGOs do not vote but act as external strategic coordinators that put issues on the ballot and help align dispersed investors’ behavior, shaping both the filing and success of shareholder proposals \citep[see also][]{brendel2024value,chuah2024shareholder,mazet2023some,mcgahan2023new}.

Third, our approach has implications for policy. For boards, early, credibility-backed campaigns are informative signals that warrant engagement rather than dismissal \citep{dimson2015active,kolbel2017media}. For investors and stewardship codes, credible NGOs supply information that helps coordinate dispersed shareholders on E\&S issues, complementing traditional voice and exit \citep{oehmke2020theory,gans2021exit,broccardo2020exit,saint2024exit,bonnefon2025moral}. For regulators, rules on AGM timing, proposal access, and disclosure shape the returns to NGO campaigns and thus the scope for private activism to substitute for, or complement, formal regulation.

The paper proceeds as follows. Section \ref{sec:theory} presents the model. Section \ref{s:data} describes the data, and Section \ref{s:strategy} outlines the empirical strategy. Sections \ref{s:media} and \ref{s:influence} examine NGO campaigns as tools for gaining media exposure and influencing firms through key stakeholders. Section \ref{s:reputation} brings these channels together by analyzing how NGO reputation enhances the success of shareholder proposals, and Section \ref{s:case} provides evidence from the quasi-experimental case study of the \emph{Fashion Transparency Index}. \auditrev{Section \ref{s:discussion} examines coordination as a mechanism and discusses implications for policy and future research}, and Section \ref{s:conclusion} concludes.

\section{A Simple Model}\label{sec:theory}

An NGO repeatedly chooses \emph{when} to campaign relative to a firm's annual general meeting (AGM), trading off \emph{exposure}, the visibility and warm glow of a high-salience AGM campaign, against \emph{influence}, the ability of an early campaign to sway votes and build a reputation for effectiveness. The trade-off produces a cutoff in the NGO's credibility: below it all NGOs pool on visible AGM campaigns; above it credible NGOs separate by campaigning early. This organizes the exposure-versus-influence distinction we take to the data; all derivations are in Online Appendix~\ref{app:proofs}.

\subsection{Setup}\label{subsec:environment}
An NGO of fixed, unobserved type $\theta\in\{C,W\}$ campaigns over discrete periods $t$. A \emph{credible} NGO ($C$) can coordinate investors and turn outreach into votes; a \emph{weak} one ($W$) cannot. The public state is $(A_t,\pi_t)$: awareness $A_t$ (issue salience among dispersed shareholders) and the belief $\pi_t\in[0,1]$ that the NGO is credible. Each period a shareholder files an AGM proposal if its perceived success covers a fixed cost $f$; the NGO chooses timing $d_t\in\{\mathrm{early},\mathrm{AGM}\}$ (early precedes the proxy vote, AGM arrives once votes are cast); an outcome $y_t\in\{0,1\}$ is realized; and the state updates. Awareness accumulates with the visibility of the chosen timing $v(d_t)$, with $v(\mathrm{AGM})>v(\mathrm{early})$, so that AGM campaigns build more of it,
\begin{equation}
A_{t+1}=\rho A_t+(1-\rho)\,v(d_t),\qquad \rho\in(0,1).
\label{eq:AwarenessLaw}
\end{equation}
 The per-period payoff trades exposure against influence,
\begin{equation}
u^\theta(d_t;\pi_t,A_t)=\underbrace{\alpha\,v(d_t)}_{\text{exposure}}+\underbrace{B\,\Gamma^\theta_{d_t}(\pi_t,A_t)}_{\text{influence}}-K,
\label{eq:utility_new}
\end{equation}
where $\alpha>0$ scales the NGO's return to visibility (warm glow, and the fundraising and legitimacy it brings) and $B>0$ its return to material influence over the firm, that is, mission fulfillment.\footnote{\label{fn:ngoobj}That NGOs weigh both public visibility and direct influence is a recurring theme in the literature on NGO objectives \citep{maxwell2012economic,yaziji2009ngos}; the visibility return reflects the donations, funding, and legitimacy that salience brings \citep{aldashev2009ngos,cooley2002ngo}, which we document in Section~\ref{s:donations}, and the influence return the capacity to change firm behavior studied in private politics \citep{baron2001private,daubanes2019rise}. Both are visible in practice: NGOs campaigning at Inditex's annual general meeting traded on visibility \citep{cleanclothes2025inditexagm}, whereas Starbucks's sustained engagement with NGOs works through direct influence \citep{argenti2004collaborating}. Section~\ref{s:case} develops one campaigner, Fashion Revolution, as a full case study of these mechanisms.} The influence technology carries the model's key restriction: only credible NGOs gain from acting early, and influence is realized through shareholders' response, so it rises with the public belief and vanishes when no one believes the NGO: $\Gamma^C_{\mathrm{early}}(\pi,A)>0$ for $\pi>0$, $\Gamma^C_{\mathrm{early}}(0,A)=0$, $\Gamma^C_{\mathrm{AGM}}\equiv0$, and $\Gamma^W\equiv0$ (Appendix~\ref{app:functional}).\footnote{Only the \emph{ranking} matters for separation: credible NGOs are more effective early than weak ones, $\Gamma^C_{\mathrm{early}}>\Gamma^W_{\mathrm{early}}\ge0$, and $\Gamma^W\equiv0$ is the cleanest case. $\Gamma^C_{\mathrm{AGM}}=0$ reflects that AGM-day campaigns arrive after most proxy votes are cast and so cannot sway the current vote. Allowing $\Gamma^W_{\mathrm{early}}>0$ adds a high-credibility region in which both types campaign early but leaves the empirically relevant transition, from pooled AGM campaigning to early campaigning by credible NGOs, unchanged.} The fixed campaign cost $K\ge0$ is common to both timings, so it drops out of the timing comparison.\footnote{Equal costs across timings are a normalization: a timing-specific cost, such as the travel an AGM-day action requires, would merely add a constant to the early-versus-AGM comparison.}

A filed proposal succeeds ($y_t=1$) or fails ($y_t=0$) with probability $S^{\theta}(A_t,\pi_t,d_t)$, the chance a type-$\theta$ NGO's campaign carries the vote; it rises with awareness, credibility, and early influence (logistic form in Appendix~\ref{app:functional}). Because shareholders do not observe $\theta$, they act on the belief-weighted success probability: a representative shareholder files only when the expected benefit covers the cost,
\begin{equation}\label{eq:FilingRule}
V_P\cdot S(A_t,\pi_t,\hat{d}_t)\ge f,\qquad S(\cdot)=\pi_t\,S^C(\cdot)+(1-\pi_t)\,S^W(\cdot),
\end{equation}
where $V_P$ is the value of a successful proposal and $\hat d_t$ the timing the shareholder anticipates. Filing therefore rises with awareness and credibility, and the realized outcome $y_t$ feeds back into the public belief $\pi_{t+1}$ about the NGO's type (Section \ref{subsec:eqbehavior_new}).

\subsection{The Timing Trade-off}\label{subsec:eqbehavior_new}
Timing and proposal outcomes update shareholders' beliefs about the NGO's type on two margins (Appendix~\ref{app:beliefupdate_new}). First, the timing choice is itself a signal: because acting early is more valuable to a credible NGO once credibility and awareness are high enough, early timing becomes favorable news about the NGO in that region, while AGM timing can be more favorable when awareness is still low. Second, the proposal outcome is a signal: at the AGM neither type has influence ($\Gamma^{\theta}_{\mathrm{AGM}}\equiv0$), so $S^{C}=S^{W}$ and the outcome is uninformative; early, the credible type is more effective ($S^{C}>S^{W}$), so a success raises $\pi_{t+1}$ while a failure lowers it. Early campaigns are therefore reputation-building experiments. Trading the immediate payoff against this evolving reputation, and discounting the future at $\beta\in(0,1)$, the NGO's optimal timing follows the action-value difference between campaigning early and at the AGM,
\begin{equation}
\Delta(\pi_t,A_t)=\underbrace{\alpha\, \big(v(\mathrm{early})-v(\mathrm{AGM})\big)+B\,\Gamma^{\theta}_{\mathrm{early}}(\pi_t,A_t)}_{\text{current wedge}}+\underbrace{\beta\, \big(\mathbb{E}_y V^\theta_{\mathrm{early}}-\mathbb{E}_y V^\theta_{\mathrm{AGM}}\big)}_{\text{continuation wedge}}.
\label{eq:Delta_new}
\end{equation}
The continuation wedge compares the NGO's expected value next period under the two timings. Here $V^\theta(\pi,A)$ is the type-$\theta$ value function, and $\mathbb{E}_y V^\theta_{d}$ is its expectation over next period's state $(\pi_{t+1},A_{t+1})$ when the current campaign is timed $d$: awareness moves deterministically according to \eqref{eq:AwarenessLaw}, while the outcome $y_t$, the model's only random element, moves the public belief to $\pi_{t+1}$, so $\mathbb{E}_y$ averages over $y_t$ under the NGO's own type. Appendix~\ref{app:proofs} gives the recursion for $V^\theta$.
This wedge, for the credible type, is strictly increasing in $\pi$ and weakly increasing in $A$ under the maintained assumptions in Appendix~\ref{app:equilibrium}: visibility dominates at low credibility ($\Delta<0$, so $d_t=\mathrm{AGM}$), while influence and reputation dominate at high credibility ($\Delta>0$, so $d_t=\mathrm{early}$).

\begin{proposition}[Separation by Campaign Timing]\label{prop:separation}
There is a unique cutoff $\bar\pi(A)\in(0,1)$, weakly decreasing in $A$. The credible NGO chooses $d^{C\star}_t=\mathrm{AGM}$ when $\pi_t<\bar\pi(A_t)$ and $d^{C\star}_t=\mathrm{early}$ when $\pi_t>\bar\pi(A_t)$, while the weak NGO chooses $d^{W\star}_t=\mathrm{AGM}$ at every state. Thus, below the cutoff both types pool on AGM campaigns; above it the credible NGO separates by campaigning early while the weak NGO stays at the AGM.
\end{proposition}

\noindent\textit{Proof.} Appendix~\ref{app:equilibrium}. 

Intuitively, at low credibility early action is unpersuasive and AGM visibility dominates for both types; as credibility rises, the credible NGO's influence and reputational payoff overtake visibility, while the weak NGO ($\Gamma^W_{\mathrm{early}}=0$) gains nothing, not even from mimicry, since its payoff does not rise with being believed credible (Appendix~\ref{app:equilibrium}), and stays. Because awareness lowers the threshold, the switch comes sooner for better-known issues, a \auditrev{life cycle} from visibility-seeking to influence.

\subsection{Predictions}\label{subsec:predictions}\label{subsec:discussion_theory}%
This \auditrev{life cycle} (young, low-credibility NGOs cluster at AGMs for visibility, mature and credible ones campaign early for influence) yields three testable predictions. AGM campaigns are about \emph{exposure}: they raise awareness, and with it the funding that salience brings, and the higher awareness makes a related proposal more likely at the \emph{next} AGM through the filing rule \eqref{eq:FilingRule}, most for credible NGOs whose perceived success already sits near the filing threshold (Section~\ref{s:media}). Early pre-AGM campaigns are about \emph{influence}: working through the credibility channel, they raise contemporaneous proposal success and subsequent E\&S performance, and their outcomes move beliefs, whereas AGM campaigns, with $\Gamma^{\theta}_{\mathrm{AGM}}\equiv0$, do not (Section~\ref{s:influence}). Finally, timing shifts over the NGO's life: as credibility and awareness accumulate the cutoff $\bar\pi(A)$ falls, so NGOs move from AGM to early campaigns, with early successes speeding the transition (Section~\ref{s:reputation}). The structure follows reputation-for-toughness models \citep{kreps1982reputation,milgrom1982limit}: early campaigning is the costly signal that trades short-run visibility for long-run influence.

 \section{Data}\label{s:data}
Our dataset integrates several sources: NGO campaigns, online attention, shareholder proposals, and firm-level news.  

First, we use data from \textit{Sigwatch}, a consulting firm tracking global NGO campaigns against corporations from 2010 to 2021 \citep{hatte2020geography}. Sigwatch compiles campaigns from the pressure groups' own websites and social media, and Appendix~\ref{ap:texts} describes its data and methodology. The data record the campaign’s public release date, cause (e.g., ``GMOs in food''), the targeted company and its prominence, and sentiment (ranging from --2 to +2, typically negative). We focus on single-NGO, single-firm, negative-sentiment campaigns where the firm appears in the headline.\footnote{Since our analysis links campaign timing with a firm’s AGM date, we exclude multi-firm campaigns as their AGM dates may differ. Appendix~\ref{app:coalitions} separately examines solo and collaborative campaigns.} The assigned campaign dataset contains 2,350 campaign records associated with 635 NGOs and 513 companies. 

Second, we collect monthly Google Trends scores for all NGOs in our sample from 2010 to 2022. Scores are normalized on a 0--100 scale, either (i) relative to the NGO’s own peak searches or (ii) relative to a benchmark NGO in the same month. We use Greenpeace (the most searched NGO) as the benchmark for comparability across organizations.  

Third, we use data from \textit{Institutional Shareholder Services} (ISS) on proposals at the Annual General Meetings (AGMs) of the 3,000 largest US-listed firms. We select 4,483 environmental and social (E\&S) proposals filed between 2010 and 2021 (836 firms in total). About one-third of these proposals are successful, either withdrawn before the AGM (after firm commitments) or passed in a vote. Both the number of E\&S proposals and their success rate increased steadily over the period.  

Fourth, we obtain firm-level news from \textit{RavenPack}, which transforms unstructured news into structured data with sentiment and event tagging.\footnote{RavenPack News Analytics assigns tone scores to news articles drawn from over 40,000 sources, such as Dow Jones Newswires, the Wall Street Journal, MarketWatch, Factiva, and Barron’s, and identifies articles linked to specific events. The validity of this score has been established in prior research \citep[e.g.,][]{bushman2024influence}. The dataset also differentiates between media-generated coverage and firm-issued press releases, enabling us to focus on the former.} We count news items related to US-listed firms from 2010 to 2020, retaining only items with a relevance score of at least 75\%. The news panel matched to AGM dates for Figure~\ref{fig:intro_fig} covers 5,045 firms.

Fifth, we obtain firm-level accounting data from Compustat and institutional ownership data from 13F filings. For each firm-year, if the reported ownership stakes sum to more than 100\%, we proportionally rescale each investor’s ownership share so that total ownership equals 100\%. We then compute the Herfindahl–Hirschman Index (HHI) of ownership concentration, along with the fraction of equity held by blockholders (stakes exceeding 5\%) and the number of such blockholders.

\begingroup\color{black}%
Lastly, we collect balance sheet and income statement data for \auditrev{U.S.} NGOs in the sample from their Internal Revenue Service Form 990, available on the National Center for Charitable Statistics website. Form 990 is the annual information return that most tax-exempt nonprofit organizations in the United States must file with the IRS to disclose their finances, governance practices, and programmatic activities. Filing obligations are primarily determined by organizational size and type: organizations subject to these filing requirements generally file the full Form 990 if gross receipts are at least \$200,000 or total assets are at least \$500,000. Our starting sample contains 635 NGOs, of which 216 are classified as \auditrev{U.S.} NGOs in Sigwatch (Table~\ref{tab:selection_us_ngos}). Of these 216 NGOs, 149 could be matched to a recorded Employer Identification Number (EIN), and 145 have financial observations in the linked IRS data. 
\endgroup%

We link Sigwatch, ISS, and RavenPack data using standard identifiers (CUSIP, ISIN, year). Table \ref{tab:summary} reports descriptive statistics. We define an AGM-day campaign as a campaign launched on the target firm’s AGM date or on the calendar day immediately preceding it. This window captures campaigns released just ahead of the meeting while still treating the AGM as the focal event. Panel (a) reports a \auditrev{mean of 3.70 campaigns per NGO}, though the median campaign count is one. Most campaigns focus on social rather than environmental issues, but the distribution is highly skewed: a small set of NGOs accounts for a disproportionate share of activity. The top 50 most active NGOs account for almost 50\% of all campaigns, with the most active ones being Sierra Club, PETA, and Greenpeace, with more than 60 campaigns each.

 Panel (b) documents substantial variation across targeted firms in size and news coverage. Targets range from very large firms with extensive media presence to smaller firms with limited coverage, suggesting that NGOs are not exclusively focused on high-visibility corporations. 1,299 campaigns (55.3\%) concern social causes and 1,051 (44.7\%) concern environmental causes. We return to these summary statistics in Section \ref{s:media}, after first providing background on AGM dates.

\begin{table}[!htb]
\color{black}
    \caption{Summary Statistics \label{tab:summary}}
\begin{center}

\resizebox{0.85\textwidth}{!}{
\begin{tabular}{lccccc}
\toprule
\midrule
\multicolumn{6}{l}{\textbf{Panel a: NGO and campaign characteristics}} \\
\midrule
 & Mean & 25\% & 50\% & 75\% & 95\% \\
Number of campaigns by NGO & \auditrev{3.70} & 1.00 & 1.00 & \auditrev{3.00} & \auditrev{13.00} \\
-- Environmental campaigns & \auditrev{1.66} & 0.00 & 1.00 & 1.00 & 5.30 \\
-- Social campaigns & \auditrev{2.05} & 0.00 & 1.00 & 2.00 & \auditrev{9.00} \\
-- On AGM date & \auditrev{0.06} & 0.00 & 0.00 & 0.00 & \auditrev{0.00} \\
Campaign sentiment & \auditrev{-1.70} & -2.00 & -2.00 & -1.00 & -1.00 \\
\auditrev{U.S.} NGO (0/1) & \auditrev{0.34} & 0.00 & 0.00 & 1.00 & 1.00 \\
\auditrev{U.S.} NGOs: number of campaigns & \auditrev{6.31} & 1.00 & \auditrev{1.50} & \auditrev{6.00} & \auditrev{30.50} \\
-- Environmental campaigns & \auditrev{2.98} & 0.00 & 1.00 & 2.00 & \auditrev{11.50} \\
-- Social campaigns & \auditrev{3.33} & 0.00 & 1.00 & 3.00 & 14.00 \\
-- On AGM date & \auditrev{0.16} & 0.00 & 0.00 & 0.00 & 1.00 \\
\auditrev{U.S.} NGOs: campaign sentiment & \auditrev{-1.67} & -2.00 & -2.00 & -1.00 & -1.00 \\
NGO total assets (USD mn) & 45.44 & 2.00 & 5.49 & 35.05 & 247.46 \\
Yearly contributions (USD mn) & 33.61 & 2.10 & 7.23 & 36.67 & 143.09 \\
Operating margin & \auditrev{0.04} & \auditrev{-0.06} & \auditrev{0.06} & \auditrev{0.20} & \auditrev{0.48} \\
NGO age (years) & 39.74 & 20.00 & 29.00 & 42.00 & 126.00 \\
\midrule
\multicolumn{6}{l}{\textbf{Panel b: Target-firm characteristics}} \\
\midrule
 & Mean & 25\% & 50\% & 75\% & SD \\
Market capitalization (USD bn) & 184.45 & 24.93 & 88.45 & 210.80 & 270.84 \\
Annual number of news articles & 5,162.23 & 1,504.25 & 2,800.00 & 5,191.00 & 6,222.46 \\
Environmental score & 5.48 & 3.70 & 5.50 & 7.10 & 2.21 \\
Social score & 4.30 & 3.20 & 4.40 & 5.50 & 1.54 \\
Shareholder HHI & 0.027 & 0.013 & 0.017 & 0.025 & 0.067 \\
Blockholding & 0.191 & 0.116 & 0.174 & 0.229 & 0.156 \\
Blockholder count & 2.520 & 2.000 & 2.000 & 3.000 & 1.457 \\
\bottomrule
\end{tabular}}
\end{center}

\begin{tablenotes}[flushleft,para]
\footnotesize\item \auditrev{Note: Campaign counts and the U.S. share use all 2,350 campaigns and 635 NGOs (216 U.S. NGOs). Totals may differ because of rounding. Operating margin is revenue minus expenses, divided by revenue. Sources: Sigwatch, IRS Form 990 / Urban Institute, MSCI, 13F, RavenPack, and Compustat. NGO = nongovernmental organization; AGM = annual general meeting; SD = standard deviation; HHI = Herfindahl--Hirschman index.}
\end{tablenotes}
\end{table}

\begingroup\color{black}%
\section{Empirical Strategy}\label{s:strategy} 
Our empirical strategy exploits the timing of NGO campaigns relative to AGMs, leveraging the fact that AGM dates are largely exogenous and stable within firms over time due to SEC regulations. During our sample period, Rule 14a-8 generally required proposals for regularly scheduled meetings to be received at least 120 calendar days before the anniversary of the previous year's proxy-statement release date, effectively anchoring AGM dates across consecutive years \citep{glac2014influence,fioretti2024shared}. Appendix~\ref{app:agmdates} illustrates this stability by comparing each firm’s AGM month across consecutive years. Most observations lie along the diagonal, indicating strong persistence of the AGM month.%
\endgroup%

This predictability is central to our identification strategy. Because AGM dates are fixed well in advance and rarely change, NGOs face an exogenous schedule of heightened public attention: they can either react immediately to external events or strategically align their campaigns with these predetermined meetings, when firms are most exposed to shareholder and media scrutiny. We put two questions to this timing: how it shapes what a campaign achieves, its visibility, the stakeholder response, and the firm's conduct, and how NGOs choose it as they gain reputation. The first is a family of event studies; the second, a regression of the timing choice on reputation.

For the first question we estimate stacked event studies \citep{baker2022much},
\begin{equation}\label{eq:eventstudy}
 Y \;=\; \sum_{\tau} \beta_\tau^{k}\,\mathbf{1}\{T_c=\tau\}\;+\;\delta\;+\;u,
\end{equation}
where $T_c$ locates campaign $c$ in event time relative to a reference event and $\delta$ collects fixed effects. Which event depends on the outcome. Most outcomes are anchored to the AGM, whether a related proposal is filed at the next meeting, whether it succeeds, or how the firm's conduct changes over the following year, so event time measures a campaign's distance to the AGM. A campaign's AGM timing is then a \emph{position} on the path: $\tau=0$ is a campaign launched at the meeting, and earlier $\tau$ a campaign launched before it. The coefficients $\beta^{k}_\tau$ trace how the outcome moves with that timing, separately for each campaign group $k$ (an NGO's early versus later campaigns), and we fix the outcome, window, groups, and fixed effects where each regression is used.

Visibility is the exception. Google searches and negative news respond to the campaign itself, in the weeks after it breaks, so there we date event time from the campaign launch rather than the AGM. Every campaign now sits at $\tau=0$ at its own launch, so its AGM timing is no longer a position on the path but an \emph{attribute} of the campaign; to compare AGM-timed campaigns with the rest, we interact an AGM indicator with the event-time dummies,
\begin{equation}\label{eq:awareness}
 Y \;=\; \sum_{\tau} \big(\beta_\tau^{AGM}\,\mathrm{AGM}_c\;+\;\beta_\tau^{Non\text{-}AGM}\big)\,\mathbf{1}\{T_c=\tau\}\;+\;\delta\;+\;u,
\end{equation}
where $\mathrm{AGM}_c$ marks campaigns launched on the target's AGM date or the preceding day: every campaign traces the benchmark path $\beta_\tau^{Non\text{-}AGM}$, and $\beta_\tau^{AGM}$ is the extra visibility an AGM-timed campaign generates over it, which the analysis of Section~\ref{s:media} reports.

The model further implies that NGOs campaign earlier as their reputation grows. Turning from the consequences of timing to the choice of it, we take a campaign's distance to the AGM as the outcome and relate it to a predetermined measure of the NGO's reputation,
\begin{equation}\label{eq:timing}
 \text{Distance to AGM}_c \;=\; \sum_{\tau} \beta_\tau\,\mathbf{1}\{R_{g,c}=\tau\}\;+\;\delta\;+\;u.
\end{equation}
For campaign-order indicators, $R_{g,c}$ identifies the focal campaign in the NGO's cumulative sequence. When used as a reputation proxy, experience counts include campaigns through the previous calendar year. Other proxies such as Google visibility or financial size are also lagged one year.

Identification rests on the exogeneity of AGM dates: $T_c$ reflects an NGO's choice of when to campaign rather than a firm response, and the fixed effects absorb the main confounders, namely time-varying issue salience, firm shocks, and NGO maturity.

\subsection{Selection}\label{s:target}
A potential concern is that NGOs might select target firms strategically to maximize either visibility or credibility rather than choosing campaign timing exogenously. For instance, NGOs could preferentially target firms that are already more visible, frequently covered in the media, or more responsive to shareholder pressure.

\begingroup\color{black}%
Such selection would leave traces in who campaigns against whom: experienced NGOs might pick different targets from first-time campaigners, and larger or \auditrev{U.S.} NGOs might target different firms. Appendix~\ref{apndx:whom} finds limited evidence of selection on observables.
\endgroup%

The evidence supports our interpretation that campaign timing reflects strategic decisions by NGOs rather than endogenous target selection. However, while our analysis finds limited evidence of selection on observables, unobserved factors may still affect campaign timing, a possibility we address in Section~\ref{s:case}, where we exploit quasi-exogenous variation in campaign dates.

These tests concern selection among NGOs that already campaign. A prior question is which nonprofits become activists at all: our sample is not the full nonprofit population, and many nonprofits, such as hospitals, universities, and churches, never campaign. Appendix~\ref{apndx:selection} compares our activist NGOs with the universe of \auditrev{U.S.} nonprofits that file Form~990 (Figure~\ref{fig:form990}) and finds they differ systematically, being younger but larger, more concentrated in Washington~DC, more often environmental and public-benefit organizations, and far more reliant on donations. These differences describe the population our results speak to.

\begingroup\color{black}%
\subsection{NGO Experience and Time-Varying Confounders}
A natural concern is that our timing patterns reflect NGO maturity or other time-varying NGO attributes (e.g., staffing, funding, or reputation) rather than strategic responses to stakeholder attention. Our proposal-success specifications in Equation~\eqref{eq:main_reg} address this by including firm-by-NGO-by-year fixed effects, which absorb any NGO-level factor common to all campaigns conducted by the same NGO in a given year, as well as annual conditions specific to its relationship with the target firm. Identification therefore comes from variation in campaign timing and issue alignment within these cells, net of topic-by-meeting-month-year fixed effects that account for changing issue salience.

In addition, the baseline campaign-order regression similarly includes NGO-by-firm-by-year and topic-by-year fixed effects. This maps directly to the model: awareness and credibility are stocks that build over an NGO's life, so these fixed effects absorb their yearly levels without ruling out differences in how reputation shapes the effectiveness of same-topic campaigns. We capture those differences through interactions between prior reputation and issue alignment in the proposal-success regressions. Separately, the event studies compare responses to young and mature NGOs, using the outcome-specific fixed effects described in their notes.

Together, these comparisons trace the model's shift from exposure to influence as reputation grows.
\endgroup%

\section{Campaign Timing and Exposure} \label{s:media}

The model predicts that campaigns timed on AGM dates generate a larger visibility shock than early campaigns. Formally, $v(\mathrm{AGM}) > v(\mathrm{early})$ in the awareness law of motion~\eqref{eq:AwarenessLaw}. Because neither the latent visibility inputs  $v(\cdot)$ nor the awareness stock $A_t$ are observed directly, we test this implication using  observable proxies that capture different components of real-world awareness formation.  Google search activity for the NGO reflects \emph{demand-based attention}, measuring how many  individuals actively seek information about the organization. Negative news coverage of the  target firm reflects a \emph{supply-side media response}, capturing how much information is  produced and disseminated about the campaign. While the model remains agnostic about the  microfoundations of awareness, it implies that AGM-timed campaigns should generate larger visibility responses in these variables than campaigns launched at other times. 

\begin{figure}[!ht]
\color{black}%
     \centering
          \caption{NGO Visibility and media coverage around AGM campaigns} 
          \label{fig:gtrends_campaigns}
     \begin{subfigure}{0.49 \linewidth}
    \centering\includegraphics[width=\textwidth]{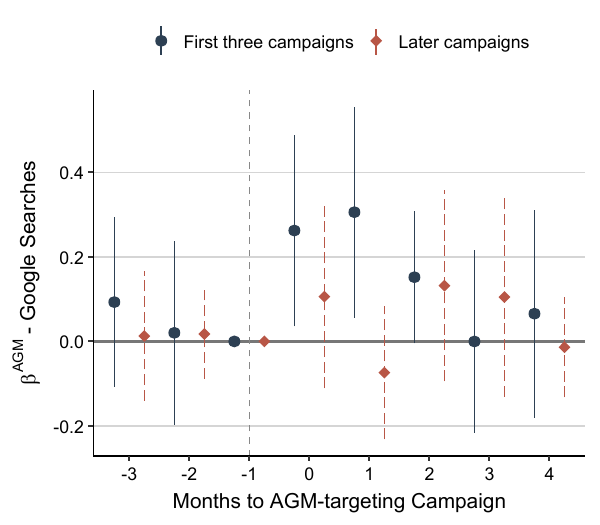}
    \subcaption{Campaign timing and campaigning NGO's  Google Trends scores}
 \end{subfigure} \hfill
  \begin{subfigure}{0.49 \linewidth}
    \centering\includegraphics[width=\textwidth]{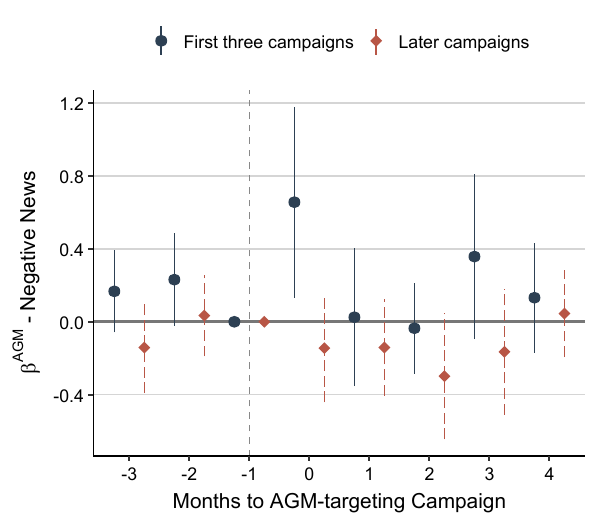}
   \subcaption{Campaign timing and number of negative E\&S news articles about target firms (standardized)}
  \end{subfigure}\hfill
   \begin{minipage}{1 \textwidth} %
 {\footnotesize Note: Each panel plots the incremental AGM interaction $\beta^{AGM}$ from Equation~\eqref{eq:awareness}, relative to the all-campaign benchmark. Panel (a) uses standardized monthly NGO Google search activity. Panel (b) uses standardized monthly negative target-firm news counts, after upper-tail winsorization at the 97.5th percentile. Blue circles denote the first three campaigns and red diamonds later campaigns, estimated separately. The window is $[-3,4]$ months relative to the campaign and month $-1$ is omitted. Both specifications include calendar-month and NGO-by-campaign-date-by-year fixed effects. Bars show 95\% confidence intervals with standard errors clustered by NGO--campaign-date pair. Data: ISS, Sigwatch, Google Trends, and RavenPack. \par}
 \end{minipage}
 \end{figure}

To test this, we estimate \eqref{eq:awareness} with these awareness proxies as the outcome and event time counting months from the campaign launch ($\tau\in[-3,4]$, reference $\tau=-1$); Figure~\ref{fig:gtrends_campaigns} plots the AGM increment $\beta^{AGM}_\tau$. As NGOs campaign repeatedly, we stack all campaign events into an NGO--month panel. NGO-by-campaign-date-by-year and month-by-year fixed effects absorb campaign-specific annual characteristics and aggregate monthly shocks, and standard errors are clustered by campaign (an NGO--campaign-date pair).

The model implies that the impact of visibility shocks on awareness depends on an NGO’s position in the $(A_t,\pi_t)$ state space. To capture this heterogeneity, we estimate this specification separately for campaigns early and late in an NGO's trajectory: its first three campaigns, launched while awareness and reputation are still building (``young''), versus its fourth and later ones (``mature'').

Panel (a) of Figure~\ref{fig:gtrends_campaigns} reports the estimated $\beta^{AGM}_\tau$ using Google Trends as a proxy for demand-based attention. AGM campaigns by young NGOs generate a sizeable rise in search activity, though the estimates are imprecise. For mature NGOs, the AGM-related coefficients are small and statistically indistinguishable from zero, consistent with saturation of awareness: once $A_t$ is high, additional visibility shocks produce little incremental response. The visibility gains are specific to AGM timing: a campaign launched at other times moves search activity only marginally, and never significantly (Appendix Figure~\ref{fig:gtrends_campaigns_app}).\footnote{For an NGO's first three campaigns, the estimated AGM increments are 0.262, 0.306, and 0.152 standard deviations in months 0, 1, and 2; the month-2 confidence interval includes zero. In the pooled comparison in Appendix Figure~\ref{fig:gtrends_campaigns_app}, the non-AGM benchmark is not significantly different from zero. This matches the model's $v(\mathrm{AGM})>v(\mathrm{early})$ with a small $v(\mathrm{early})$.}

Panel (b) turns to a supply-side proxy for awareness: the volume of negative E\&S news about the target firm, winsorized at the 97.5$^{\text{th}}$ percentile. Here the contrast across NGO maturity is sharper. AGM campaigns by young NGOs produce a clear spike in negative media coverage in the campaign month, while mature NGOs exhibit no response. This pattern suggests that media outlets react more strongly to campaigns by less-established organizations (whose actions may be more ``newsworthy'').

The contrast between early and later campaigns also sheds light on the persistence of awareness. If awareness evolves according to $A_{t+1}=\rho A_t + (1-\rho)v(d_t)$, then the first few campaigns produce the largest increases in $A_t$ unless $\rho$ is very close to one. The empirical pattern (strong AGM effects for young NGOs and no detectable response for mature NGOs) accordingly points to an awareness process that converges relatively quickly.\footnote{To see this formally, suppose an NGO always chooses $d_t=\mathrm{AGM}$ and normalize $v(\mathrm{AGM})=1$, with initial awareness $A_0=0$.  Then awareness follows $A_{t+1}=\rho A_t + (1-\rho)$, which solves to $A_t = 1 - \rho^{\,t}$. Awareness therefore converges quickly when $\rho$ is moderate: for example, with $\rho=0.5$, the first three campaigns raise $A_t$ to $1-0.5^3 = 0.875$, leaving little room for additional campaigns to generate detectable increases in visibility. This illustrates why strong AGM effects appear for early campaigns but not for later ones.} Given the \auditrev{mean of 3.70 campaigns per NGO} in Table~\ref{tab:summary}, treating NGOs with four or more campaigns as ``mature'' is thus a reasonable way to proxy for having reached a high-awareness region of the state space.

Note that the estimates are obtained within NGOs: the sample split is not based on the total number of \auditrev{campaigns between 2010 and 2021}, but on each NGO’s early versus later campaigns: that is, all NGOs appearing in the red regressions also appear in the blue regressions. Thus, early in their \auditrev{life cycle}, NGOs benefit more from AGM timing to build visibility, but the returns to this tactic decline as their public recognition grows.

\paragraph{Robustness.} These results are \auditrev{robust to three main concerns}. First, the stacked event-study framework implicitly treats campaign timing as simultaneous across NGOs, whereas in reality campaigns are staggered across organizations and calendar months. This misalignment can distort estimates when treatment effects are heterogeneous across groups or over time. 

Second, some NGOs launch multiple campaigns within the same year, which may confound the analysis if two campaigns fall within five months of each other. To address both issues, Appendix \ref{apndx:specs} replicates the analysis using the staggered difference-in-differences estimator of \citet{sun2021estimating}, which explicitly accounts for variation in treatment timing, as well as a distributed-lag specification that flexibly controls for past campaigns within a year. Both approaches yield similar results, with substantially larger gains for AGM campaigns by younger NGOs. The same appendix replicates the exposure analyses on the full sample, without the maturity split.

\begingroup\color{black}%
Third, across these analyses we find no evidence of pre-trends: outcomes move
only after campaigns, not in anticipation of them, consistent with our identifying assumption that campaign timing is not driven by the outcomes studied.
\endgroup%

Next, we turn to the reaction of stakeholders to NGOs.

\subsection{Stakeholders' Reactions}\label{s:stakeholders}
We next examine how key stakeholders respond to NGO campaigns. In the model, shareholders react by filing proposals when awareness and credibility increase (Section \ref{s:shareholders}), while donors respond through warm-glow motives that translate into higher contributions (Section \ref{s:donations}). In addition, although consumer behavior lies outside the model, we explore whether campaign timing affects target firms’ revenues as a proxy for potential consumer backlash (Section \ref{s:consumers}).

\subsubsection{Shareholders} \label{s:shareholders}
We begin with shareholders. We test whether they are more likely to submit proposals on the same cause when an NGO campaigned around the firm’s AGM date in the previous year, as suggested by the filing rule \eqref{eq:FilingRule}. A proposal is defined as ``related'' when it addresses the same cause as the NGO campaign. As detailed in Appendix~\ref{apndx:match}, we identify related proposals by matching each campaign topic in the Sigwatch data to the topic of an AGM proposal at the same target firm in the following year, using ISS topic definitions. Whenever such a match occurs, we classify the shareholder proposal as being on the same topic as the earlier NGO campaign.

\begingroup\color{black}%
To examine this, we estimate the event study \eqref{eq:eventstudy} with an indicator for a related proposal at firm $n$'s next AGM as the outcome and the campaign's distance to that AGM as event time ($\tau\in[-15,-3]$, reference $\tau=-3$). Month $-3$ is the omitted category because, by then, the Rule 14a-8 deadline and typical advance-notice bylaw windows have generally passed, so later campaigns are generally too late to affect the focal filing. NGO-by-calendar-month and firm-by-year-by-topic fixed effects absorb time-varying NGO activity and firm-topic conditions, and standard errors are clustered by firm.
\endgroup%

\begin{figure}[!htbp]
\color{black}%
 \centering
 \caption{NGO campaigns and shareholder proposals}
     \label{fig:campaign_proposal}
 \begin{subfigure}[t]{0.49\textwidth} %
   \centering
   \includegraphics[width=\textwidth]{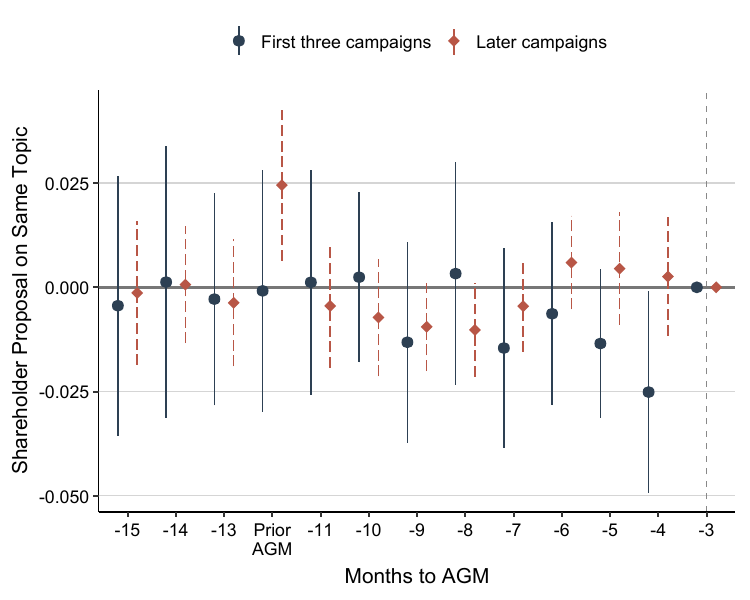}
   \subcaption{By past experience}
   
 \end{subfigure}
 \begin{subfigure}[t]{0.49\textwidth}
   \centering
   \includegraphics[width=\textwidth]{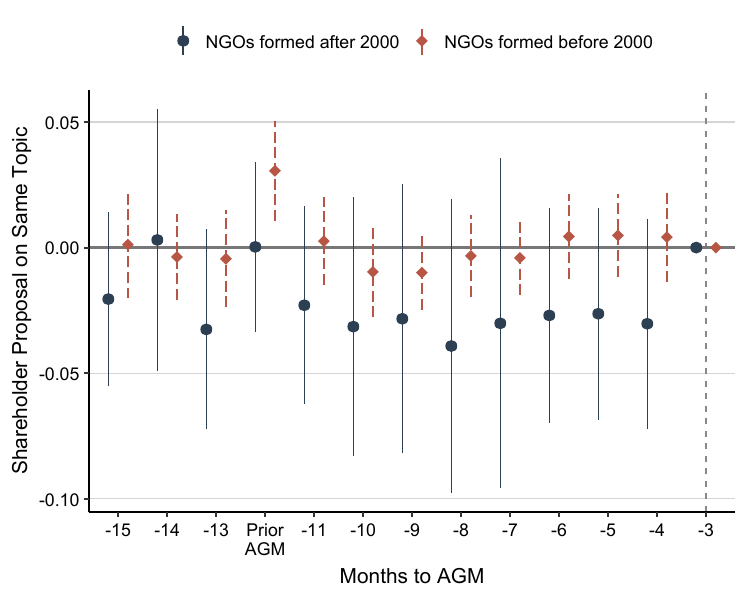}
   \subcaption{By NGO age}
 \end{subfigure}
 \begin{minipage}{\textwidth} %
   {\footnotesize Notes: The dependent variable indicates a shareholder proposal on the campaign topic. Event time is campaign timing relative to the prospective AGM; month $-12$ corresponds to the previous AGM. The window is $[-15,-3]$, with month $-3$ omitted. Both panels include NGO-by-calendar-month and firm-by-year-by-topic fixed effects, with firm-clustered standard errors. Panel (a) splits the sample into the NGO's first three and later campaigns. Panel (b) splits \auditrev{U.S.} NGOs by formation after 2000 versus in or before 2000. Error bars show 95\% confidence intervals. Data: ISS, Sigwatch, and IRS Form 990 / Urban Institute. \par}
 \end{minipage}
\end{figure}

Panel (a) of Figure~\ref{fig:campaign_proposal} compares an NGO’s first three campaigns (blue dots) with its later campaigns (red diamonds). AGM-timed campaigns (identified at $\beta_{-12}$) by more established NGOs increase the probability of a related shareholder proposal by roughly 2.45 percentage points, whereas AGM campaigns by younger NGOs generate no significant response.\footnote{The lag to the next meeting is also consistent with the campaigns NGOs launch just after the AGM, which can announce the actions they will bring to the following year (Appendix~\ref{ap:postagm}).} 

This pattern is consistent with the success function $S(\cdot)$ in \eqref{eq:FilingRule}: for more mature NGOs, beliefs $\pi_t$ are higher, so the expected success probability $\pi_t S^C + (1-\pi_t) S^W$ lies closer to the credible-type benchmark $S^C$, making shareholders more responsive to visibility shocks. Although fully credible NGOs would eventually shift entirely to early campaigns in the model, the mature NGOs in our data need not have reached that limiting region. AGM-timed campaigns by higher $\pi$ NGOs are therefore consistent with NGOs still transitioning toward higher credibility and awareness, for whom AGM visibility remains sufficiently informative to affect filing decisions. In contrast, the youngest NGOs are not credible enough to encourage shareholders to submit proposals.\footnote{Appendix Figure \ref{fig:campaign_proposal_all} replicates the analysis on the full sample, suggesting that more ``mature'' NGOs drive the effect we observe in the full sample.}

Panel (b) focuses on the subset of U.S.\ NGOs for which we observe balance-sheet data and 
incorporation year, allowing us to compare younger and more mature organizations based on year of incorporation. We classify as ``young'' (blue dots) those founded after 2000, corresponding to NGOs younger than ten years at the start of the sample. The pattern is similar to Panel~(a): a 3-percentage-point increase in filing rates appears at \auditrev{$\tau=-12$} for mature NGOs (red diamonds), whereas effects for younger NGOs are smaller and imprecisely estimated. Note that, as discussed earlier (see also Appendix~\ref{apndx:selection}), these NGOs are generally larger than those in Panel (a), which explains why we see some actions around the AGM also for younger NGOs. 

\subsubsection{Donors}
\label{s:donations}

We next ask whether AGM-timed campaigns improve the financial position of the campaigning NGO, using the \auditrev{U.S.} NGOs for which Form~990 balance sheets are available, and whether the response differs with NGO maturity. For these organizations, donations account for a large share of revenue. Because an AGM campaign can move both the funds an NGO raises and the costs it incurs (travel to the meeting, on-site action), and because we cannot allocate costs to specific months, we measure its financial effect by the operating margin, $(\text{Revenue}-\text{Expenses})/\text{Revenue}$. As we observe donations only annually and many NGOs run several campaigns per year, precluding the estimation of \eqref{eq:awareness}, we estimate the event study \eqref{eq:eventstudy} at the NGO--year level, with the operating margin as the outcome and the campaign's distance to the AGM as $T_c$ ($\tau\in[-5,5]$). No month is omitted: indicators for every distance enter jointly, so each coefficient compares NGO--years with a campaign at that distance to years without one. The regression controls for lagged log assets, absorbs NGO and year fixed effects, and clusters standard errors by NGO and year.\footnote{We retain years before 2022 with positive revenue and expenses, winsorize operating margin at the pooled 2.5th and 97.5th percentiles, and then require annual revenue of at least \$50,000.}

\begin{figure}[!t]
 \centering
 \caption{NGO campaign timing and financial gains to young and old NGOs}
   \label{fig:campaign_donations}
   
   \centering
   \begin{subfigure}{0.49\textwidth}
   \includegraphics[width=\linewidth]{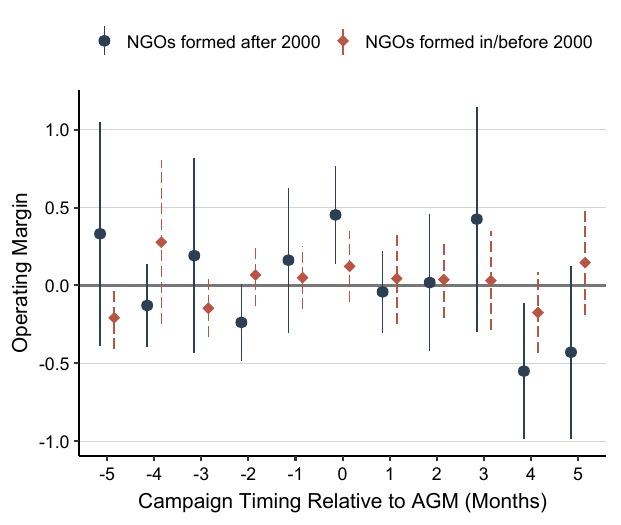}
   \subcaption{Operating margin by NGO age}
   \end{subfigure}\hfill
   \begin{subfigure}{0.49\textwidth}
   \includegraphics[width=\linewidth]{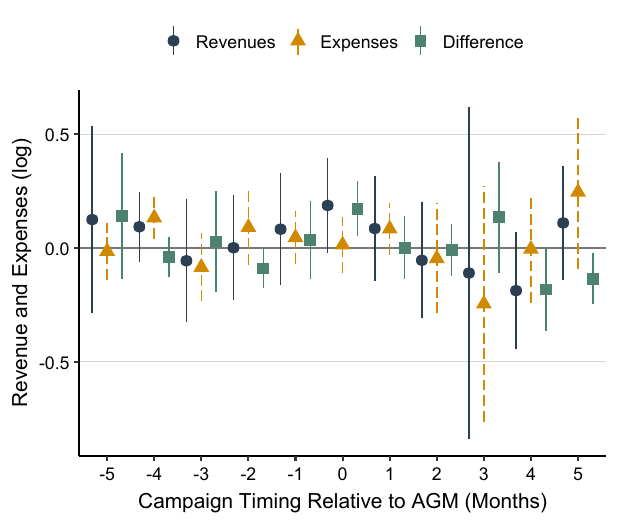}
   \subcaption{Revenue, expenses, and their difference for young NGOs}
   \end{subfigure}
 \begin{minipage}{\textwidth} %
   {\footnotesize Notes: This figure presents event-study coefficients for NGO financial performance by campaign timing relative to the target's AGM (Month 0), from Equation~\eqref{eq:eventstudy}. Panel (a) uses standardized operating margin (revenue minus expenses, divided by revenue), winsorized at the 2.5th and 97.5th percentiles, for NGOs formed after 2000 (blue circles) and in or before 2000 (red diamonds). Panel (b) uses young NGOs and plots log revenue (blue circles), log expenses (orange triangles), and their difference (green squares). All distance indicators enter jointly, with no omitted month; the $\pm5$ bins pool campaigns five or more months from the AGM. Regressions control for lagged log assets and include NGO and year fixed effects. Error bars show 95\% confidence intervals with standard errors clustered by NGO and year. Data: ISS, Sigwatch, and IRS Form 990 / Urban Institute. \par}
 \end{minipage}
\end{figure}

Panel~(a) of Figure~\ref{fig:campaign_donations} shows that young NGOs experience a \auditrev{0.453-standard-deviation increase in operating margin for campaigns in the AGM-month bin, equivalent to 14.9 percentage points}. The estimate for mature NGOs is not significant. Panel~(b) reports AGM-month \auditrev{coefficients of 0.1874 for log revenue and 0.0140 for log expenses}, corresponding to approximately 20.6\% and 1.4\%, respectively. Although neither is significant, the green series tests the difference between these coefficients by regressing log revenue minus log expenses on the same controls and fixed effects. \auditrev{At the AGM bin this difference is 0.173 log points and is statistically significant.}\footnote{\label{fn:donor_extval}We compare all \auditrev{U.S.} NGOs with the other NGOs to assess how far this evidence extends (Appendix Table~\ref{tab:selection_us_ngos}). Foreign NGOs campaign on \auditrev{U.S.} firms less often, and less often on AGM dates, than \auditrev{U.S.} NGOs, with similar campaign sentiment. This fits their tending to be more established (e.g., \emph{Which?}, UK, 1957; \emph{Clean Clothes Campaign}, Netherlands, 1989) and less focused on \auditrev{U.S.} firms, which in our model implies less AGM timing, although we cannot test this directly without their financials. \auditrev{Most differences in target characteristics become imprecise with fixed effects (Table~\ref{tab:selection_us_ngos}, Panel~b).}} As operating margin nets expenses, the estimates suggest a positive financial return to campaigns near the AGM for younger NGOs, which helps explain why the tactic concentrates early in an NGO's life rather than being adopted universally. These patterns provide reduced-form evidence on how the exposure created by campaign timing translates into financial support, not estimates of the model's warm-glow term $\alpha v(d_t)$: as Figure~\ref{fig:gtrends_campaigns} shows, AGM campaigns generate larger visibility shocks for young NGOs, so the financial response at the AGM is concentrated among them.

\subsubsection{Consumers} \label{s:consumers}

We last ask whether NGO campaigns provoke a consumer response, a channel outside the model. Restricting the sample to business-to-consumer industries, where any consumer reaction should be most visible, we estimate the event study \eqref{eq:eventstudy} with the target firm's log quarterly sales as the outcome.\footnote{\auditrev{We define business-to-consumer industries using 2-digit SIC codes}, following the classification at \url{https://www.osha.gov/data/sic-manual}; the current B2C sample contains 668 campaign events. We find the same null effect in the full sample.} Appendix Figure~\ref{fig:campaign_sales} shows no meaningful change in sales around either AGM-timed or other campaigns, so we find no detectable consumer response; the influence we document operates through shareholders and donors.

\medskip
Overall, the evidence in this section indicates that AGM timing delivers meaningful visibility gains for younger NGOs. We now turn to whether this timing advantage also translates into differences in shareholder proposal outcomes and firm responses.

\section{Campaign Timing and Influence}\label{s:influence}

The model predicts that early campaigns offer NGOs greater scope to influence contemporaneous AGM outcomes through the term $\Gamma^\theta_{d_t}$ in the proposal success probability entering the filing rule \eqref{eq:FilingRule}. Because most shareholders cast their votes by proxy before the meeting, campaigns launched on the AGM date have limited ability to affect voting outcomes (i.e., $\Gamma^\theta_{\mathrm{AGM}}=0$). For example, more than 85\% of vote-eligible shares for Microsoft’s 2023 AGM were submitted in advance by proxy \citep{microsoft2023}, illustrating why visibility at the AGM does not necessarily translate into influence over the vote itself.

\begin{figure}[t]
     \centering
          \caption{NGOs' campaigns before the AGM help E\&S shareholder proposals succeed \label{fig:ngo_support}} 
     \begin{subfigure}{0.49 \linewidth}
    \centering\includegraphics[width=\textwidth]{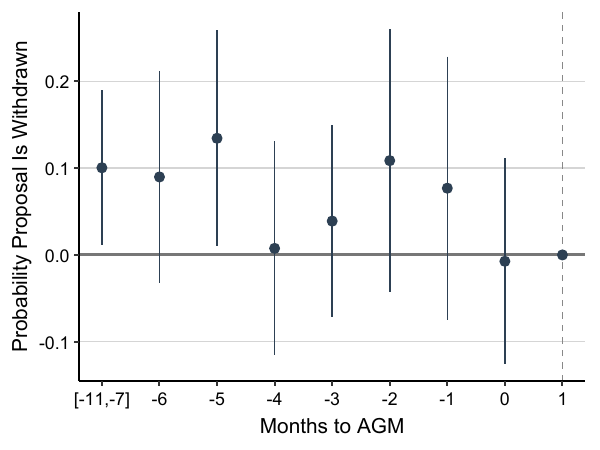}
    \subcaption{Proposal withdrawn}
  \end{subfigure} \hfill
  \begin{subfigure}{0.49 \linewidth}
    \centering\includegraphics[width=\textwidth]{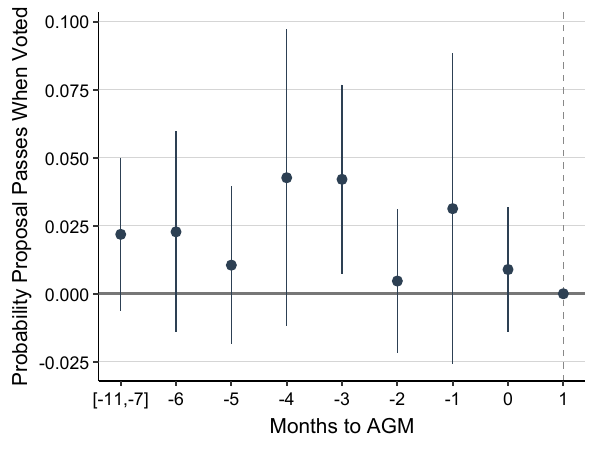}
    \subcaption{Proposal passes when voted %
    }
  \end{subfigure}\hfill
   \begin{minipage}{1 \textwidth} %
 {\footnotesize Note: Panel (a) uses an indicator for proposal withdrawal (3,241 regression observations); Panel (b) uses an indicator for more than 50\% voting support (1,556 regression observations). Coefficients interact same-topic campaigning with campaign timing relative to the AGM. Months $-11$ to $-7$ are pooled; month $+1$ is omitted. Proposals without a same-topic campaign provide controls. The specifications include a same-topic-campaign indicator, firm and topic-by-year fixed effects; Panel (b) additionally includes management voting-recommendation fixed effects and retains ``For'' and ``Against'' recommendations. Error bars show 95\% confidence intervals with firm-clustered standard errors. Coefficients are in probability units. Data: ISS and Sigwatch. \par}
 \end{minipage}
 \end{figure}

We distinguish two outcomes in Figure~\ref{fig:ngo_support}: withdrawal before the AGM and passage when a proposal goes to a vote. We estimate the event study \eqref{eq:eventstudy} using indicators for a same-topic campaign interacted with its timing relative to the AGM. Campaigns 7 to 11 months before the AGM are pooled in the earliest bin, and month $+1$ is the omitted category. The sample also includes proposals without a same-topic campaign as controls. Both regressions include firm and topic-by-year fixed effects and a same-topic-campaign indicator.\footnote{The sample is not split by NGO maturity because matching campaigns and proposals by topic already reduces the number of observations.}

Because AGMs recur annually, campaigns in the $[-11,-7]$ window fall 1 to 5 months after the previous meeting: the two distances label the same campaign on two clocks. The estimate is anchored to the focal meeting, as the regressor is the campaign's distance to that AGM and the outcomes are withdrawal and vote passage there; the previous meeting enters only as a possible prompt of the campaign's timing, and the campaigns launched just after a meeting are overwhelmingly AGM-related, often referring to the next one (Appendix~\ref{ap:postagm}). The fixed effects further limit what the previous meeting could contribute directly: the within-firm, within-cause-year comparison absorbs a firm's persistent contentiousness and a cause's activism wave in a given year, the natural paths from one meeting's aftermath to the next meeting's outcome.

\begingroup\color{black}%
Panel (a) of Figure~\ref{fig:ngo_support} plots the estimated ${\beta}_\tau$ for proposal withdrawal. Relative to the omitted month after the AGM, the coefficients are 10.0 percentage points for campaigns seven to eleven months before the meeting and 13.4 percentage points for campaigns five months before it, both statistically significant. The estimates at six and two months before the AGM are 9.0 and 10.9 percentage points, respectively, but their 95\% confidence intervals include zero.
\endgroup%

\begingroup\color{black}%
To distinguish engagement from voting outcomes, we re-estimate \eqref{eq:eventstudy} for voted proposals, using an indicator for more than 50\% support as the dependent variable.\footnote{The regression controls for management voting recommendations and retains proposals with a ``For'' or ``Against'' recommendation.} Panel (b) of Figure~\ref{fig:ngo_support} reports increases in the passage probability of approximately 4.2 percentage points three and four months before the AGM. The plotted 95\% confidence interval excludes zero at three months, but includes zero at four months.
\endgroup%

\auditrev{These results suggest that early campaigns may induce the target firm to address the issue highlighted in both the campaign and the shareholder proposal.}

Note that the subset of shareholder proposals going to a vote is not random, a point raised, for instance, by \cite{flammer2015does}. Given the generally low support for shareholder proposals going to a vote, managers may let proposals that are the least likely to succeed go to a vote. However, those votes can still be helpful in disciplining firms as strong shareholder support for a cause may be hard to conceal for managers \citep{levit2011nonbinding}. This reinforces the importance of analyzing all E\&S proposals, including those withdrawn as in Panel (a).

Taken together, the withdrawal and vote-passage patterns in Figure~\ref{fig:ngo_support} are consistent with different stages of influence. Earlier campaigns likely reflect engagement and bargaining when concessions remain feasible, while campaigns closer to the AGM likely reflect last-mile mobilization in support of the vote.

\subsection{Campaign Timing and Prosocial Changes}
\label{s:ES}

\begingroup\color{black}%
Established NGOs that campaign before the AGM improve several measures of firms' subsequent conduct. We show this by estimating the event study \eqref{eq:eventstudy} for three forward outcomes of the target firm: the one-year change in its E\&S score,\footnote{Acknowledging the limits of ESG ratings as measures of corporate impact \citep[e.g.,][]{kotsantonis2019four,allcott2022economic}, we rely on MSCI data to mitigate signal risk, given their consistency over time and widespread use among sustainable investors \citep{berg2021history,berg2024economic}.} an indicator for extensive ESG reporting, and the share of independent board members, splitting campaigns by NGO maturity as in earlier analyses. All three specifications absorb firm-by-AGM-date fixed effects; because MSCI E\&S scores are already industry-adjusted, that regression uses month-by-year fixed effects, while the ESG-reporting and board-independence regressions add the industry dimension with month-by-year-by-industry fixed effects. 
\endgroup%

\begin{figure}[htbp]
\color{black}%
\centering\caption{Campaign Timing and Changes in E\&S Scores}\label{fig:campaign_esg_score}
    \centering\includegraphics[width=.49\linewidth]{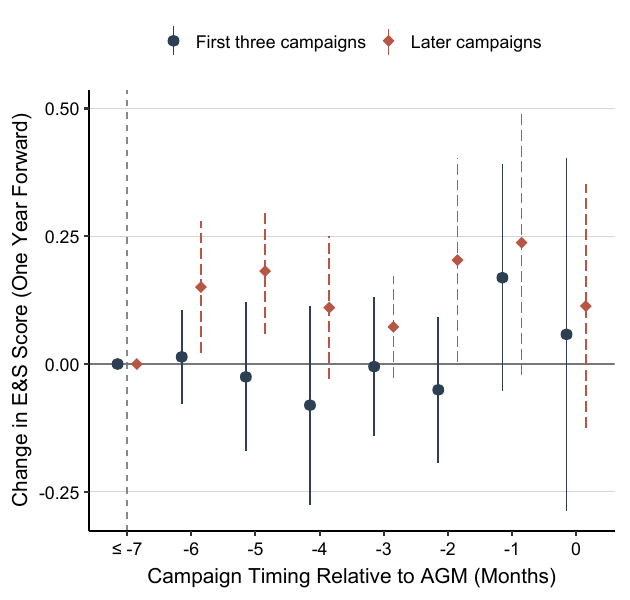}
\begin{minipage}{1\textwidth}
{\footnotesize
Note: This figure presents event study coefficients of the one-year forward change in E\&S scores on the timing of an NGO campaign relative to the AGM month (Month 0) from Equation~\eqref{eq:eventstudy}. The regression also includes firm-by-AGM-date and month-by-year fixed effects. The blue dots refer to an NGO's first three campaigns while red diamonds refer to later campaigns. Error bars show 95\% confidence intervals based on standard errors clustered by firm and calendar month. Data combine MSCI ESG scores, NGO campaigns from Sigwatch, firm characteristics from Compustat, and AGM dates from ISS. \par}
\end{minipage}
\end{figure}

\begingroup\color{black}%
The estimates broadly support this pattern. Established NGOs' campaigns six, five, and two months before the AGM are associated with E\&S-score increases of 0.151, 0.182, and 0.203 score points over the following year, respectively---about 3--5\% of the estimation-sample mean of 4.46 (Figure~\ref{fig:campaign_esg_score}). Their campaigns four months before the AGM make extensive ESG reporting 6.9 percentage points more likely (Appendix Figure~\ref{fig:campaign_other_dep}, Panel~a). The reporting estimate at the AGM is also positive, at 4.9 percentage points. The responses to younger NGOs' campaigns are generally indistinguishable from zero and imprecisely estimated at \auditrev{$\tau=-1$}.
\endgroup%

\begingroup\color{black}%
Board independence tells the same story, more sharply. Using the index of \cite{hsu2025eco}, we ask whether a campaign increases the share of independent board members over the following year. Campaigns by established NGOs in the months before the AGM increase this share, with the effect approximately 1.55 percentage points and significant four months before the meeting; younger NGOs' campaigns point the same way but never reach significance (Appendix Figure~\ref{fig:campaign_other_dep}, Panel~b). These estimates suggest that NGOs with the credibility of a track record are better able to move a firm's governance, with the clearest board-independence estimate occurring before the vote. Taken together, these results favor influence over shaming: several firm-outcome responses occur ahead of the AGM and are concentrated among established NGOs. It is the same margin that lifts shareholder-proposal success in Figure~\ref{fig:ngo_support}, now visible in what firms do and what they disclose.
\endgroup%

\section{Exposure vs.\ Influence}\label{s:reputation}
The model predicts a tradeoff between the visibility benefits of AGM-timed campaigns 
and the influence and reputation gains from early campaigns, implying that NGOs should shift toward earlier actions as their visibility and credibility accumulate. In this section, we test these implications by examining whether NGOs with greater accumulated exposure, measured by past campaign experience (Section~\ref{s:timing_past_and_current}) or current visibility (Section~\ref{s:timing_visibility_current}), time their subsequent campaigns earlier in the AGM cycle, and whether this accumulated reputation increases NGOs’ effectiveness in shaping shareholder proposal outcomes (Section~\ref{sec:success}).

\begingroup\color{black}%
\subsection{Choosing when to Campaign based on Past Campaigns}\label{s:timing_past_and_current}
We first examine whether NGOs shift from targeting AGMs to launching earlier campaigns as they campaign more. We estimate the timing regression \eqref{eq:timing} with the NGO's cumulative campaign count as the reputation proxy $R_{g,c}$, absorbing NGO-by-firm-by-year and topic-by-year fixed effects and clustering by NGO. We use the cumulative count of past campaigns as the maturity proxy for two reasons. First, it rises monotonically within an NGO, so the same organization appears as young in its early campaigns and as mature in its later ones, and the \auditrev{life cycle} is identified within NGOs rather than across different organizations, which a non-monotonic size measure cannot deliver. Second, the count is observed for every NGO, whereas size, age, and the other balance-sheet proxies come from Form~990 and exist only for \auditrev{U.S.} NGOs.
Identification uses repeated campaigns within NGO--firm--year cells, net of topic-by-year effects.
\endgroup%

Panel (a) of Figure \ref{fig:timing_exp_gt} shows that subsequent campaigns are launched earlier relative to the first campaign, which is the omitted category. The coefficient for the tenth and subsequent campaigns is approximately $-5.45$ months. All point estimates are negative, but turn significant after the fifth campaign. This pattern is consistent with NGOs initially using AGMs to gain visibility and later exploiting their visibility to influence voting outcomes.

These patterns are examined under alternative explanations in Appendix~\ref{app:lifecycle}. Results rule out the mechanical ones: the shift is unchanged for NGOs whose first campaign came after the AGM, so it does not reflect one-time learning that AGM-day campaigns cannot move votes already cast; it survives proxying maturity by age or size where Form 990 data permit, so it does not hinge on the campaign count or its 2010 start; and it holds with fixed effects that also use variation across years, so it is not built into the specification. Nor is it target selection: size, ESG scores, and ownership concentration do not systematically covary with an NGO’s past AGM exposure (Appendix~\ref{app:selection}).

\subsection{Choosing When to Campaign Based on NGO Visibility}\label{s:timing_visibility_current}

\begin{figure}[t]
\color{black}%
     \centering
          \caption{NGO reputation and campaign timing \label{fig:timing_exp_gt}} 
     \begin{subfigure}{0.49 \linewidth}
    \centering\includegraphics[width=\textwidth]{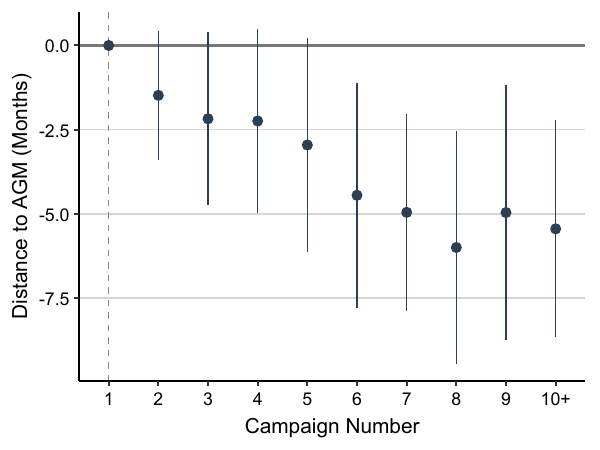}
    \subcaption{NGOs tend to campaign before the target's AGM date as they gain experience \label{fig:timing_experience}}
  \end{subfigure} \hfill
  \begin{subfigure}{0.49 \linewidth}
    \centering\includegraphics[width=\textwidth]{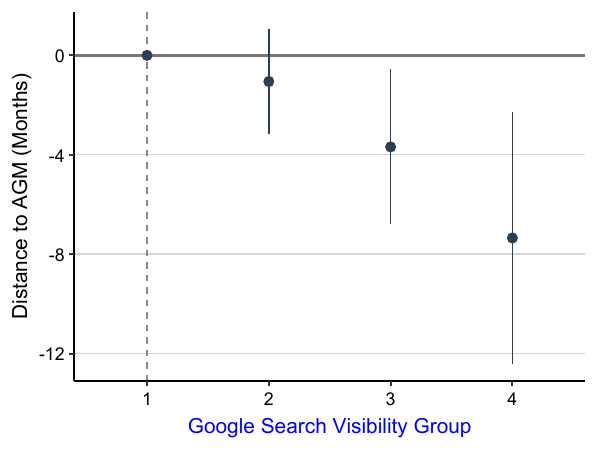}
    \subcaption{More visible NGOs campaign before the target's AGM date \label{fig:timing_gtrends}}
  \end{subfigure}\hfill
   \begin{minipage}{1 \textwidth} %
 {\footnotesize Note: The dependent variable is campaign distance to the AGM in rounded months, with dates more than one month after the nearest AGM mapped to the next annual cycle. Panel (a) uses campaign-order indicators, omits the first campaign, pools the tenth and later campaigns, and includes NGO-by-firm-by-year and topic-by-year fixed effects. Panel (b) uses lagged relative Google search activity: zero searches form the omitted group and positive scores are divided into three within-year groups. Its fixed effects are NGO-by-firm and topic-by-year. Error bars show 95\% confidence intervals with NGO-clustered standard errors. Data: ISS, Sigwatch, and Google Trends. \par}
 \end{minipage}
 \end{figure}

\begingroup\color{black}%
We next proxy reputation with NGO visibility. For each NGO $g$ we compute a \emph{Relative Google Score} (its average search score in the previous year relative to Greenpeace, the most searched NGO in our sample) and form four groups: zero measured searches and three within-year groups of positive search scores. We estimate \eqref{eq:timing} with these search-activity groups as $R_{g,y-1}$. Because the proxy varies only across NGOs and years, we use NGO-by-firm and topic-by-year fixed effects; standard errors are clustered by NGO.\footnote{The Relative Google Score is constructed from the previous year's Google Trends data, generating variation only across NGOs and years but not within NGO--year cells.}

Panel~(b) of Figure~\ref{fig:timing_exp_gt} shows that only the most visible NGOs, those in the upper groups of the relative search distribution, campaign several months before AGMs, relative to NGOs with zero measured searches, with coefficients of approximately $-3.68$ and $-7.35$ months in the two highest groups.
These results highlight a systematic shift: NGOs start by targeting AGMs to maximize exposure, and as their reputation grows, they move campaigns earlier in the AGM cycle to increase their chances of influencing firm decisions.  
\endgroup%

\subsection{NGO Reputation and AGM Influence}\label{sec:success}

We now examine how the evolution of NGO reputation shapes their ability to influence shareholder proposal outcomes, tying together the patterns documented in Sections~\ref{s:timing_past_and_current} and~\ref{s:timing_visibility_current}. In the model, proposal success depends jointly on awareness and credibility through $S(A_t,\pi_t,d_t)$ in \eqref{eq:FilingRule}. NGOs that rely heavily on AGM-timed campaigns accumulate visibility, but the marginal credibility content of additional AGM actions eventually declines. By contrast, NGOs that shift toward earlier, influence-oriented campaigns build credibility over time. The model therefore implies that overusing visibility-driven tactics should diminish the effectiveness of AGM campaigns for otherwise credible NGOs, whereas NGOs with stronger reputations should be more effective at shaping proposal outcomes.

To test this implication, we estimate the following specification:
\begingroup\color{black}
\begin{equation}\label{eq:main_reg}
\begin{aligned}
\text{Success}_{p,g,c} ={}& \beta_1\,\text{SameTopic}_{p,c}
+\beta_2\,D_c+\beta_3\,\text{SameTopic}_{p,c}D_c\\
&+\beta_4\,\text{SameTopic}_{p,c}R_{g,y-1}
+\alpha_{n,g,y}+\alpha_{i,m(p)}+u_{p,g,c}.
\end{aligned}
\end{equation}
Here success indicates withdrawal or passage, $D_c$ is standardized raw campaign distance to the AGM in days, and $R_{g,y-1}$ is experience through the previous calendar year. The sample includes campaigns eight to one month before the AGM. Firm-by-NGO-by-year and topic-by-meeting-month-year fixed effects absorb annual experience levels, so only their same-topic interactions are identified. Standard errors are clustered by firm. Columns with quadratic experience add $\text{SameTopic}_{p,c}R_{g,y-1}^{2}$; the indicator specification separates any prior AGM and non-AGM experience.\footnote{We focus on campaigns occurring 1 to 8 months before the AGM, since this is the window in which early campaigning is associated with increasing shareholder support (Figure~\ref{fig:ngo_support}). Repeating the analysis for campaigns launched within one month of the AGM yields consistently null effects.}
\par\endgroup

Column~(1) of Table~\ref{ngo_experience_success} shows that campaigning on the same cause as a shareholder proposal has little effect on its own ($\beta_1 \approx 0$). Instead, proposals are more likely to succeed when the campaigning NGO has greater accumulated experience: a one-standard-deviation increase in past campaigns is associated with a 4.6-percentage-point increase in the same-topic effect, significant at the 10\% level.

\begin{table}[!htb]
\color{black}
\begin{center}
   \caption{NGOs' strategies and support for related shareholder proposals} \label{ngo_experience_success}
   \resizebox{ \textwidth}{!}{
      \begin{tabular}{lcccc}
\toprule
Dependent variable: & \multicolumn{4}{c}{Proposal success (0/1)} \\
Model: & (1) & (2) & (3) & (4) \\
Distance to AGM (std.) & -0.008 & -0.006 & -0.007 & -0.006 \\
 & (0.021) & (0.021) & (0.021) & (0.021) \\
Same topic $\times$ Distance to AGM & -0.016 & -0.018 & -0.017 & -0.017 \\
 & (0.019) & (0.020) & (0.019) & (0.020) \\
Campaign on same topic & -0.052 & -0.096* & -0.042 & -0.014 \\
 & (0.042) & (0.057) & (0.049) & (0.043) \\
Same topic $\times$ Past Campaigns & 0.046* &  &  &  \\
 & (0.024) &  &  &  \\
Same topic $\times$ Past AGM Campaigns (0/1) &  & 0.112* &  &  \\
 &  & (0.062) &  &  \\
Same topic $\times$ Past Non-AGM Campaigns (0/1) &  & 0.046 &  &  \\
 &  & (0.028) &  &  \\
Same topic $\times$ Past Non-AGM Campaigns &  &  & 0.068 &  \\
 &  &  & (0.043) &  \\
Same topic $\times$ Squared Past Non-AGM Campaigns &  &  & -0.012 &  \\
 &  &  & (0.017) &  \\
Same topic $\times$ Past AGM Campaigns &  &  &  & 0.137*** \\
 &  &  &  & (0.050) \\
Same topic $\times$ Squared Past AGM Campaigns &  &  &  & -0.035*** \\
 &  &  &  & (0.012) \\
\midrule
Firm $\times$ NGO $\times$ Year FE & $\checkmark$ & $\checkmark$ & $\checkmark$ & $\checkmark$ \\
Topic $\times$ Month-Year FE & $\checkmark$ & $\checkmark$ & $\checkmark$ & $\checkmark$ \\
\midrule
Observations & 2,361 & 2,361 & 2,361 & 2,361 \\
R$^2$ & 0.63923 & 0.63930 & 0.63927 & 0.63966 \\
\bottomrule
\end{tabular}}
   \end{center}
   \begin{tablenotes}[flushleft,para]
\footnotesize\item Note: The dependent variable equals one for a withdrawn or passed proposal and zero otherwise. Experience is measured through December 31 of the previous calendar year. Column (1) uses the standardized count of all past campaigns. Total and non-AGM counts are winsorized at the 97.5th percentile before standardization; AGM counts are not winsorized. Column (2) uses indicators for any past AGM and non-AGM campaigns. Columns (3) and (4) use the standardized counts and squared standardized counts of past non-AGM and AGM campaigns, respectively. Each campaign-history measure is interacted with the indicator for a campaign on the same topic. Experience main effects are absorbed by Firm $\times$ NGO $\times$ Year fixed effects. Topic $\times$ Month-Year refers to the proposal meeting month and year. The sample includes campaigns 240 to 30 days before the AGM; distance is the signed raw number of days from the AGM to the campaign date, standardized across the common Table~\ref{ngo_experience_success} input sample before fixed-effect singleton removal. Standard errors are clustered by firm. Standard errors in parentheses. * p $<$ 0.10; ** p $<$ 0.05; *** p $<$ 0.01.
\end{tablenotes}
   \end{table}

The next columns distinguish past AGM and non-AGM experience. Column~(2) uses indicators for any prior AGM and any prior non-AGM campaign. Their same-topic interactions are 0.112 and 0.046, respectively, with only the former significant at the 10\% level.

This pattern is consistent with the model's interpretation of AGM-day actions as visibility-generating. Column~(3) uses standardized non-AGM counts and their square; neither interaction is statistically significant. Column~(4) shows a positive linear interaction with standardized AGM experience (0.137) and a negative squared interaction ($-0.035$), both significant at the 1\% level. This supports diminishing marginal returns to prior AGM experience, rather than a constant effect per additional campaign.

Importantly, these mechanism patterns are not driven by NGO maturity: all specifications include NGO-by-year fixed effects, so the results compare campaigns by the same NGO within the same year, holding fixed any time-varying NGO-level factors.

\paragraph{Other reputation measures.} Column~(1) of Appendix Table~\ref{ngo_size_success} replaces the campaign count with lagged standardized Google search popularity. The linear interaction is positive and the quadratic interaction negative, but neither is statistically significant.

\begingroup\color{black}%
Column~(2) uses NGO power as reported by Sigwatch in raw units; its linear and squared same-topic interactions are positive and negative, respectively, both significant at the 1\% level.\footnote{As that measure of NGO power is mostly \auditrev{static---for more than 95\% of NGOs in the sample---}it does not allow its use to illustrate the life cycle results.} Focusing on U.S. NGOs, the remaining columns measure reputation with lagged assets (Column~3), lagged total revenue (Column~4), and lagged expenditures (Column~5), manifestations of the standing that places an NGO higher in the $(A_t,\pi_t)$ state space.\footnote{A large literature documents that NGO reputation is closely tied to organizational scale, resources, and operational capacity. Larger and better-resourced NGOs are viewed as more credible, legitimate, and effective by external stakeholders \citep[e.g.,][]{baur2011moral,cooley2002ngo,mitchell2017reputations}.} The linear interactions are positive and the quadratic interactions negative across these proxies, but precision differs. Overall, the results are consistent with diminishing marginal returns to reputation, with the strongest support coming from prior AGM experience.
\endgroup%

\subsection{Accounting for Selection}\label{s:case}
A remaining concern is that NGOs may choose campaign dates strategically, potentially confounding the timing effects documented above. To address this, we turn to a setting in which \textit{both} campaign timing and target identities are anchored to an external event and thus provide quasi-exogenous variation. The relevant exhibits are in Appendix \ref{apndx:additional_selection}.

\paragraph{Institutional background.}
\textit{Fashion Revolution} was founded in the aftermath of the Rana Plaza factory collapse on April~24,~2013, and campaigns for greater transparency, workers' rights, and environmental sustainability in global fashion supply chains \citep{koenig2022effects}. While our empirical analysis centers on the organization’s annual \emph{Fashion Transparency Index} (FTI), it is important to note that the Index is closely tied to \emph{Fashion Revolution Week}, a recurring campaign period held each year around the Rana Plaza anniversary. In its early years, the FTI was released to coincide with this anniversary (on April~23$^\text{rd}$ or April~24$^\text{th}$ between 2016 and 2019), aligning the publication with the Week’s peak public attention. Because this timing is fixed in the calendar and unrelated to firm-specific events, we treat publication dates as plausibly exogenous, providing a natural setting to test whether the exposure--influence mechanisms identified earlier persist when timing is not strategically chosen by the NGO.\footnote{Appendix Table~\ref{fti_balance_check} verifies that AGM timing relative to Index publication is not correlated with key observable characteristics such as a target's capitalization, geographic origin, and its first FTI score, which we take as a direct measure of the firm's prosocial stance.}

\paragraph{Data.} We collect daily stock prices (from LSEG), AGM dates, and data on environmental and social (E\&S) shareholder proposals (from ISS) for 97 listed firms included in the Index between 2016 and 2019.\footnote{The first report included 40 brands globally, and its coverage had increased to 200 by 2020, located in 16 different countries. We focus on the firms that have been listed between 2016 and 2019.} 

\begingroup\color{black}%
\paragraph{Stock market reactions.} For each firm, we download daily returns on its national stock index (e.g., FTSE 100 for U.K. companies, CAC40 for French companies) and compute its daily excess return over the index. We then cumulate those returns and plot the results in Appendix Figure \ref{fig:fti_cer}, depending on the report's publication timing relative to the company's AGM (Panel a), and depending on the company's initial FTI score (above or below the within-year median, in Panel b). The patterns are consistent with the model’s \emph{visibility channel}: when firms are most exposed to scrutiny, credible activism amplifies the financial costs from a deteriorating reputation \citep{kolbel2017media}.
\endgroup%

\begingroup\color{black}%
\paragraph{Shareholder outcomes.} Appendix Figure~\ref{fig:fti_proposals} shows that the share of firm-years with at least one withdrawn or passed E\&S proposal is about 6~percentage points higher when the Index is published before the AGM. Average voting support for E\&S proposals is about 9~percentage points higher. \auditrev{These magnitudes are broadly consistent with the shareholder-outcome evidence in Section~\ref{s:influence} (Figure~\ref{fig:ngo_support})}, reinforcing that early, credibility-driven actions enhance influence through coordinated shareholder responses.
\endgroup%

\paragraph{Shareholder proposals.}
Finally, Appendix Figure~\ref{fig:fti_regs} replicates the timing tests from earlier sections. \auditrev{Publication in the month leading up to the AGM is associated with a positive estimated change in the probability of a related proposal at the next AGM}, whereas publication two months before the AGM (analogous to early campaigning) leads to roughly 8-point improvements in firms’ FTI scores over the following year.  

These results mirror the model’s main mechanism while accounting for target selection: AGM-timed actions mobilize shareholders, while earlier actions directly influence corporate behavior.

\section{Discussion and Mechanisms}
\label{s:discussion}

\begingroup\color{black}%
Our findings show that \emph{when} NGOs campaign can be as important as \emph{whether} they campaign. Early in their life cycle, NGOs concentrate actions on AGM dates, when firms are most visible to investors and the media \citep{baron2001private,egorov2017private,fioretti2024shared}: these campaigns raise media coverage and searches, and campaigns around the AGM improve younger NGOs' operating margins through higher revenues. AGM campaigns by more experienced NGOs also predict related proposals at the next AGM, while contemporaneous voting and E\&S-score responses at the AGM are limited. As reputation accumulates, campaigns move to the months before the meeting, raising the withdrawal and vote-passage probabilities of related proposals and preceding improvements in E\&S scores \citep[e.g.,][]{cabral2010dynamics,baur2011moral}. Timing thus reveals both the intended audience and the likely margin of influence, a dimension that asking only \emph{whether} NGOs campaign misses.
\endgroup%

Conceptually, the paper opens the black box of NGO activism by exploiting one of the few actions that must be public to be effective: campaigns. NGOs are otherwise opaque organizations, with limited standardized reporting and scarce data on internal decisions \citep{maxwell2012economic}. Our framework models the impact of NGO campaigns through awareness and credibility, and summarizes influence-oriented effectiveness by $\Gamma_{\text{early}}(\pi,A)$. This reduced-form approach is deliberate: in practice, NGO influence can arise from distinct micro-channels that are difficult to observe directly. 

Two channels are particularly relevant for shareholder outcomes. One is \emph{information provision}: NGOs can generate and disseminate research, due diligence, ratings, or arguments that shift investors’ beliefs about the merits of a proposal, potentially interacting with intermediaries such as proxy advisors \citep{malenko2016role}. The other is \emph{coordination}: campaigns can help dispersed shareholders overcome collective-action frictions by making support salient and mutually reinforcing, and by increasing the perceived likelihood that others will also act. Both channels map into $\Gamma_{\text{early}}(\pi,A)$; the question is which better rationalizes the cross-sectional patterns in campaign effectiveness.

\subsection{Mechanisms: Coordination vs.\ Information Provision}
\label{sec:mechanisms}

A useful discriminator is ownership structure. Coordination frictions are naturally more salient when ownership is neither so concentrated that pivotal holders can act unilaterally, nor so dispersed that free-riding makes mobilization infeasible \citep{levit2011nonbinding}. Under this view, campaign effectiveness should display a non-monotonic pattern across concentration, strongest at intermediate levels. By contrast, if campaigns primarily transmit information to pivotal investors, the ownership prediction is less sharp ex ante: concentration could attenuate effects if campaigns mostly reach non-pivotal investors, but could also amplify them if a small number of large holders can process and act on credible information efficiently \citep{malenko2016role}.

The data align more closely with the coordination-frictions interpretation. First, when we augment the proposal-success specification \eqref{eq:main_reg} with interactions between same-topic campaigning and shareholder concentration (HHI, blockholding measures, and the number of blockholders), Table~\ref{ngo_dispersion_success} shows that the marginal effect of campaigning is weaker when ownership is more concentrated. Second, when we revisit the filing analysis and estimate Equation~\eqref{eq:eventstudy} separately by terciles of concentration, AGM-timed campaigns predict additional related filings at the next AGM most strongly for firms in the middle tercile (Table~\ref{campaign_proposal_concentration}). This combination of attenuation with concentration on outcomes and a peak in the middle tercile on filing is difficult to reconcile with \auditrev{an information-based account alone} and is especially consistent with a coordination-based reading of $\Gamma_{\text{early}}(\pi,A)$.\footnote{A natural concern is endogenous sorting: experienced NGOs might campaign where ownership structure makes success more likely. As mentioned in Section~\ref{s:target}, we find little evidence of such sorting.}

\begin{table}[!htb]
\color{black}
   \caption{Firms' ownership concentration and support for related shareholder proposals} \label{ngo_dispersion_success}
   \begin{center}
   \resizebox{ 0.95\textwidth}{!}{
\begin{tabular}{lccc}
\toprule
Dependent variable: & \multicolumn{3}{c}{Proposal success (0/1)} \\
Model: & \auditrev{(1)} & \auditrev{(2)} & \auditrev{(3)} \\
Ownership measure & Shareholder HHI & Blockholding & Blockholder count \\
\midrule
Distance to AGM (std.) & \auditrev{-0.005} & \auditrev{-0.006} & \auditrev{-0.008} \\
 & \auditrev{(0.021)} & \auditrev{(0.021)} & \auditrev{(0.021)} \\
Same topic $\times$ Distance to AGM & \auditrev{-0.009} & \auditrev{-0.001} & \auditrev{0.001} \\
 & \auditrev{(0.016)} & \auditrev{(0.017)} & \auditrev{(0.017)} \\
Campaign on same topic & \auditrev{0.083*} & \auditrev{0.114**} & \auditrev{0.128**} \\
 & \auditrev{(0.044)} & \auditrev{(0.053)} & \auditrev{(0.054)} \\
Same topic $\times$ Ownership measure & \auditrev{-0.025**} & \auditrev{-0.029**} & \auditrev{-0.031***} \\
 & \auditrev{(0.011)} & \auditrev{(0.013)} & \auditrev{(0.011)} \\
\midrule
Firm $\times$ NGO $\times$ Year FE & $\checkmark$ & $\checkmark$ & $\checkmark$ \\
Topic $\times$ Month-Year FE & $\checkmark$ & $\checkmark$ & $\checkmark$ \\
\midrule
Observations & \auditrev{2,361} & \auditrev{2,361} & \auditrev{2,361} \\
R$^2$ & \auditrev{0.64005} & \auditrev{0.64027} & \auditrev{0.64070} \\
\bottomrule
\end{tabular}}
   \end{center}
   \begin{tablenotes}[flushleft,para]
\footnotesize\item Note: The dependent variable equals one for a withdrawn or passed proposal and zero otherwise. Ownership is lagged one year and divided into deciles; the regressions include its same-topic interaction. The sample includes campaigns 240 to 30 days before the AGM. Distance is the signed raw number of days from the AGM to the campaign date (standardized). Fixed effects are Firm $\times$ NGO $\times$ Year and Topic $\times$ proposal meeting Month-Year. Standard errors are clustered by firm. Standard errors in parentheses. * p $<$ 0.10; ** p $<$ 0.05; *** p $<$ 0.01.
\end{tablenotes}
   \end{table}

\begin{table}[!htb]
\color{black}
   \caption{Firms' ownership concentration and filing of related shareholder proposals} \label{campaign_proposal_concentration}
   \begin{center}
   \resizebox{ 0.7\textwidth}{!}{
\begin{tabular}{lccc}
\toprule
\multicolumn{4}{l}{\textbf{Panel a: Shareholder HHI}} \\
\midrule
Tercile & Low & Middle & High \\
Campaign at previous AGM & 0.004 & 0.034* & 0.017 \\
 & (0.009) & (0.019) & (0.012) \\
Firm FE & $\checkmark$ & $\checkmark$ & $\checkmark$ \\
Month $\times$ NGO $\times$ Topic $\times$ Industry FE & $\checkmark$ & $\checkmark$ & $\checkmark$ \\
\midrule
Observations & 2,722 & 2,634 & 2,816 \\
R$^2$ & 0.68637 & 0.73417 & 0.62643 \\
\midrule
\multicolumn{4}{l}{\textbf{Panel b: Blockholding}} \\
\midrule
Tercile & Low & Middle & High \\
Campaign at previous AGM & -0.003 & 0.038** & 0.018 \\
 & (0.005) & (0.016) & (0.015) \\
Firm FE & $\checkmark$ & $\checkmark$ & $\checkmark$ \\
Month $\times$ NGO $\times$ Topic $\times$ Industry FE & $\checkmark$ & $\checkmark$ & $\checkmark$ \\
\midrule
Observations & 2,443 & 2,799 & 2,960 \\
R$^2$ & 0.71602 & 0.77057 & 0.61897 \\
\midrule
\multicolumn{4}{l}{\textbf{Panel c: Blockholder count}} \\
\midrule
Tercile & Low & Middle & High \\
Campaign at previous AGM & -0.004 & 0.039** & 0.024* \\
& (0.005) & (0.019) & (0.013) \\
Firm FE & $\checkmark$ & $\checkmark$ & $\checkmark$ \\
Month $\times$ NGO $\times$ Topic $\times$ Industry FE & $\checkmark$ & $\checkmark$ & $\checkmark$ \\
\midrule
Observations & 2,375 & 2,444 & 2,797 \\
R$^2$ & 0.72290 & 0.70602 & 0.66581 \\
\bottomrule
\end{tabular} }
   \end{center}
   \begin{tablenotes}[flushleft,para]
\footnotesize\item Note: The dependent variable indicates a shareholder proposal on the campaign topic. Each panel uses lagged ownership terciles formed within year. The regressor equals one for a campaign twelve months before the prospective AGM; the comparison comprises other months from -15 to -1. All models include firm fixed effects. All models also include Calendar-month $\times$ NGO $\times$ Topic $\times$ Industry fixed effects. Standard errors are clustered by firm. Standard errors in parentheses. * p $<$ 0.10; ** p $<$ 0.05; *** p $<$ 0.01.
\end{tablenotes}
   \end{table}

The institutional setting reinforces this reading. Proposals in our data are typically filed by small shareholders, individuals and religious groups rather than large institutions, so a pivotal-investor information story sits awkwardly: it would predict effectiveness at least as strong where a few large holders can act decisively, the opposite of what we find. And conditional on a proposal being filed, more dispersed ownership is associated with greater success, consistent with campaign-induced coordination among many small shareholders when no dominant holder can block the outcome. NGO credibility thus operates, at least in part, by facilitating shareholder coordination; informational and engagement channels may coexist and are central in broader theories of private politics \citep{baron2001private,egorov2017private}.

Appendix~\ref{app:coalitions} extends the analysis to collaborative campaigns. Early solo campaigns show the clearest attention gains, and prior solo AGM experience predicts coalition proposal success. We find no evidence of collaborative-history benefits.

\subsection{Implications}

NGO activism also sits differently in the ecology of corporate activism from the channels it is usually compared with. It usually carries no equity stake, unlike the hedge-fund campaigns built on Schedule~13D positions \citep{brav2008hedge}, and it casts no vote, unlike the shareholders who file resolutions. Its influence is instead upstream and coordinating. A campaign works partly by moving others: the shareholders it helps to file and pass proposals and the donors whose support it draws. This accords with a broader, system-level view of activism as interdependent actors rather than isolated events \citep{chuah2024shareholder}, in which the boundaries between channels blur. Its effect is therefore diffused across these actors, hard to attribute to any single agent or to read off the campaign count alone.

Beyond mechanisms, our results speak to stakeholder salience and the evolution of organizational objectives. How firms decide which stakeholders matter remains an open question; the large literature on stakeholder prioritization is mostly qualitative \citep[e.g.,][]{freeman1984strategic,mitchell1997toward}. Because campaign timing is anchored to the exogenous AGM calendar yet chosen by NGOs with modeled incentives, it offers a quantitative handle on how firms form and update their objectives and whose welfare reaches the agenda, complementing recent empirical work on stakeholder weights \citep[e.g.,][]{fioretti2022caring}.

\auditrev{The policy implications outlined in the Introduction depend on which channel is active. The ownership tests suggest that engagement with NGOs may be particularly useful when it helps shareholders coordinate; visibility alone does not establish an ability to influence the current vote.}

Future work could separate information from coordination by combining campaign timing with micro-level evidence on how support forms (investor-level votes, proxy-advisor recommendations, investor communication), tracing how the two channels' importance evolves over an NGO's \auditrev{life cycle}. Effectiveness may also vary with NGOs' internal governance and professionalization, which shape credibility and strategic capacity, and, at the firm level, with shareholder monitoring, which conditions whether external pressure becomes coordinated action and substantive change \citep[e.g.,][]{cooley2002ngo,baur2011moral,mitchell2017reputations,edmans2014blockholders,dimson2015active}.

\section{Conclusion}\label{s:conclusion}

\begingroup\color{black}%
This paper opens the black box of NGO behavior by showing how campaign timing reflects NGOs’ underlying objectives. Empirically, we use data on 2{,}350 campaigns to show that AGM-day actions generate large visibility shocks, raise media coverage and search activity, and improve younger NGOs' operating margins. Experienced NGOs' AGM campaigns predict related proposals at the next AGM, while having little effect on contemporaneous votes or firms’ E\&S performance. By contrast, campaigns launched in the months before the AGM increase withdrawal and vote-passage probabilities and are followed by improvements in E\&S outcomes, particularly for established NGOs.
\endgroup%

A simple  model in which NGOs trade off two forces explains these patterns. The model features an \emph{awareness} channel, whereby AGM visibility builds issue salience, and a \emph{credibility} channel, whereby earlier, costlier actions update beliefs and directly influence shareholder behavior by coordinating dispersed shareholders. 

Together, the model and evidence reveal NGOs as strategic actors whose objectives evolve with external incentives, vary across contexts, and reflect the views of multiple stakeholders who shape NGOs’ reputational payoffs. These insights matter for policy, showing that NGO campaigns mobilize different classes of investors with distinct implications for firms’ outcomes, whether by exerting direct influence through support for shareholder votes or indirectly, by  prompting new proposals at future AGMs. Overall, the results demonstrate how civil-society pressure can help firms internalize social and environmental externalities.

\bibliographystyle{ecca-mod}
\bibliography{main_RR1_v3_references}	
\clearpage\newpage
\appendix

\section*{\Huge{Online Appendix}}

\section{Proofs}\label{app:proofs}
\setcounter{table}{0}\renewcommand{\thetable}{A\arabic{table}}
\setcounter{equation}{0}\renewcommand{\theequation}{A\arabic{equation}}
\setcounter{figure}{0}\renewcommand{\thefigure}{A\arabic{figure}}
\renewcommand{\theproposition}{A\arabic{proposition}}\setcounter{proposition}{0}
\renewcommand{\thelemma}{A\arabic{lemma}}\setcounter{lemma}{0}
\renewcommand{\theassumption}{A\arabic{assumption}}\setcounter{assumption}{0}

This appendix collects the structure relegated from Section~\ref{sec:theory}. Appendix~\ref{app:functional} records the functional forms and the maintained assumptions used throughout; Appendix~\ref{app:beliefupdate_new} derives belief updating and its consistency; Appendix~\ref{app:equilibrium} proves Proposition~\ref{prop:separation} in four steps: a unique value function, single-crossing of $\Delta$, the cutoff curve, and uniqueness of the Markov policy induced by the maintained belief system.

\subsection{Functional Forms and Maintained Assumptions}\label{app:functional}
\paragraph{Influence technology.} Only credible NGOs gain from acting early, and influence is realized through shareholders' response, which scales with the public belief:
\begin{equation}
\begin{gathered}
\Gamma^{C}_{\mathrm{early}}(\pi,A)>0\ \text{ for }\pi>0,\qquad \Gamma^{C}_{\mathrm{early}}(0,A)=0,\qquad \frac{\partial\Gamma^{C}_{\mathrm{early}}}{\partial \pi}>0,\\
\frac{\partial\Gamma^{C}_{\mathrm{early}}}{\partial A}\ge0,\qquad \Gamma^{C}_{\mathrm{AGM}}\equiv0,\qquad \Gamma^{W}\equiv0.
\end{gathered}
\label{eq:GammaProps}
\end{equation}
At $\pi=0$ no shareholder responds, so even a credible NGO converts an early campaign into nothing; a higher public belief amplifies what the same campaign achieves.
With unobserved type and belief $\pi_t$, expected influence is
\begin{equation*}
\Gamma_{d_t}(\pi_t,A_t)=\pi_t\cdot \Gamma_{d_t}^C(\pi_t,A_t)+(1-\pi_t)\cdot \Gamma_{d_t}^W(\pi_t,A_t).
\end{equation*}

\paragraph{Success probability and outcome.} When a proposal is on the ballot, success is logistic,
\begin{equation}
S^\theta(A_t,\pi_t,d_t)
=\frac{\exp\big(h_0+ h_A A_t+ h_\pi \pi_t+ h_\Gamma\,\Gamma_{d_t}^\theta(\pi_t,A_t)\big)}
{1+\exp\big(h_0+ h_A A_t+ h_\pi \pi_t+ h_\Gamma\,\Gamma_{d_t}^\theta(\pi_t,A_t)\big)},
\qquad h_A,h_\pi,h_\Gamma>0,
\label{eq:SuccessFunction}
\end{equation}
and the realized outcome is the model's only random element,
\begin{equation*}
y_t=\begin{cases}1,&\text{w.p. }S^\theta(A_t,\pi_t,d_t),\\ 0,&\text{w.p. }1-S^\theta(A_t,\pi_t,d_t).\end{cases}
\end{equation*}

\paragraph{Timing signal.} The informativeness of the timing choice is
\begin{gather*}
\ln\frac{p^C(d_t\mid \pi_t,A_t)}{p^W(d_t\mid \pi_t,A_t)}
=\Pi(\pi_t,A_t)\cdot \big(\mathbf 1_{\{d_t=\mathrm{early}\}}-\mathbf 1_{\{d_t=\mathrm{AGM}\}}\big),\\
\Pi(\pi_t,A_t)=\eta_R\cdot \pi_t-\eta_A\cdot \big(v(\mathrm{AGM})-A_t\big),
\end{gather*}
with $\eta_R,\eta_A>0$: the reputation motive ($\eta_R\pi_t$) makes early campaigning more revealing as credibility rises, while the awareness-seeding motive favors AGM visibility when awareness is low ($A_t<v(\mathrm{AGM})$). We maintain signal sensitivities in the range for which the induced posterior kernel is monotone in the current state. We treat  $p^\theta(d_t\mid\pi_t,A_t)$ as interior choice probabilities generated by small, privately observed payoff shocks (e.g., logit): each type best-responds in pure strategies to its realized shock, so the cutoff policy of Appendix~\ref{app:equilibrium} is deterministic state by state, while the observed choice probabilities remain interior and the timing signal finite. We use this as a reduced-form belief-updating device rather than solving a full dynamic signaling refinement.

\paragraph{Visibility normalization.} We set $v(\mathrm{AGM})=1$ and $v(\mathrm{early})=\bar v\in(0,1)$.

\paragraph{Assumptions.} Two maintained assumptions carry the equilibrium analysis of Appendix~\ref{app:equilibrium}: the first anchors the timing wedge at full credibility, the second makes credibility and awareness substitutes.

\begin{assumption}[Influence advantage at full credibility]\label{ass:boundary}
With $\underline\Gamma\equiv\inf_A\Gamma^C_{\mathrm{early}}(1,A)>0$ and $\bar G\equiv\sup_A\partial\Gamma^C_{\mathrm{early}}(1,A)/\partial A$,
\[
\alpha(1-\bar v)\;+\;\frac{\beta(1-\rho)}{1-\beta\rho}\,(1-\bar v)\,B\,\bar G\;<\;B\,\underline\Gamma.
\]
\end{assumption}
The left side is the cost of campaigning early at $\pi=1$, the visibility given up today plus the discounted cost of the lower awareness it leaves behind; the right side is the influence advantage. When influence does not vary with the awareness stock ($\bar G=0$) it reduces to $\alpha(1-\bar v)<B\,\underline\Gamma$. Appendix~\ref{app:equilibrium} uses it to sign $\Delta(1,A)>0$, where the function $\Delta(\pi_t,A_t)$ was defined in \eqref{eq:Delta_new}.

\begin{assumption}[Decreasing differences and awareness concavity]\label{ass:incdiff}
The credible type's value function $V(\pi,A)$ has decreasing differences in $(\pi,A)$ and is concave in $A$.
\end{assumption}
\noindent Assumption~\ref{ass:incdiff} means that an extra unit of awareness is worth less once the NGO is already credible, because credibility is already doing part of the work that awareness would otherwise do: the two crowd each other out.\footnote{This is a sufficient condition for the continuation wedge to be monotone, not the only possible one. One could instead impose a primitive dominance condition requiring the current influence gain from a higher belief to dominate the continuation cost of lower awareness; impose separability between credibility and awareness in the continuation value, which removes the cross-effect; or assume the single-crossing property of $\Delta(\pi,A)$ directly. We use decreasing differences because it matches the empirical saturation pattern: AGM visibility raises attention for young NGOs but contributes little once NGOs are already established.} This is the empirical pattern in Section~\ref{s:media}: AGM visibility has its largest return for young NGOs, while the incremental attention from AGM timing fades for mature NGOs. It is used \emph{only} in Step~2 of Appendix~\ref{app:equilibrium}; everything else follows from Equations \eqref{eq:AwarenessLaw} and \eqref{eq:GammaProps} and standard contraction arguments.

\subsection{Belief Updating and Consistency}\label{app:beliefupdate_new}
Observers update $\pi_t$ from the timing choice and, when a proposal is on the ballot, the outcome. By Bayes' rule, in log-odds form,
\begin{equation}
\ln\!\left(\frac{\pi_{t+1}}{1-\pi_{t+1}}\right)-\ln\!\left(\frac{\pi_{t}}{1-\pi_{t}}\right)
=\underbrace{\ln\frac{p^C(d_t\mid \pi_t,A_t)}{p^W(d_t\mid \pi_t,A_t)}}_{\text{timing signal}}
+\underbrace{\ln\frac{p^C(y_t\mid d_t,\pi_t,A_t)}{p^W(y_t\mid d_t,\pi_t,A_t)}}_{\text{outcome signal, Lemma~\ref{lem:bern_llr}}}.
\label{eq:PosteriorOdds}
\end{equation}

\paragraph{Outcome likelihood ratio.}
Conditional on $(A_t,\pi_t,d_t)$, $y_t$ is Bernoulli with success probability \eqref{eq:SuccessFunction}, so $p^\theta(y_t\mid d_t,A_t,\pi_t)=[S^\theta]^{y_t}[1-S^\theta]^{1-y_t}$.

\begin{lemma}[Bernoulli outcome log--likelihood ratio]
\label{lem:bern_llr}
For the two types $C$ and $W$,
\[
\ln\frac{p^C(y_t\mid d_t,A_t,\pi_t)}{p^W(y_t\mid d_t,A_t,\pi_t)}
= y_t\,\ln\frac{S^C(A_t,\pi_t,d_t)}{S^W(A_t,\pi_t,d_t)}
+(1-y_t)\,\ln\frac{1-S^C(A_t,\pi_t,d_t)}{1-S^W(A_t,\pi_t,d_t)}.
\]
\end{lemma}
\begin{proof}
By the Bernoulli likelihoods, $\dfrac{p^C}{p^W}=\dfrac{[S^C]^{y_t}[1-S^C]^{1-y_t}}{[S^W]^{y_t}[1-S^W]^{1-y_t}}$ with $S^\theta=S^\theta(A_t,\pi_t,d_t)$; taking logs gives $y_t(\ln S^C-\ln S^W)+(1-y_t)(\ln(1-S^C)-\ln(1-S^W))$, the stated expression.
\end{proof}
\noindent\emph{Remarks.} (i) If $S^C=S^W$ (e.g., at $d_t=\mathrm{AGM}$ where $\Gamma_{\mathrm{AGM}}\equiv0$) the term is $0$ and outcomes do not move beliefs. (ii) When $S^C>S^W$ (early), a success raises and a failure lowers the posterior odds of type $C$. (iii) The outcome signal is informative whenever $S^C\ne S^W$; under the credible type's law the expected log-likelihood ratio is nonnegative (a Kullback--Leibler divergence), zero iff $S^C=S^W$, and under the weak type's law it is nonpositive.

\paragraph{Consistency.} Iterating \eqref{eq:PosteriorOdds},
\begin{equation*}
\operatorname{logit}(\pi_t)=\operatorname{logit}(\pi_0)
+\sum_{\tau=0}^{t-1}\;\!\left[\ln\frac{p^C(d_\tau\mid \pi_\tau,A_\tau)}{p^W(d_\tau\mid \pi_\tau,A_\tau)}
+y_\tau\ln\frac{S^C}{S^W}+(1-y_\tau)\ln\frac{1-S^C}{1-S^W}\right].
\end{equation*}
With (i) a non-degenerate prior $\pi_0\in(0,1)$; (ii) signals observed each period; (iii) interior smoothed choice probabilities, so early campaigns ($S^C\ne S^W$) occur infinitely often almost surely; and (iv) bounded expected increments, a law of large numbers on the informative (early) subsequence gives $\operatorname{logit}(\pi_t)\to+\infty$ if $\theta=C$ and $\to-\infty$ if $\theta=W$. Hence beliefs are consistent: $\pi_t\to1$ a.s.\ under $C$ and $\pi_t\to0$ under $W$. The optimal timing rule is deterministic given the state; the smoothed probabilities are used only to keep the belief system finite and informative off the deterministic path. Stochastic outcomes nonetheless make belief trajectories, and with them the timing of the AGM-to-early transition, random and path-dependent.

\subsection{Existence and Uniqueness of the Cutoff Policy}\label{app:equilibrium}
This appendix proves Proposition~\ref{prop:separation}: existence and uniqueness of the cutoff $\bar\pi(A)$ and of the Markov policy induced by the maintained belief system. The argument has four steps: (i) a unique bounded continuous value function; (ii) single-crossing of $\Delta(\pi,A)$; (iii) existence and monotonicity of $\bar\pi(A)$; (iv) uniqueness of the cutoff policy. We solve the credible type's problem and drop the $\theta$ superscript from $V^\theta$; the weak type is handled after Step~3. Recall the timing wedge \eqref{eq:Delta_new}, here for the credible type,
\[
\Delta(\pi,A)=\underbrace{\alpha\big(v(\mathrm{early})-v(\mathrm{AGM})\big)+B\,\Gamma^{C}_{\mathrm{early}}(\pi,A)}_{\text{current wedge}}+\underbrace{\beta\big(\mathbb{E}_y V_{\mathrm{early}}-\mathbb{E}_y V_{\mathrm{AGM}}\big)}_{\text{continuation wedge}},
\]
the credible type's payoff advantage from campaigning early rather than at the AGM, so the NGO campaigns early when $\Delta>0$.

\paragraph{Step 1: Value function.} Let $\mathcal{B}$ be the space of bounded continuous $V:(0,1)\times[0,1]\to\mathbb{R}$ under the sup norm, with Bellman operator
\begin{equation}
(\mathcal{T}V)(\pi,A)=\max_{d\in\{\mathrm{early},\mathrm{AGM}\}}\Big\{u^\theta(d;\pi,A)+\beta\,\mathbb{E}[V(\pi',A')\mid \pi,A,d]\Big\}.
\label{eq:Bellman_T}
\end{equation}
Because $u$ is bounded and continuous, $0<\beta<1$, and the transition kernel is well defined and independent of $V$, $\mathcal{T}$ is a contraction of modulus $\beta$. Under the maintained monotonicity of the posterior kernel (for moderate signal sensitivities), and since $u$ is increasing in $(\pi,A)$ by \eqref{eq:GammaProps}, the fixed point $V^\star$ is increasing in each argument. This gives existence and uniqueness of the value function.

\paragraph{Step 2: Single-crossing in $\pi$, monotonicity in $A$.}
\begin{lemma}[Single-Crossing and Monotonicity]
\label{lem:SC_equilibrium}
For the credible type, $\Delta(\pi,A)$ in \eqref{eq:Delta_new} is continuous, strictly increasing in $\pi$, and weakly increasing in $A$.
\end{lemma}
\begin{proof}
\emph{Current wedge.} $\alpha\big(v(\mathrm{early})-v(\mathrm{AGM})\big)+B\,\Gamma^{C}_{\mathrm{early}}(\pi,A)$: the first term is constant in $(\pi,A)$ and, by \eqref{eq:GammaProps}, $\Gamma^{C}_{\mathrm{early}}(\pi,A)$ is strictly increasing in $\pi$ and weakly increasing in $A$, so the current wedge is strictly increasing in $\pi$ and weakly increasing in $A$.

\emph{Continuation wedge.} Write $\Delta^{\mathrm{cont}}(\pi,A)\equiv\mathbb{E}[V(\pi_{t+1},A_{t+1})\mid\mathrm{early}]-\mathbb{E}[V(\pi_{t+1},A_{t+1})\mid\mathrm{AGM}]$. With $E,G$ denoting next-period states under early/AGM,
\begin{equation*}
\Delta^{\mathrm{cont}}(\pi,A)=\underbrace{\mathbb{E}[V(\pi^E_{t+1},A^G_{t+1})]-\mathbb{E}[V(\pi^G_{t+1},A^G_{t+1})]}_{\text{belief channel}}
+\underbrace{\mathbb{E}[V(\pi^E_{t+1},A^E_{t+1})]-\mathbb{E}[V(\pi^E_{t+1},A^G_{t+1})]}_{\text{awareness channel}}.
\end{equation*}
The belief channel is nondecreasing in $(\pi,A)$ under the credible type's law: as $\pi$ or $A$ rises, the posterior distribution following early campaigning shifts upward relative to the AGM branch in expectation, and $V$ is increasing in $\pi$. For the awareness channel, $A^E_{t+1}-A^G_{t+1}=(1-\rho)(v(\mathrm{early})-v(\mathrm{AGM}))\equiv\kappa<0$; writing $D(\pi',a)\equiv V(\pi',a+\kappa)-V(\pi',a)$, the term is $\mathbb{E}[D(\pi^E_{t+1},A^G_{t+1})\mid\mathrm{early}]$. By Assumption~\ref{ass:incdiff}, decreasing differences make the awareness loss from early timing smaller at higher credibility, so $D$ is nondecreasing in $\pi'$, while concavity in $A$ makes $D$ nondecreasing in $a$. Hence the awareness channel is nondecreasing in $(\pi,A)$. Combining, $\Delta^{\mathrm{cont}}$ is nondecreasing in $(\pi,A)$, and with the strictly increasing current wedge, $\Delta(\pi,A)$ is continuous, strictly increasing in $\pi$, weakly increasing in $A$.
\end{proof}
\noindent\emph{Intuition.} At low $(\pi,A)$ the timing signal can be negative and early outcomes only weakly informative, so the visibility term dominates and both types pool on AGM. As $\pi$ or $A$ rises the signal strengthens and the continuation wedge grows; with the strictly increasing current wedge this yields a single crossing, at a lower belief threshold when awareness is higher.

\paragraph{Step 3: Cutoff curve.}
\begin{proposition}[Existence and Shape of the Cutoff Curve]
\label{prop:cutoff_equilibrium}
For each $A\in[0,1]$ there is a unique continuous $\bar\pi(A)\in(0,1)$ such that the NGO campaigns early iff $\pi>\bar\pi(A)$; the cutoff is weakly decreasing in $A$.
\end{proposition}
\noindent\emph{Proof.} By Lemma~\ref{lem:SC_equilibrium}, $\Delta$ is continuous and strictly increasing in $\pi$. At $\pi=0$ influence is worthless now and, the belief being absorbing, forever \eqref{eq:GammaProps}, so the value is independent of the state and $\Delta(0,A)=\alpha(\bar v-1)<0$. At $\pi=1$, Assumption~\ref{ass:boundary} makes $\Delta(1,A)>0$ for all $A$.\footnote{At $\pi=1$ the value $V(1,\cdot)$ is Lipschitz in $A$ with constant at most $B\bar G/(1-\beta\rho)$: the flow payoff moves with $A$ at rate at most $B\bar G$, and next-period awareness inherits differences at rate $\rho$. Since the awareness gap between the two timings is $(1-\rho)(1-\bar v)$, Assumption~\ref{ass:boundary} bounds the discounted continuation cost of early timing below the current influence advantage $B\underline\Gamma$, giving $\Delta(1,A)>0$.} The intermediate value theorem gives a unique $\bar\pi(A)$ with $\Delta(\bar\pi(A),A)=0$, and $\partial\Delta/\partial A\ge0$ with the implicit function theorem gives $\bar\pi'(A)\le0$. \qed

\paragraph{The weak type.} No condition is needed for the weak type: $\Delta^W<0$ follows from the primitives. Since $\Gamma^W\equiv0$ and the exposure term $\alpha v(d)$ does not depend on beliefs, the weak type's flow payoff is independent of $(\pi,A)$; the unique bounded solution of its Bellman equation is the constant $V^W=(\alpha-K)/(1-\beta)$, attained by campaigning at the AGM every period. 
	
A better reputation therefore buys the weak type nothing, and the timing signal creates no mimicry incentive: $\Delta^W(\pi,A)=\alpha(\bar v-1)<0$ everywhere, so the weak type never campaigns early and the cutoff of Proposition~\ref{prop:cutoff_equilibrium} separates exactly as stated in Proposition~\ref{prop:separation}: below $\bar\pi(A)$ both types pool on AGM timing; above it the credible type campaigns early while the weak type stays.

This insulation reflects a modeling choice: the weak type values visibility, not being believed credible. If the exposure return also rose with $\pi$, for instance because donors reward credibility, mimicry incentives would appear, and separation would additionally require the expected outcome penalty of early failure to dominate them; by Lemma~\ref{lem:bern_llr}, under the weak type's law the expected outcome log-likelihood ratio from early campaigning is nonpositive, so failures would erode any mimicked reputation.

\paragraph{Step 4: Policy uniqueness.} Because \eqref{eq:Bellman_T} is a contraction, $V^\star$ is unique; the cutoff is unique for each $A$ and the policy $d^\star(\pi,A)$ is unique off the measure-zero set $\{\pi=\bar\pi(A)\}$.
\begin{proposition}[Existence and Uniqueness of the Markov Policy]
\label{prop:policy_equilibrium}
Under the maintained assumptions and belief system, the model admits a unique bounded continuous $V^\star(\pi,A)$ solving \eqref{eq:Bellman_T} and a unique cutoff policy $\bar\pi(A)$ (Proposition~\ref{prop:cutoff_equilibrium}); the induced Markov policy is unique up to a measure-zero indifference set.
\end{proposition}
\noindent Steps~1--4 together establish Proposition~\ref{prop:separation}: single-crossing (Lemma~\ref{lem:SC_equilibrium}) makes timing incentives monotone, Proposition~\ref{prop:cutoff_equilibrium} gives the cutoff, and Proposition~\ref{prop:policy_equilibrium} makes it part of a unique Markov policy under the maintained belief system.

\clearpage

\section{Data and Measurement}\label{app:data}
\setcounter{table}{0}\renewcommand{\thetable}{B\arabic{table}}
\setcounter{equation}{0}\renewcommand{\theequation}{B\arabic{equation}}
\setcounter{figure}{0}\renewcommand{\thefigure}{B\arabic{figure}}

This appendix documents how the raw sources become the variables used in the analysis: how Sigwatch detects and records campaigns, how we match campaigns to shareholder proposals by topic, and the institutional anchor of the research design, the stability of AGM dates within firms.

\subsection{Sigwatch Data and Methodology}\label{ap:texts}
Sigwatch is a consulting firm that has tracked pressure-group campaigns worldwide since 2010 \auditrev{\citepsecvthree{koenig2017notes,hatte2020geography}}.\footnote{We thank Pamina Koenig for making the Sigwatch data available to us, and Robert Blood, the founder of Sigwatch, for kindly answering our questions.} It monitors more than 13,000 \auditrev{nonprofit} organizations, and its primary sources are the groups' own public-facing websites and social media, rather than news coverage of the companies they target \citepsecvthree{sigwatch2025methodology}. Campaigns are collected with proprietary software and validated by analysts, who record each campaign's date, cause, target firm, the firm's prominence, its sentiment, and its source URL. Prominence runs from zero to four, where four means the firm is named in the campaign's headline; sentiment runs from $-2$ (strong criticism) to $+2$ (strong praise). Our sample keeps the negative-sentiment campaigns in which a single firm appears in the headline, that is, those with prominence four.\footnote{These data are widely used in economics and finance \citepsecvthree{koenig2019social,alfarourena2022responsible,mazet2024brown,brendel2024value,koenig2025local}.}

We checked the source URLs against the original Sigwatch files, recovering at least one link for 2,348 of the 2,350 campaigns. Among the 2,149 primary sources classified with high confidence, 1,990 (92.6\%) are NGO-linked.\footnote{We use the first available numbered Sigwatch link as each campaign's primary source. NGO-linked sources comprise NGO-owned websites and official accounts, related chapters or affiliates, and partner- or coalition-controlled pages, including campaign-platform pages where such control is identified. The denominator excludes broken or missing URLs, unclear classifications, and sources with lower confidence.} Campaigns thus enter the data predominantly through NGOs' own communications or those of affiliated groups rather than through coverage of the target firm. Text annotations cover all 2,350 campaigns, with complete AGM or shareholder-proposal mentions and recent-event classifications for 2,332 (99.2\%). We use GPT-4o to classify the trigger for each campaign based on its text: the firm's AGM or a shareholder proposal, another company-specific event, or a broader event such as a sector-wide development or policy change. If multiple triggers are identified, we assign the most specific one, prioritizing AGM or shareholder-proposal mentions first and other company-specific events second; remaining campaigns are classified as other.

\begin{figure}[!htb]
\color{black}%
	\centering
	\caption{Triggers of NGO Campaigns \label{fig:news_reason}}
		\includegraphics[width=.65\linewidth]{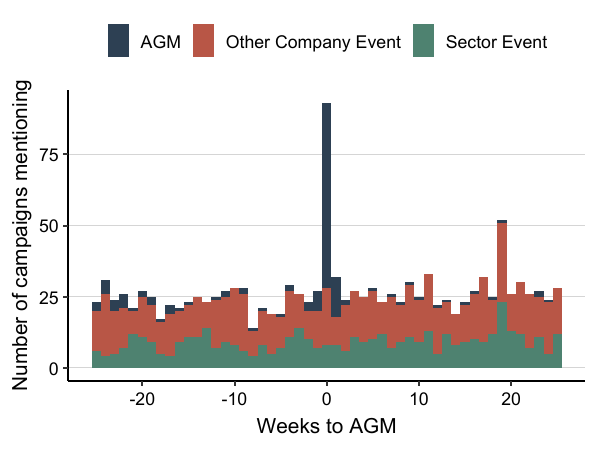}
	\begin{minipage}{1\textwidth}
		{\footnotesize
			Note: The figure plots weekly counts of NGO campaigns by classified trigger relative to the nearest AGM. Records classified as Other are omitted. The plotted categories are \textit{AGM or shareholder-proposal mentions}, \textit{other company events}, and \textit{sector or policy events}. The annotation procedure prioritizes AGM or proposal mentions over other event types. Data from Sigwatch (campaigns) and ISS (AGM dates).
			\par}
	\end{minipage}
\end{figure}

These texts show that the clustering of campaigns on AGM dates in Figure~\ref{fig:intro_fig} reflects deliberate timing by NGOs, not a recording artifact. Because campaigns are detected almost solely from the NGOs' own channels, whether one is recorded is largely independent of the target firm's media coverage, which peaks during the AGM week (Figure~\ref{fig:intro_fig}, Panel~b). Figure~\ref{fig:news_reason} bears this out: campaigns that name the AGM or a shareholder proposal cluster sharply in the AGM week, whereas campaigns triggered by other company events or sector events stay flat around the meeting date. A recording artifact tied to firm coverage, such as the coverage generated by quarterly earnings calls, would have raised the other-event groups too.

\auditrev{We observe public campaigns but not the private engagement that may precede them.} A public campaign can be read as an escalation of a private conversation that has broken down, and the timing of going public is then the exposure-versus-influence choice in our model: an NGO credible enough to sway the coming vote escalates before the AGM to influence it, whereas an NGO that cannot move the vote, or is not credible enough to be heard privately, shames the firm on the AGM date, when public scrutiny peaks. Choosing that moment over continued private pressure is a deliberate bid for exposure, so the possibility of private engagement reinforces rather than weakens our interpretation.

\subsection{The Stability of AGM Dates}\label{app:agmdates}
\begin{figure}[!htb]
	\centering
	\caption{Persistence of AGM month over time \label{fig:AGM}}
	\includegraphics[width=.5\linewidth]{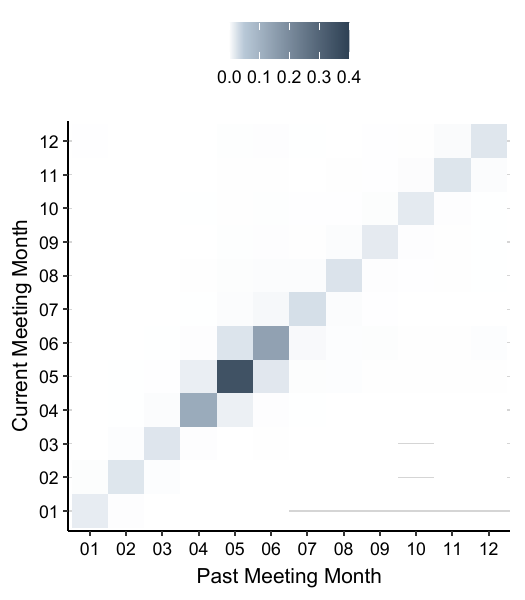}
	\begin{minipage}{1\textwidth}
		{\footnotesize 
			Note: The heatmap shows the joint distribution of the current AGM month (vertical axis) and the previous year's AGM month (horizontal axis). Cell values are shares of all matched meeting observations, not conditional shares within a previous-month category. Darker cells indicate a larger share. The sample contains 38,471 matched meeting observations for 5,745 CUSIPs from 2010 to 2020, with a prior-year match and a CUSIP present among \auditrev{U.S.} Compustat firms. Data: ISS and Compustat. \par}
	\end{minipage}
\end{figure}

The design in Section~\ref{s:strategy} takes AGM dates as a predictable schedule. Figure~\ref{fig:AGM} verifies this premise: a firm's AGM month is highly persistent from one year to the next, so campaigns can be timed to it well in advance.

\subsection{Variable Definitions}
Table~\ref{tab:variables} summarizes the construction and data source of the variables used in the paper.

\begingroup\color{black}\small\setlength{\tabcolsep}{4pt}\renewcommand{\arraystretch}{1.15}
\begin{longtable}{p{0.20\textwidth}p{0.55\textwidth}p{0.19\textwidth}}
\caption{Variable Definitions}\label{tab:variables}\\
\toprule
\textbf{Variable} & \textbf{Construction} & \textbf{Data source} \\
\midrule\endfirsthead
\multicolumn{3}{l}{\textit{Table \thetable{} continued}}\\\toprule
\textbf{Variable} & \textbf{Construction} & \textbf{Data source} \\
\midrule\endhead
\midrule\multicolumn{3}{r}{\textit{Continued on next page}}\\\endfoot
\bottomrule\endlastfoot
Campaign & A dated NGO action against a target firm. & Sigwatch \\
Campaign topic & Campaign issue mapped to the 17 environmental and social categories used to match shareholder proposals; see Table~\ref{tab:proposals}. & Sigwatch; author classification \\
Environmental campaign & Indicator that the assigned campaign topic belongs to an environmental category. & Sigwatch; author classification \\
Campaign sentiment & Sigwatch campaign sentiment score; from -2 (most negative) to 2 (most positive). & Sigwatch \\
AGM date & Target firm annual general meeting date. & ISS \\
AGM-timed campaign & Indicator for a campaign on the matched AGM date or the preceding day. & Sigwatch; ISS \\
Distance to AGM & Signed campaign date minus AGM date in days. & Sigwatch; ISS \\
Campaign order & Chronological campaign number within NGO. & Sigwatch \\
Past campaigns & Number of AGM-matched campaigns through the previous calendar year in proposal-success models. Table~\ref{ngo_experience_success} winsorizes at the 97.5th percentile. & Sigwatch; ISS \\
Past AGM campaigns & Past campaigns on the AGM date or preceding day, accumulated through the previous calendar year. & Sigwatch; ISS \\
Past non-AGM campaigns & Past campaigns minus past AGM campaigns. Table~\ref{ngo_experience_success} winsorizes at the 97.5th percentile. & Sigwatch; ISS \\
Same topic & Indicator that the campaign and shareholder proposal concern the same assigned topic. & Sigwatch; ISS; author classification \\
Proposal filing & Indicator for a shareholder proposal concerning the campaign topic at the prospective AGM. & ISS; Sigwatch topic match \\
Proposal withdrawal & Indicator that the proposal is recorded as withdrawn before the AGM. & ISS \\
Proposal passage & Indicator for more than 50\% voting support. & ISS \\
Proposal success & Indicator that a proposal is withdrawn or passed. & ISS \\
Voting support & Reported percentage of votes supporting a proposal. & ISS; Diligent \\
Google searches & Monthly NGO Google Trends search score, standardized in the attention regressions. & Google Trends \\
Relative Google score & Previous-year mean NGO search score relative to Greenpeace. & Google Trends \\
NGO power & Sigwatch NGO power variable, ranging from 0.5 to 2.75. & Sigwatch \\
\auditrev{U.S.} NGO & \auditrev{Country codes US/USA identify U.S. NGOs; other country codes identify non-U.S. NGOs.} & Sigwatch \\
NGO age & Observation year minus reported formation year. & IRS Form 990 / Urban Institute \\
NGO total assets & End-of-year total assets. & IRS Form 990 / Urban Institute \\
NGO total revenue & Annual total revenue. & IRS Form 990 / Urban Institute \\
NGO total expenses & Annual total expenses. & IRS Form 990 / Urban Institute \\
Contributions and grants & Annual total contributions and grants; displayed in monetary units, as a share of assets, or as a revenue share, depending on the exhibit. & IRS Form 990 / Urban Institute \\
Operating margin & \auditrev{Total revenue minus total expenses, divided by total revenue. Margin is winsorized at the pooled 2.5th and 97.5th percentiles.} & IRS Form 990 / Urban Institute \\
Revenue-source share & Component revenue divided by positive total revenue. Electronic filings separate government grants from other contributions; paper filings combine them. & IRS Form 990 / Urban Institute \\
E-filing & Indicator that the Form 990 record is an electronic filing. & IRS Form 990 / Urban Institute \\
Nonprofit sector & Reported NTEE major-group classification. & IRS Form 990 / Urban Institute \\
Nonprofit location & Reported organization-address state. & IRS Form 990 / Urban Institute \\
Market capitalization & Firm equity market value, computed as share price times number of shares outstanding. & Compustat \\
News count & Count of target-firm news items with relevance at least 75. Selection tables use log annual news count. & RavenPack \\
Negative news & Monthly count of negative target-firm news, winsorized at the 97.5th percentile and standardized in the attention regressions. & RavenPack \\
Environmental score & Firm industry-adjusted environmental pillar score, ranging from 0 to 10. & MSCI \\
Social score & Firm industry-adjusted social pillar score, ranging from 0 to 10. & MSCI \\
E\&S score change & One-year-forward change in the equally weighted environmental and social score. & MSCI \\
ESG reporting scope & Share of the firm's activities covered by its CSR report. & LSEG \\
Board independence change & One-year-forward change in the share of independent board members, following \cite{hsu2025eco}. & BoardEx \\
Sales & Target-firm quarterly sales, entered in logs in the consumer-response regressions. & Compustat \\
Shareholder HHI & Sum of squared institutional ownership shares. & 13F \\
Blockholding & Fraction of equity held by institutional blockholders with ownership stakes exceeding 5\%. & 13F \\
Blockholder count & Number of institutional holders with stakes exceeding 5\%. & 13F \\
Collaborative campaign & Campaign listing more than one distinct NGO among recorded participant fields; a solo campaign has one participant. & Sigwatch \\
Collaboration size & Number of distinct NGOs participating in the current campaign. & Sigwatch \\
Past solo campaigns & Prior-year campaigns in which an NGO participated alone. & Sigwatch; ISS \\
Past collaborative campaigns & Prior-year campaigns in which an NGO participated with other NGOs. & Sigwatch; ISS \\
Exact-coalition history & Number of prior-year AGM-matched campaigns involving the identical set of NGO participants. & Sigwatch; ISS \\
FTI score & Firm score in the Fashion Transparency Index. & Fashion Revolution \\
Cumulative excess return & Daily firm stock return minus its national stock-index return, accumulated around first FTI publication; daily excess returns are winsorized within event day. & LSEG; Fashion Revolution \\
\end{longtable}
\noindent\footnotesize Note: Definitions describe the variables used in the paper. Timing windows, omitted bins, sample restrictions, quantile groupings, fixed effects and standard-error clustering are specified in the relevant exhibit notes. Standardization subtracts the sample mean and divides by the sample standard deviation. Squared terms and same-topic interactions are constructed from the corresponding regressors.
\par\endgroup

\clearpage
\subsection{Matching Proposals and NGO Campaigns}\label{apndx:match}
This section outlines the process of matching each shareholder proposal in the ISS dataset with a relevant NGO campaign cause.

\begingroup\color{black}%
Sigwatch assigns ``Generic Topics'' to campaigns in its dataset. For campaigns with missing topics (primarily in earlier years), we infer them from the more granular ``Issue names''. We then retain the first assigned topic for each campaign. To streamline the analysis, we combine certain causes when they are difficult to distinguish or when one category has few campaigns and can reasonably fit within a broader category. Specifically, we merge ``Labor Rights'' with ``Human \& Labor Rights," ``Children'' with "Human \& Labor Rights," ``CSR / ESG'' with ``CSR," and ``Biodiversity'' and ``Wildlife'' with "Environment." For campaigns under the ``Safety'' cause, we examine each campaign’s specific purpose and classify it as either ``Product Safety'' or ``Food Safety."
\endgroup%

Using these consolidated ``Generic Cause Names'', we then categorize each shareholder proposal based on its short description. When a shareholder proposal is submitted, it includes a brief, one-sentence description, which is provided in the ISS dataset and serves as the foundation for cause classification. We start by assigning proposals to causes based on keywords in their descriptions and then manually classify any remaining proposals. 

\begin{table}[h!]
\color{black}
\caption{Proposal Causes} \label{tab:proposals}
\begin{center}
\resizebox{\textwidth}{!}{
\begin{tabular}{lcr}
\toprule
NGO Campaign  & Category & Example of Shareholder Proposals  \\ 
Topic Classification & (E or S)  & \\
  \midrule
Climate Change & E & Encourage and Support Carbon Tax to Address Climate Change \\
Environment & E & Stop financing mountain top removal coal mining \\
Pollution & E & Report on Efforts to Reduce Pollution From Products and Operations \\
Rainforest & E & Report on impact on rainforest \\
Sustainability & E & Establish board committee on sustainability \\
Waste Generation & E & Do not increase radioactive waste production \\
Water & E & Report on supply chain water impacts \\
\midrule
Animal Rights \& Welfare & S & Review suppliers' animal slaughter methods \\
CSR & S & Disclose Charitable Contributions \\
Food Safety & S & Label Genetically Modified Ingredients\\
Health & S & Reduce Tobacco Harm to Health \\
Human \& Labor Rights & S & End Sri Lanka Trade Partnerships Until Human Rights Improve \\
Indigenous People & S & Adopt Global Policy Regarding the Rights of Indigenous People \\
Intellectual Property Rights & S & Implement a Process to Fully Vet Whether a \\ 
& & Supervisor is a Bona Fide Inventor Before a Patent is Filed \\
Politics & S & Affirm Political Nonpartisanship \\
Product Safety & S & Report on Nanomaterial Product Safety \\
Worker Safety & S & Report on Reducing Risks of Accidents \\
 \bottomrule
\end{tabular}}
\end{center}
\begin{tablenotes}
\footnotesize \vspace{-1.5em}
\item Note: This table lists the 17 consolidated topics used to classify shareholder proposals and match them to NGO campaigns. Shareholder proposal data are from ISS. The topics consolidate Sigwatch campaign categories using the author classification described in Appendix~\ref{apndx:match}. 
\end{tablenotes}
\end{table}

\subsection{Daily campaign timing around the AGM}\label{ap:postagm}
\begin{figure}[!htb]
\color{black}%
	\centering

	\caption{Daily NGO Campaigns Around AGMs \label{fig:campaigns_daily}}
	\begin{subfigure}{0.49\linewidth}
		\centering
		\includegraphics[width=\linewidth]{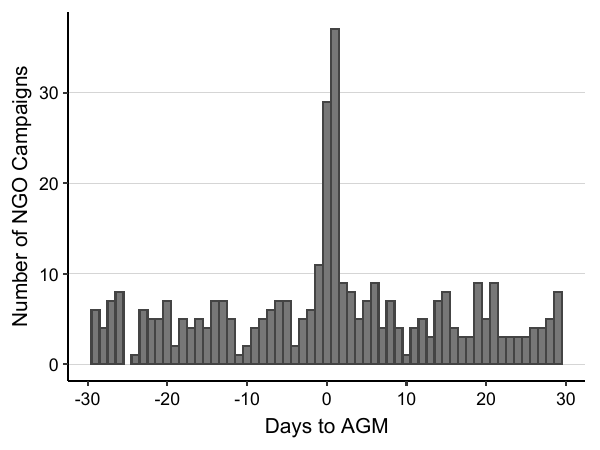}
		\caption{Campaign Timing around AGMs}
	\end{subfigure}
	\begin{subfigure}{0.49\linewidth}
		\centering
		\includegraphics[width=\linewidth]{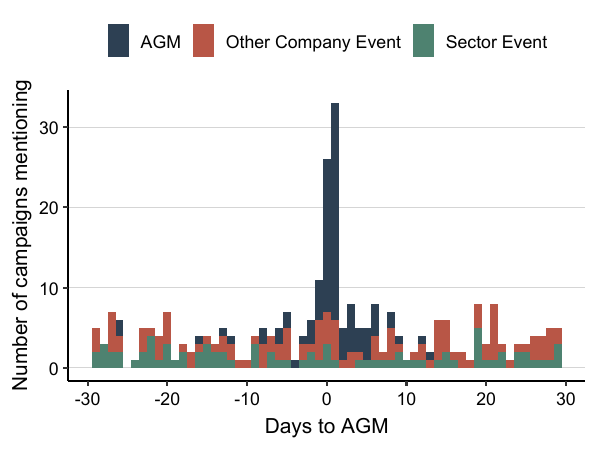}
		\caption{Campaign Triggers around AGMs}
	\end{subfigure}
	\begin{minipage}{1\textwidth}
		{\footnotesize
			Note: The figure plots the daily number of NGO campaigns around target firms' AGMs. Panel (a) reports the overall timing of campaigns relative to the AGM date. Panel (b) reports classified campaign triggers over the same window. \auditrev{Both panels use observations from days $-29$ to $+29$.} Panel (b) omits records classified as Other, so its counts need not sum to Panel (a). Data from Sigwatch (campaigns) and ISS (AGM dates).
			\par}
	\end{minipage}
\end{figure}

Zooming to the daily level, the clustering on the AGM date is accompanied by a second feature: campaigns also spike in the days just after the meeting (Figure~\ref{fig:campaigns_daily}, Panel~a), and these later campaigns are overwhelmingly AGM-related (Panel~b). Their texts often refer to a \textit{future} AGM: NGOs react to the meeting that just closed and announce the actions they will bring to next year's. These campaigns are the reactive, forward-looking phase of the same engagement cycle rather than a separate strategy: having pressed their case at one meeting, NGOs immediately begin building toward the next.

\color{black}

\clearpage

\section{Selection into Activism and Targeting}\label{app:selection}
\setcounter{table}{0}\renewcommand{\thetable}{C\arabic{table}}
\setcounter{equation}{0}\renewcommand{\theequation}{C\arabic{equation}}
\setcounter{figure}{0}\renewcommand{\thefigure}{C\arabic{figure}}

This appendix asks whether selection, rather than the strategic timing the paper documents, could explain our results. It proceeds at three levels: which nonprofits become activists at all, whom NGOs target, and whether targets drift as NGOs gain experience. Activist NGOs differ from the broader nonprofit population, which defines the population to which our results apply; among activist NGOs, however, we find little evidence that target selection on observables explains the timing patterns. Section~\ref{s:case} addresses selection on unobservables with quasi-exogenous timing.

\subsection{Who Becomes an Activist}\label{apndx:selection}
We start from the entire universe of \auditrev{U.S.} non-profits filing Form 990 at the IRS between 2010 and 2021, for a total of 983,726 non-profits. We combine both e-filings and paper filings for all entities during that period and categorize as \textit{activist} any \auditrev{nonprofit} identified as being in Sigwatch, regardless of the company targeted or campaign sentiment, for a total of 957 NGOs matched by EIN to the \auditrev{U.S.} NGO roster. Combining the two matters because electronic filing became mandatory only with the Taxpayer First Act of 2019 (Pub.\ L.\ No.\ 116-25), so paper filings were still prevalent in the earlier years of the sample.

\begin{figure}[htbp]
\color{black}%
    \centering
            \caption{Comparison between Activist and Other Nonprofits}
        \includegraphics[width=\linewidth]{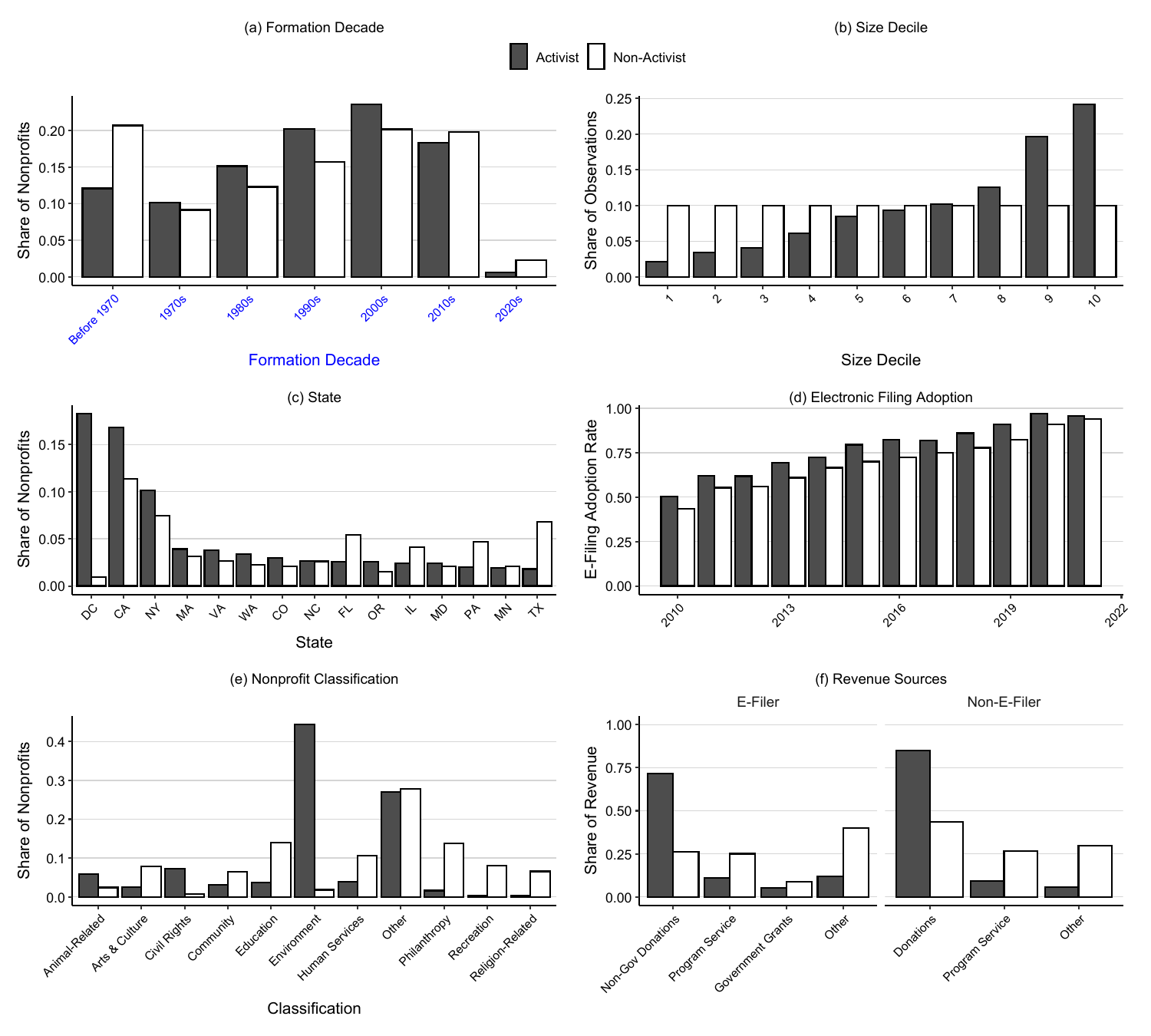}
        \label{fig:form990}
	\begin{minipage}{1\textwidth}
	{\footnotesize
	Note: The sample covers IRS Form 990 / Urban Institute observations from 2010 to 2021, including electronic and paper filings. Activist status is assigned by matching EINs to the broader \auditrev{U.S.} Sigwatch NGO roster, not just the 216 \auditrev{U.S.} NGOs in the main campaign sample. \auditrev{Panel (a) groups formation years into decades: the first bin contains years before 1970, followed by 1970--1979, 1980--1989, and subsequent decades.} Formation-year, state, and sector panels use distinct combinations of NGO and reported organizational characteristics; asset-decile and e-filing panels use filing observations. Panel (f) reports average revenue shares separately for e-filers and paper filers. For e-filers, non-government donations equal total contributions and grants minus government grants. Missing contributions and program-service revenues are set to zero; observations with nonpositive revenue or component shares outside $[0,1]$ are excluded. Data: Urban Institute Form 990 files. \par}
\end{minipage}
\end{figure}

Figure~\ref{fig:form990} compares activist and non-activist nonprofits across several observable characteristics. Overall, the figure suggests that activist nonprofits are not simply a random subset of the nonprofit sector. Instead, they appear to differ systematically from non-activist organizations in terms of age, sector, geography, size, filing behavior, and revenue structure.

\paragraph{Age.} Panel (a) shows that activist nonprofits are more concentrated among \auditrev{organizations founded in the 1970s through the 2000s}, especially those founded in the 2000s. This pattern suggests that activist nonprofits may be relatively newer organizations, or that activism has become more common among nonprofits formed in recent decades. By contrast, non-activist nonprofits are more represented among \auditrev{organizations founded before 1970 and since 2010}.

\paragraph{Size.} Panel (b) shows that activist nonprofits are more concentrated in the upper size deciles, especially deciles 8 through 10. This suggests that activist organizations tend to be larger than non-activist organizations. Larger organizations may have more staff capacity, financial resources, and organizational infrastructure to engage in advocacy, public campaigns, or political activity. Non-activist nonprofits, by contrast, are more evenly distributed across size deciles.

\paragraph{Location.} Panel (c) reports the distribution across states. Activist nonprofits are disproportionately located in a small number of politically or organizationally important locations, particularly the District of Columbia, California, and New York. This geographic concentration may reflect proximity to policymakers, donors, advocacy networks, media markets, or large nonprofit ecosystems. In contrast, non-activist nonprofits appear more geographically dispersed, with relatively larger shares in states such as Texas, Florida, and Pennsylvania.

\paragraph{Filing behavior.} Panel (d) shows that e-filing adoption generally increased over time for both activist and non-activist nonprofits. However, activist nonprofits have higher e-filing adoption rates in every year. This may indicate that activist organizations are more administratively professionalized, more likely to comply with digital reporting requirements, or more likely to have the resources needed to adopt electronic filing earlier. More broadly, e-filing status may proxy for an NGO's technological sophistication, a dimension that future work on nonprofit behavior could exploit.

\paragraph{Sector and mission.} Panel (e) shows substantial differences in NTEE classification. Activist nonprofits are especially concentrated in environmental, civil-rights, and animal-related categories. This is consistent with the idea that nonprofit activism is more prevalent in issue areas where policy advocacy, public mobilization, and regulatory debates are central to organizational missions. Non-activist nonprofits, by contrast, are more represented in categories such as education, philanthropy, human services, recreation, and religion, suggesting that many non-activist organizations may focus more on service provision, grantmaking, or community-based activities.

\paragraph{Revenue structure.} Panel (f) shows that activist nonprofits rely more heavily on contributions and grants than on fees, program services, or other earned income. Among e-filers, non-government donations account for 71.7\% of activist NGOs' revenue on average and government grants for 5.4\%, compared with 26.1\% and 8.9\% for non-activists. Most contribution revenue for activists thus comes from non-government sources. Among paper filers, contributions and grants account for 84.9\% of activists' revenue, compared with 43.6\% for non-activists; government contributions cannot be separately identified. This reliance on contributions may reflect the nature of advocacy work, which often produces public or collective benefits rather than fee-for-service outputs. Non-activists, by contrast, receive larger revenue shares from program services and other sources.

\clearpage
\subsection{Whom NGOs Target}\label{apndx:whom}
\begin{figure}[!htbp]
    \centering
    \caption{How do target characteristics affect the decision to campaign on AGM dates?}
    \includegraphics[width=0.6\textwidth]{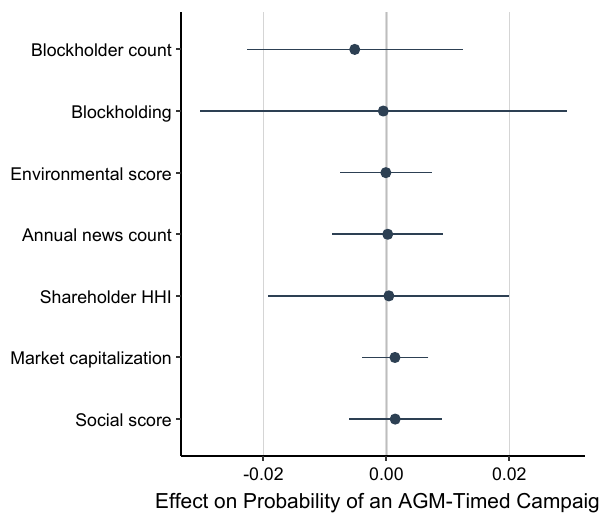}
    \label{fig:selection}
    \begin{minipage}{1 \textwidth} %
 {\footnotesize Note: This figure reports coefficients from a linear probability model relating NGO campaign timing to target-firm characteristics:
\begin{equation*}
\mathbf{1}_{\{\text{AGM-timed campaign}_c\}}=\alpha+\mathbf{X}_c\beta+\delta_{y(c)}+\varepsilon_c.
\end{equation*}
The dependent variable equals one for a campaign on the AGM date or preceding day. The vector $\mathbf{X}_c$ contains standardized market-capitalization and news-count deciles, environmental and social scores, and ownership HHI, blockholding, and blockholder-count deciles. The specification includes year fixed effects ($\delta_{y(c)}$). Error bars show 90\% and 95\% confidence intervals based on standard errors clustered by firm. Data combine Sigwatch, ISS, Form 990, MSCI, RavenPack, and 13F filings. \par}
 \end{minipage}
 \end{figure}

\begingroup\color{black}%
Section~\ref{s:target} argues that strategic targeting would leave traces in who campaigns against whom; this appendix examines them one at a time. The first would be larger or more experienced NGOs gravitating toward bigger, more covered, or more concentrated firms. Table~\ref{tab:selection} shows little evidence of a uniform pattern. With fixed effects, NGO size has little association with most target characteristics, although larger NGOs target firms with lower shareholder HHI and blockholding at the 10\% significance level (Panel~a). Past AGM experience is also largely unrelated to target characteristics, except for a lower blockholder count (Panel~b). Nor is there clear evidence that the NGO itself changes as its record grows: larger NGOs campaign more in the cross section, but within an NGO, associations of assets, contributions/assets, and expenses/assets with accumulated campaign counts are imprecise once NGO and year fixed effects are included (Panel~c).

A second trace would be geographic: \auditrev{U.S.} and foreign NGOs differ in resources and proximity to \auditrev{U.S.} firms, so if targets were chosen to fit the campaigner, the two groups should pick different firms. They do campaign differently, with \auditrev{U.S.} NGOs campaigning more often and more often on AGM dates; their higher environmental campaign share is not statistically significant (Table~\ref{tab:selection_us_ngos}, Panel~a). \auditrev{U.S.} NGOs target smaller, less-covered firms with more concentrated ownership in the cross section, but these differences become imprecise with fixed effects. Only the lower blockholder count remains significant at the 10\% level (Panel~b). The clearest differences therefore concern how intensively the two groups campaign, with limited evidence of systematic differences in whom they pursue after fixed effects.

The last trace would sit with the firm: some characteristic that attracts AGM-timed campaigns. Putting all candidates on a common scale, a regression of an AGM-timing indicator on standardized firm characteristics finds that none significantly predicts when the campaign lands (Figure~\ref{fig:selection}). The observable profile of the firm therefore offers little explanation for campaign timing.
\endgroup%

\begin{table}[!ht]
\color{black}
    \caption{Selection on Observables \label{tab:selection}}
\begin{center}
\vspace{-0.5cm}

\resizebox{0.5\textheight}{!}{
\begin{tabular}{lccccc}
\toprule
\midrule
\multicolumn{6}{l}{\textbf{Panel a: Target characteristics and NGO assets decile}} \\
\multicolumn{6}{c}{\textcolor{black}{$z_c=\beta_0+\beta_1\,\textit{Assets Decile}_{g(c),y(c)}+\varepsilon_c$}} \\
\midrule
Dependent variable: & Coef., no FE & SE & Coef., FE & SE & N: no FE / FE \\
Market capitalization (decile) & \auditrev{0.002} & \auditrev{(0.008)} & \auditrev{0.010} & \auditrev{(0.012)} & \auditrev{750} / 611 \\
Annual number of news articles (log) & \auditrev{-0.052**} & \auditrev{(0.021)} & \auditrev{-0.001} & \auditrev{(0.012)} & \auditrev{733} / 589 \\
Environmental score & \auditrev{-0.036} & \auditrev{(0.043)} & \auditrev{-0.053} & \auditrev{(0.066)} & \auditrev{743} / 599 \\
Social score & \auditrev{-0.006} & \auditrev{(0.025)} & \auditrev{0.076} & \auditrev{(0.079)} & \auditrev{743} / 599 \\
Shareholder HHI (decile) & \auditrev{0.045} & \auditrev{(0.045)} & \auditrev{-0.166*} & \auditrev{(0.087)} & \auditrev{767} / 622 \\
Blockholding (decile) & \auditrev{0.035} & \auditrev{(0.042)} & \auditrev{-0.176*} & \auditrev{(0.091)} & \auditrev{767} / 622 \\
Blockholder count (decile) & \auditrev{0.029} & \auditrev{(0.040)} & \auditrev{-0.159} & \auditrev{(0.104)} & \auditrev{767} / 622 \\
\midrule
\multicolumn{6}{l}{\textbf{Panel b: Target characteristics and AGM experience (0/1)}} \\
\multicolumn{6}{c}{\textcolor{black}{$z_c=\beta_0+\beta_1\,\textit{AGM Experience}_c+\varepsilon_c$}} \\
\midrule
Dependent variable: & Coef., no FE & SE & Coef., FE & SE & N: no FE / FE \\
Market capitalization (decile) & \auditrev{-0.089*} & \auditrev{(0.045)} & \auditrev{-0.012} & \auditrev{(0.038)} & \auditrev{1,732} / 1,351 \\
Annual number of news articles (log) & \auditrev{-0.174} & \auditrev{(0.117)} & \auditrev{-0.028} & \auditrev{(0.036)} & \auditrev{1,710} / 1,321 \\
Environmental score & \auditrev{-0.056} & \auditrev{(0.219)} & \auditrev{-0.177} & \auditrev{(0.211)} & \auditrev{1,710} / 1,326 \\
Social score & \auditrev{0.064} & \auditrev{(0.163)} & \auditrev{0.013} & \auditrev{(0.176)} & \auditrev{1,710} / 1,326 \\
Shareholder HHI (decile) & \auditrev{0.546**} & \auditrev{(0.250)} & \auditrev{-0.262} & \auditrev{(0.211)} & \auditrev{1,761} / 1,370 \\
Blockholding (decile) & \auditrev{0.438*} & \auditrev{(0.228)} & \auditrev{-0.349} & \auditrev{(0.220)} & \auditrev{1,761} / 1,370 \\
Blockholder count (decile) & \auditrev{0.276} & \auditrev{(0.208)} & \auditrev{-0.541**} & \auditrev{(0.257)} & \auditrev{1,761} / 1,370 \\
\midrule
\multicolumn{6}{l}{\textbf{Panel c: NGO characteristics and number of past campaigns}} \\
\multicolumn{6}{c}{\textcolor{black}{$z_{gy}=\beta_0+\beta_1\,\textit{\# Past Campaigns}_{gy}+\varepsilon_{gy}$}} \\
\midrule
Dependent variable: & Coef., no FE & SE & Coef., FE & SE & N: no FE / FE \\
NGO total assets (log) & \auditrev{0.038***} & \auditrev{(0.007)} & \auditrev{-0.004} & \auditrev{(0.003)} & \auditrev{768} / 726 \\
Contributions / Total assets & \auditrev{0.001} & \auditrev{(0.011)} & \auditrev{-0.001} & \auditrev{(0.008)} & \auditrev{752} / 711 \\
Expenses / Total assets & \auditrev{-0.002} & \auditrev{(0.010)} & \auditrev{0.001} & \auditrev{(0.008)} & \auditrev{768} / 726 \\
\bottomrule
\end{tabular}}
\end{center}

\begin{tablenotes}[flushleft,para]
\footnotesize\color{black}
\item Note: This table reports selection-on-observables tests across three panels. Each regression row reports $\beta_1$ and its standard error from the panel-header equation; $z$ denotes the characteristic listed in that row, $c$ the campaign, $g$ the NGO, and $y$ the year.\\
\item \textbf{Panel (a).} Change in targets' characteristics as an NGO grows. Each row regresses a target characteristic (e.g., \textit{Market capitalization decile}) on the NGO's size decile, based on total assets. Columns (3)--(4) include \textit{NGO}, \textit{Topic-by-Year}, and \textit{Firm} fixed effects. Standard errors are clustered by firm.\\
\item \textbf{Panel (b).} Selection-on-observables test comparing targets of NGOs with and without AGM campaign experience. Each row regresses a target characteristic on an indicator for at least one AGM campaign in the NGO's cumulative history. Columns (3)--(4) include the same fixed effects as Panel (a), with standard errors clustered by firm.\\
\item \textbf{Panel (c).} Change in NGO characteristics as an NGO campaigns multiple times. Each row regresses an NGO characteristic (e.g., \textit{Total assets (log)}) on its cumulative campaign count. Columns (3)--(4) include \textit{NGO (EIN)} and \textit{Year} fixed effects. Standard errors are clustered by NGO name.\\
Separate sample sizes are reported for regressions without and with fixed effects. Data combine Compustat, Sigwatch, IRS Form 990 / Urban Institute, MSCI, 13F ownership, and RavenPack. * p $<$ 0.10; ** p $<$ 0.05; *** p $<$ 0.01.
\end{tablenotes}
\end{table}
\begin{table}[!htbp]
    \caption{\textbf{Differences Between \auditrev{U.S.} and \auditrev{non-U.S.} NGOs }}\label{tab:selection_us_ngos}
\begin{center}

\resizebox{0.95\textwidth}{!}{
\begin{tabular}{lccccc}
\toprule
\midrule
\multicolumn{6}{l}{\textbf{Panel a: NGO-level campaign activity by \auditrev{U.S.} status}} \\
\midrule
 & \auditrev{U.S.} NGOs (N=216) &  & Other NGOs (N=419) &  &  \\
 & Mean & SE & Mean & SE & p-value \\
Number of campaigns & 6.315 & 0.892 & 2.353 & 0.144 & 0.000 \\
Number of AGM campaigns & 0.157 & 0.042 & 0.014 & 0.006 & 0.001 \\
Environmental campaign share & 0.496 & 0.031 & 0.440 & 0.023 & 0.142 \\
Campaign sentiment & -1.671 & 0.031 & -1.741 & 0.023 & 0.071 \\
\midrule
\multicolumn{6}{l}{\textbf{Panel b: Target characteristics and \auditrev{U.S.} NGO status (0/1)}} \\
\multicolumn{6}{c}{\textcolor{black}{$z_c=\beta_0+\beta_1\,\textit{\auditrev{U.S.} NGO}_c+\varepsilon_c$}} \\
\midrule
Dependent variable: & Coef., no FE & SE & Coef., FE & SE & N: no FE / FE \\
Market capitalization (decile) & -0.117*** & (0.035) & -0.016 & (0.010) & 1,732 / 1,626 \\
Annual number of news articles (log) & -0.396*** & (0.120) & 0.023 & (0.016) & 1,710 / 1,599 \\
Environmental score & -0.345 & (0.245) & 0.019 & (0.064) & 1,710 / 1,603 \\
Social score & -0.137 & (0.156) & -0.059 & (0.055) & 1,710 / 1,603 \\
Shareholder HHI (decile) & 0.549** & (0.251) & -0.039 & (0.065) & 1,761 / 1,648 \\
Blockholding (decile) & 0.302 & (0.230) & -0.098 & (0.085) & 1,761 / 1,648 \\
Blockholder count (decile) & 0.112 & (0.205) & -0.167* & (0.098) & 1,761 / 1,648 \\
\bottomrule
\end{tabular}}
\end{center}

\begin{tablenotes}[flushleft,para]
\footnotesize\color{black}
\item Note: This table reports summary statistics and selection-on-observables tests across two panels. \auditrev{U.S. status follows the NGO country recorded by Sigwatch: codes US and USA identify U.S. NGOs; other country codes identify non-U.S. NGOs.}\\
\item \textbf{Panel (a).} Difference in NGO campaign activity between \auditrev{U.S.} and \auditrev{non-U.S.} NGOs. The sample includes all 2,350 campaigns, and gives equal weight to each NGO. Entries report means, standard errors, and two-sided Welch-test p-values. AGM campaigns occur on the meeting date or preceding day; campaigns without an AGM match are not counted as AGM campaigns.\\
\item \textbf{Panel (b).} Selection-on-observables test comparing targets of \auditrev{U.S.} versus \auditrev{non-U.S.} NGOs. Each row reports $\beta_1$ from the panel-header regression of a target characteristic $z_c$ (e.g., \textit{Market capitalization decile}) on a dummy indicating that the campaigning NGO is based in the \auditrev{U.S.}. The sample retains campaigns within 200 days of the nearest AGM. Columns (3)--(4) add \textit{Topic-by-Year} and \textit{Firm} fixed effects. Standard errors are clustered by firm. Separate sample sizes are reported without and with fixed effects. * p $<$ 0.10; ** p $<$ 0.05; *** p $<$ 0.01.
\end{tablenotes}
\end{table}

\begin{figure}
    \caption{Past campaigns and target firm's size}
    \centering
    \includegraphics[width=0.66\linewidth]{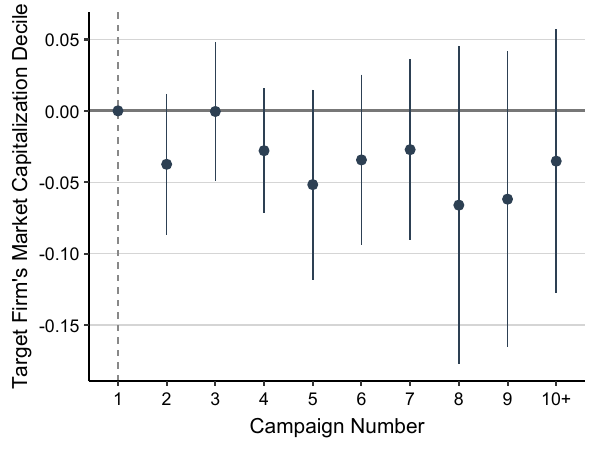}
    \label{ngo_campaign_experience_mc}
    
\begin{minipage}{1\textwidth}%
{\footnotesize Note: This figure plots coefficients from a regression of the target firm's market capitalization on indicators for the focal NGO's campaign order, following the approach in Equation~\eqref{eq:timing}. Market capitalization is measured in deciles. The first campaign is omitted, and the tenth and later campaigns are pooled. The specification includes NGO-by-year, topic-by-year, and firm fixed effects. Error bars denote 95\% confidence intervals based on standard errors clustered by firm. Data combine ISS (AGM dates), Compustat (market capitalization), and Sigwatch (NGO campaigns). \par}
\end{minipage}

\end{figure}

\begin{figure}
\caption{Past campaigns and target firm's E\&S characteristics} \label{ngo_campaign_experience_esg}
    \centering

\begin{subfigure}[t]{0.66\textwidth}
 \centering
 \subcaption{Target firm's E score}
 \includegraphics[width= \linewidth]{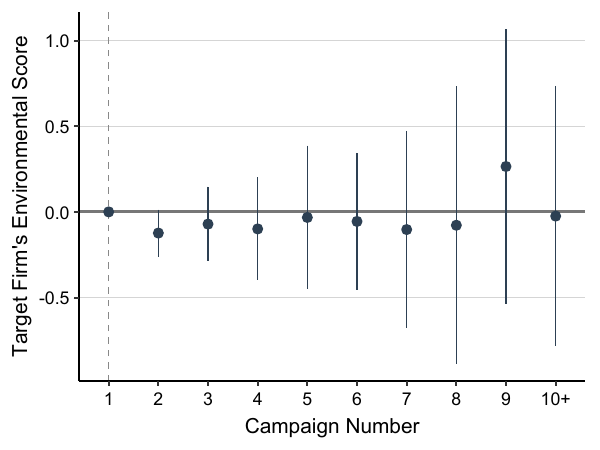}
\end{subfigure}

\begin{subfigure}[t]{0.66\textwidth}
 \centering
 \subcaption{Target firm's S score}
 \includegraphics[width= \linewidth]{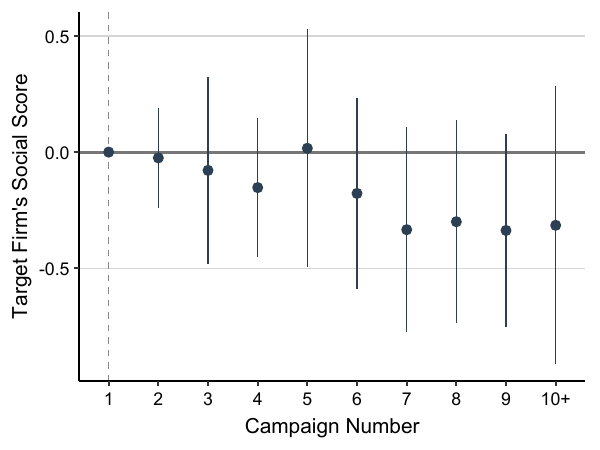}
\end{subfigure}

\begin{minipage}{1\textwidth}%
{\footnotesize Note: This figure plots coefficients from regressions of the target firm's standardized environmental (Panel~a) and social (Panel~b) scores on indicators for the focal NGO's campaign order, following the approach in Equation~\eqref{eq:timing}. The panels restrict campaigns to environmental and social topics, respectively; campaign order is measured across all campaigns. The first campaign is omitted, and the tenth and later campaigns are pooled. The specification includes NGO-by-year, topic-by-year, and firm fixed effects, as in Figures~\ref{ngo_campaign_experience_mc} and~\ref{ngo_campaign_experience_concentration}. Error bars denote 95\% confidence intervals based on standard errors clustered by firm. Data combine ISS, MSCI ESG scores, and Sigwatch. \par}
\end{minipage}
\end{figure}

\begin{figure}
\color{black}%
    \caption{Past campaigns and target firm's ownership concentration}
    \centering
    \includegraphics[width= \linewidth]{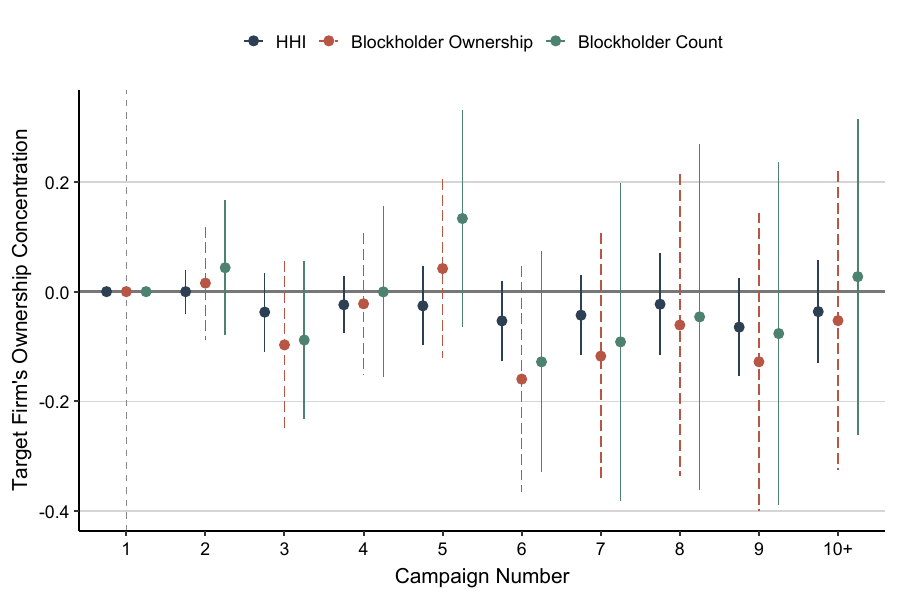}
    \label{ngo_campaign_experience_concentration}
    
\begin{minipage}{1\textwidth}%
{\footnotesize Note: This figure plots coefficients from regressions of the target firm's ownership concentration on indicators for the focal NGO's campaign order, following the approach in Equation~\eqref{eq:timing}. Ownership concentration is measured by the HHI of institutional holdings (blue circles), blockholding share (red circles), and blockholder count (green circles), using lagged holdings and standardized outcomes. The first campaign is omitted, and the tenth and later campaigns are pooled. The specification includes NGO-by-year, topic-by-year, and firm fixed effects. Error bars denote 95\% confidence intervals based on standard errors clustered by firm. Data combine 13F ownership filings and Sigwatch. \par}
\end{minipage}
\end{figure}

\begingroup\color{black}%
\clearpage
\subsection{Experience and Target Characteristics}
\begingroup\color{black}
Finally, because the \auditrev{life cycle} results compare campaigns across an NGO's career, we examine whether targets drift as experience accumulates. Neither target size (Figure~\ref{ngo_campaign_experience_mc}), E\&S scores (Figure~\ref{ngo_campaign_experience_esg}), nor ownership concentration (Figure~\ref{ngo_campaign_experience_concentration}) shows a systematic progression with campaign order. These patterns offer limited evidence of a systematic change in targets as NGOs gain experience.
\endgroup
\endgroup%

\clearpage

\section{Robustness of the Exposure Results}\label{apndx:specs}
\setcounter{table}{0}\renewcommand{\thetable}{D\arabic{table}}
\setcounter{equation}{0}\renewcommand{\theequation}{D\arabic{equation}}
\setcounter{figure}{0}\renewcommand{\thefigure}{D\arabic{figure}}

This appendix tests the robustness of the exposure results of Section~\ref{s:media} in four steps: alternative estimators, for heterogeneous campaign timing and overlapping campaigns; the benchmark of campaigns launched away from the AGM; the full sample, without the maturity split; and the consumer margin, where we find no response.

\subsection{Staggered Difference-in-Differences Specification}
We first replicate Figure \ref{fig:gtrends_campaigns}, applying the staggered difference-in-differences approach to account for heterogeneous campaign timing across NGOs \citep{sun2021estimating}. 

\begin{figure}[h]
\color{black}%
     \centering
          \caption{Staggered difference-in-differences specification} 
          \label{fig:gtrends_campaigns_sunab_first_later}
    \centering\includegraphics[width= 0.5\textwidth]{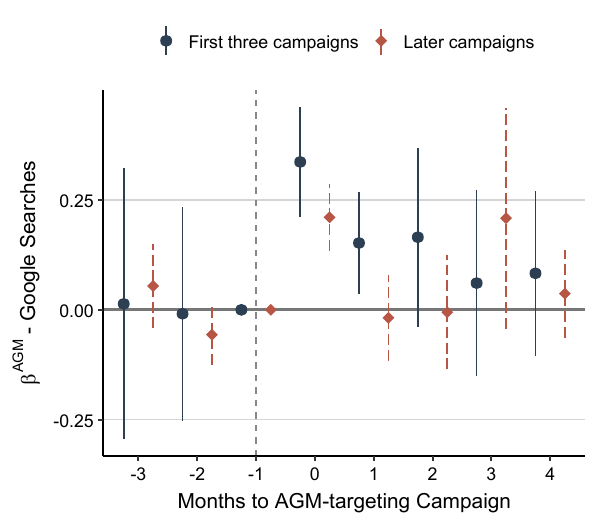}
   \begin{minipage}{1 \textwidth} %
 {\footnotesize Note: This figure plots event-study coefficients $\beta^{AGM}$ estimating the impact of NGO campaigns on public attention similar to Equation~\eqref{eq:awareness}, using the staggered difference-in-differences estimator of \cite{sun2021estimating}. The dependent variable is standardized monthly Google search activity for the focal NGO. Blue circles represent estimates for the first three campaigns and red diamonds later campaigns. The specification accounts for cohort-specific treatment effects to ensure robustness against heterogeneous treatment effects in a staggered setting. Error bars denote 95\% confidence intervals based on standard errors clustered at the campaign level. Data are sourced from Google Trends, ISS, and Sigwatch.}
 \end{minipage}
 \end{figure}

\begingroup\color{black}
Estimated coefficients $\beta_{\tau}^{AGM}$ are reported in Figure~\ref{fig:gtrends_campaigns_sunab_first_later}. The results are consistent with our baseline findings: AGM-timed campaigns generate sizeable visibility gains for young NGOs, raising standardized Google search intensity by 0.336 in the campaign month and 0.152 in the following month. By contrast, later campaigns generate a smaller increase of 0.210 standard deviations in the campaign month, with no significant post-campaign persistence at the 5\% level.
\endgroup

\subsection{Distributed Lags Specification}
In the baseline event-study, if an NGO launches several campaigns in close succession, their effects may overlap in calendar time. For example, if one campaign takes place at $t=0$ and another at $t=2$, the months around $t=2$ capture the lingering impact of both events. This overlap can bias the estimated effects in the previous analyses. To address this, we estimate a distributed-lag specification that explicitly includes \textit{leads} and \textit{lags} of both AGM and non-AGM campaigns:
\begingroup\color{black}%
\begin{equation}
    \begin{aligned}\label{eq:dist_lag}
\text{Google Trends Score}_{g,m,y}  & = \sum^{\tau=5}_{\substack{\tau=-3}} \left(\beta_{\tau}^{AGM} \cdot \text{AGM Campaign}_{g,m-\tau} \right) \\ 
&\quad + \sum^{\tau=5}_{\substack{\tau=-3}} \left(\beta_{\tau}^{Non-AGM} \cdot \text{Non-AGM Campaign}_{g,m-\tau}\right)
\\  & \quad  + \alpha_{m,y} + \alpha_{g,y}+ u_{g,m,y},
 \end{aligned}
\end{equation}
\endgroup%
\begingroup\color{black}%
By doing so, the regression attributes observed variation in search activity in each month to the full sequence of campaigns occurring in that window, rather than assigning the entire effect to whichever campaign happens to be closest in event time. Put differently, the distributed-lag approach disentangles the impact of overlapping campaigns by allowing each campaign to contribute separately to visibility over several months. The estimated $\beta^{AGM}_\tau$ therefore measures the response to an AGM campaign month, net of other campaign months in the window; multiple campaigns of the same type within a month share one indicator.
\endgroup%

 \begin{figure}[!htb]
\color{black}%
     \centering
          \caption{Distributed lags specification} 
          \label{fig:gtrends_campaigns_distr_lags}
     \begin{subfigure}{0.49 \linewidth}
    \centering\includegraphics[width=\textwidth]{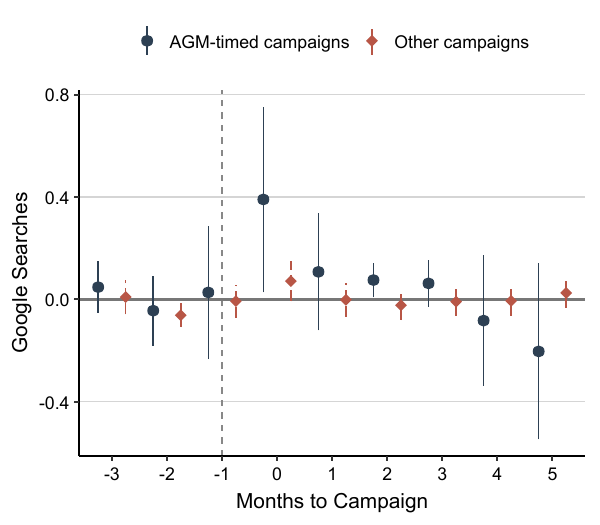}
    \subcaption{First three campaigns}
 \end{subfigure} \hfill
  \begin{subfigure}{0.49 \linewidth}
    \centering\includegraphics[width=\textwidth]{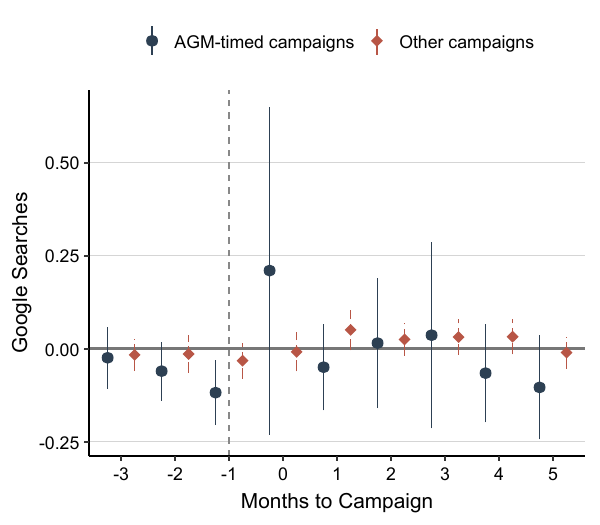}
   \subcaption{Later campaigns}
  \end{subfigure}\hfill
   \begin{minipage}{1 \textwidth} %
 {\footnotesize Note: This figure plots coefficients for standardized monthly NGO Google search activity using the distributed-lag specification in Equation~\eqref{eq:dist_lag}. Blue circles represent $\beta^{AGM}$ for AGM campaign months, and red diamonds represent $\beta^{Non-AGM}$ for other campaign months. Panels (a) and (b) use indicators for the first three and later campaigns, respectively. All leads and lags from $-3$ to $+5$ enter jointly, with no omitted month; positive values denote months after the campaign. The specification includes NGO-by-year and calendar-month fixed effects. Error bars denote 95\% normal-approximation confidence intervals based on standard errors clustered by NGO. Data combine Google Trends, Sigwatch, and ISS. \par}
 \end{minipage} %
 \end{figure}

Figure \ref{fig:gtrends_campaigns_distr_lags} reports the estimated coefficients. When restricting attention to an NGO’s first three campaigns, the results confirm that AGM-timed campaigns lead to significant short-run increases in visibility (blue dots), while later campaigns generate little or no effect. Whether early or late, non-AGM campaigns have a limited effect on an NGO's visibility.

\begin{figure}[!ht]
\color{black}%
     \centering
          \caption{Visibility Gains at and Away from AGMs} 
          \label{fig:gtrends_campaigns_app}
    \centering\includegraphics[width=0.6\textwidth]{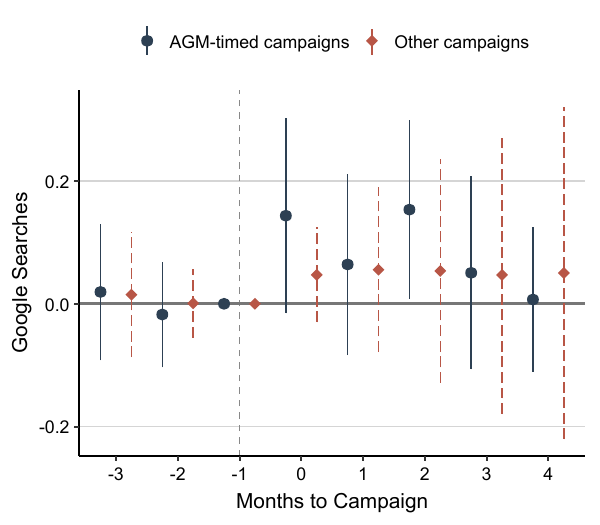}
   \begin{minipage}{1 \textwidth} %
 {\footnotesize Note: This figure plots the event-study coefficients $\beta^{AGM}$ (blue circles) and $\beta^{Non-AGM}$ (red diamonds) from Equation~\eqref{eq:awareness}, pooling all campaigns. The dependent variable is standardized monthly NGO Google search activity. Month 0 is the campaign month, and month $-1$ is omitted. The red series is the all-campaign benchmark; the blue series is the additional response associated with AGM timing, so their sum gives the fitted AGM path. The specification includes calendar-month and NGO-by-campaign-date-by-year fixed effects. Error bars denote 95\% confidence intervals based on standard errors clustered by NGO--campaign-date pair. Data combine ISS, Sigwatch, and Google Trends. \par}
 \end{minipage}
 \end{figure}

\begingroup\color{black}%
\clearpage
\subsection{Campaigns away from the AGM}
\begingroup\color{black}
Figure~\ref{fig:gtrends_campaigns_app} compares the AGM increment with the all-campaign benchmark from the baseline event study \eqref{eq:awareness}. The benchmark remains statistically indistinguishable from zero, while the additional AGM response is positive and significant two months after the campaign. This contrast underlies the discussion in Section~\ref{s:media}; the total AGM path is the sum of the two series.
\endgroup
\endgroup%

\subsection{The Full Sample}
\begin{figure}[htbp]
\color{black}%
     \centering
          \caption{Increased NGO visibility around an AGM campaign (Full sample)} 
          \label{fig:gtrends_campaigns_all}
     \begin{subfigure}{0.49 \linewidth}
    \centering\includegraphics[width=\textwidth]{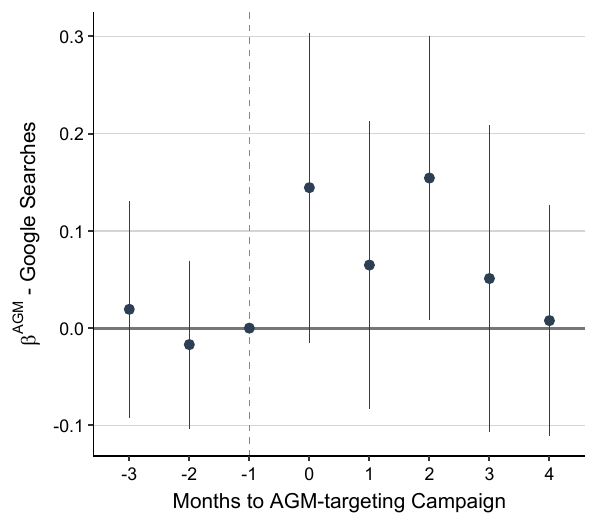}
    \subcaption{Campaign timing and campaigning NGO's  Google Trends scores}
 \end{subfigure} \hfill
  \begin{subfigure}{0.49 \linewidth}
    \centering\includegraphics[width=\textwidth]{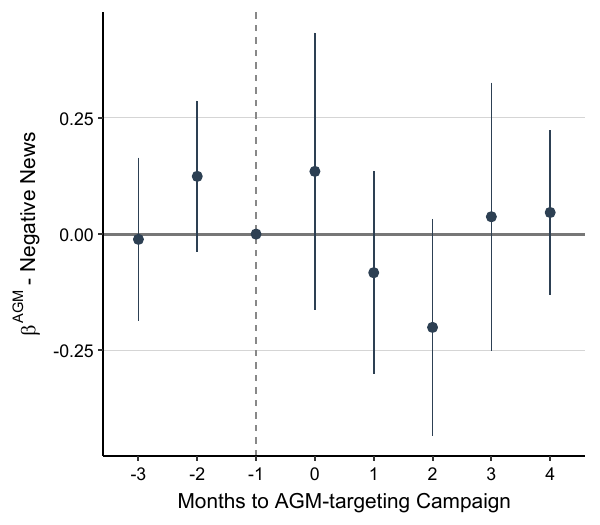}
   \subcaption{Campaign timing and number of negative news articles about target firms (standardized)}
  \end{subfigure}\hfill
   \begin{minipage}{1 \textwidth} %
 {\footnotesize Note: This figure plots the incremental AGM interaction coefficients $\beta^{AGM}$ from Equation~\eqref{eq:awareness}, estimated on all campaigns without the maturity split; month $-1$ is omitted. Panel (a) uses standardized monthly NGO Google search activity. Panel (b) uses standardized monthly negative target-firm news counts, upper-tail winsorized at the 97.5th percentile. The specification includes NGO-by-campaign-date-by-year and calendar-month fixed effects.  Error bars denote 95\% confidence intervals based on standard errors clustered at the campaign level. Data are sourced from ISS (AGM dates), Sigwatch (campaign history), Google Trends (search metrics), and RavenPack (media coverage). \par}
 \end{minipage}
 \end{figure}

\begin{figure}[!htbp]
 \centering
 \caption{NGO campaigns and shareholder proposals (Full Sample)}
     \label{fig:campaign_proposal_all}
   \includegraphics[width=0.66 \textwidth]{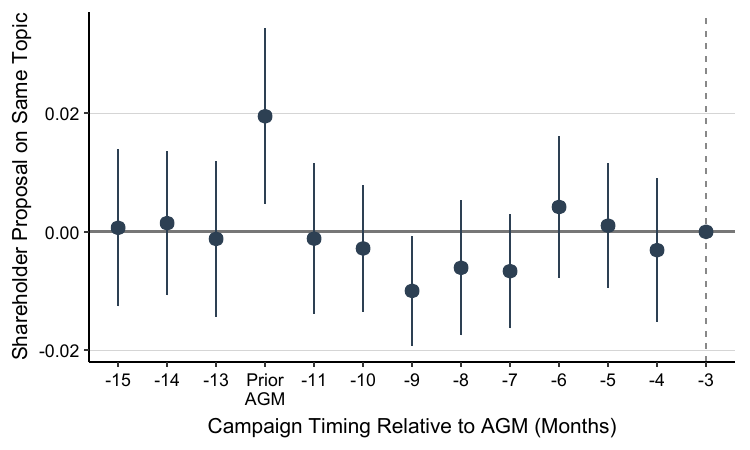}
   \subcaption{All NGOs}

 \begin{minipage}{\textwidth} %
   {\footnotesize Notes: This figure plots event-study coefficients for the probability of a related shareholder proposal at the target firm's prospective AGM, following Equation~\eqref{eq:eventstudy} and pooling all campaigns. The dependent variable equals one if a proposal addresses the same topic as the campaign. Event time measures the campaign's distance to the prospective AGM, from $-15$ to $-3$ months, with $-3$ omitted. The specification includes NGO-by-calendar-month and firm-by-year-by-topic fixed effects. Error bars denote 95\% confidence intervals based on standard errors clustered by firm. Data combine ISS (proposals and AGM dates) and Sigwatch (campaigns). \par}
 \end{minipage}
\end{figure}

\begin{figure}[!htbp]
 \centering
 \caption{NGO campaign timing and financial performance (Full sample)}
   \label{fig:campaign_donations_all}
   \begin{minipage}[t]{0.49\textwidth}
     \centering
     \includegraphics[width=\linewidth]{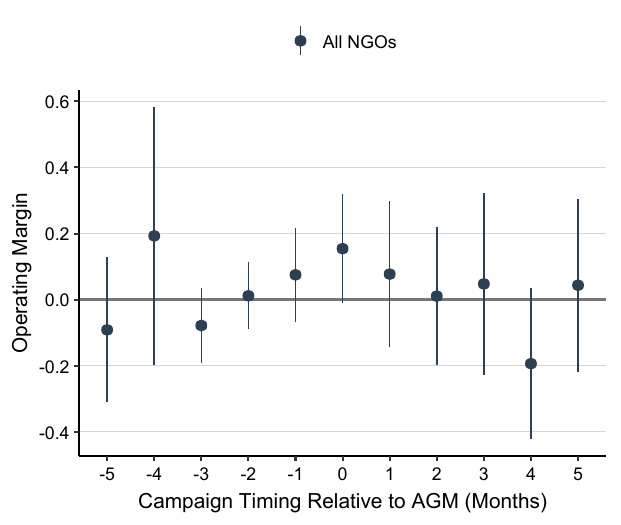}
     \par\smallskip
     {\footnotesize Panel (a): Operating margin}
   \end{minipage}\hfill
   \begin{minipage}[t]{0.49\textwidth}
     \centering
     \includegraphics[width=\linewidth]{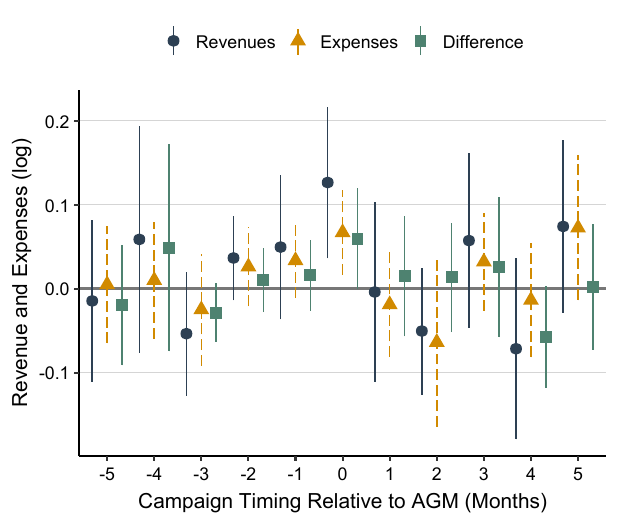}
     \par\smallskip
     {\footnotesize Panel (b): Revenue, expenses, and their difference}
   \end{minipage}
 \begin{minipage}{\textwidth} %
   {\footnotesize Notes: This figure replicates Figure~\ref{fig:campaign_donations} without the maturity split. Panel (a) uses standardized operating margin (revenue minus expenses, divided by revenue), winsorized at the 2.5th and 97.5th percentiles. Panel (b) plots log revenue (blue circles), log expenses (orange triangles), and their difference (green squares). All distance indicators enter jointly, with no omitted month; the $\pm5$ bins pool campaigns five or more months from the AGM. Regressions control for lagged log assets and include NGO and year fixed effects. Error bars show 95\% confidence intervals with standard errors clustered by NGO and year. Data: ISS, Sigwatch, and IRS Form 990 / Urban Institute. \par}
 \end{minipage}
\end{figure}

\begingroup\color{black}
The remaining exhibits replicate the main analyses without the maturity split. Search visibility rises after AGM campaigns, by less than for young NGOs alone: the AGM increment is 0.144 standard deviations in month zero and 0.154 in month two, with only the latter significant at the 5\% level (Figure~\ref{fig:gtrends_campaigns_all}). The pooled negative-news response is imprecise. The spike in related proposals at month $-12$ remains, at 1.95 percentage points (Figure~\ref{fig:campaign_proposal_all}). In the financial regression, AGM-month \auditrev{log revenue rises by 0.1264, or about 13.5\%}, and is statistically significant (Figure~\ref{fig:campaign_donations_all}). The pooled sample thus retains evidence of exposure-related responses, although their size and precision differ across outcomes.
\endgroup

\clearpage
\subsection{Consumer Response}
\begin{figure}[!htbp]
 \centering
 \caption{NGO campaign timing and sales}
   \label{fig:campaign_sales}
 \begin{subfigure}[t]{0.49\textwidth} %
   \centering
   \includegraphics[width=\textwidth]{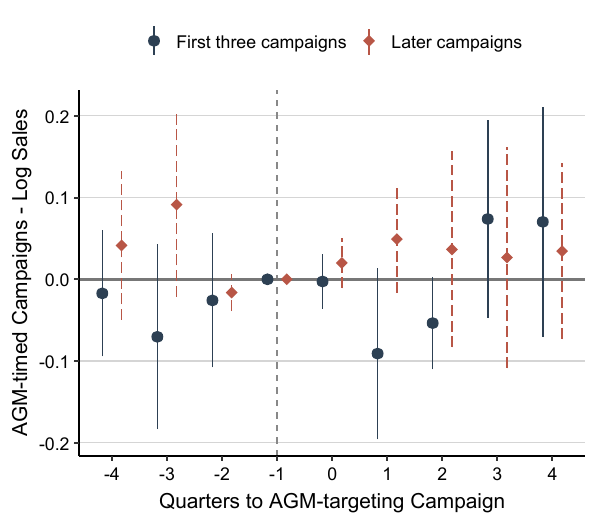}
   \subcaption{NGO Campaigns on AGM dates}
   
 \end{subfigure}
 \begin{subfigure}[t]{0.49\textwidth}
   \centering
   \includegraphics[width=\textwidth]{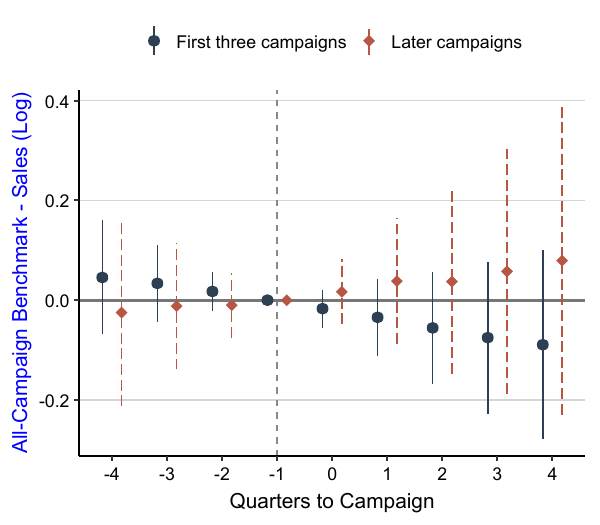}
   \subcaption{\auditrev{All-Campaign Benchmark}}
 \end{subfigure}
 \begin{minipage}{\textwidth} %
   {\footnotesize Notes: This figure presents event-study coefficients evaluating the dynamic response of target firm sales to NGO activism, focusing on firms in \auditrev{B2C industries (SIC codes 41, 45, 48, 49, 52--59, 70, 72, 75, 76, 78--80, 82--84, 86, and 88)}. The dependent variable is the logarithm of quarterly sales. The coefficients are estimated from the following specification:
\begin{equation*}
\begin{aligned}
log\left(\text{Sales}_{n,q,y} \right) & = \sum^{\tau=4}_{\substack{\tau=-4 \\ \tau\neq -1}} \left(\beta_{\tau}^{AGM} \cdot \text{AGM Campaign}_{c}\cdot \mathbf{1}_{\{M_{c,q,y}=\tau\}}\right) + \sum^{\tau=4}_{\substack{\tau=-4 \\ \tau\neq -1}} \left(\beta_{\tau}^{Non-AGM} \cdot \mathbf{1}_{\{M_{c,q,y}=\tau\}}\right)  \\ & + \alpha_{c,y} + \alpha_{j,q,y} + u_{n,q,y}.
 \end{aligned}
\end{equation*}
where $n$ denotes the firm, $q$ the quarter, and $y$ the year. The indicators $\mathbf{1}_{\{M_{c,q,y}=\tau\}}$ identify the quarters relative to the campaign launch (Quarter 0), with the quarter preceding the campaign ($\tau = -1$) serving as the omitted baseline. Both paths come from the same regression, the quarterly analogue of Equation~\eqref{eq:awareness}: Panel~(a) reports the AGM increments $\beta_{\tau}^{AGM}$ and Panel~(b) the benchmark path $\beta_{\tau}^{Non-AGM}$ traced by all campaigns. The specification includes campaign-by-year fixed effects ($\alpha_{c,y}$) and industry-by-quarter fixed effects ($\alpha_{j,q,y}$). Blue dots (red diamonds) refer to coefficients from a regression run on the first three campaigns (later campaigns). Error bars denote 95\% confidence intervals based on standard errors clustered by campaign. Data are sourced from ISS (AGM dates), Sigwatch (campaign timing) and Compustat (quarterly financial data).

\par}
 \end{minipage}
\end{figure}

Section~\ref{s:consumers} reports no consumer response to campaigns. Figure~\ref{fig:campaign_sales} shows the underlying event study for business-to-consumer industries, separately for AGM-timed and other campaigns: target firms' sales do not move around either.

\clearpage

\section{Additional Results: Influence and the NGO Lifecycle}\label{app:lifecycle}
\setcounter{table}{0}\renewcommand{\thetable}{E\arabic{table}}
\setcounter{equation}{0}\renewcommand{\theequation}{E\arabic{equation}}
\setcounter{figure}{0}\renewcommand{\thefigure}{E\arabic{figure}}

Sections~\ref{s:influence} and~\ref{s:reputation} read as a chain: early campaigns by established NGOs change what firms do, and NGOs shift toward early campaigning as reputation accumulates. This appendix stress-tests the chain link by link: that firms actually change, and not only in ratings; that the timing shift is neither a specification artifact nor one-time learning; and that the \auditrev{life cycle} does not hinge on measuring reputation by campaign counts.

\subsection{Firm Outcomes}
\begin{figure}[htbp]
\centering\caption{Campaign Timing and Changes in E\&S Scores (Full Sample)}\label{fig:campaign_esg_score_all}
    \centering\includegraphics[width=.6\linewidth]{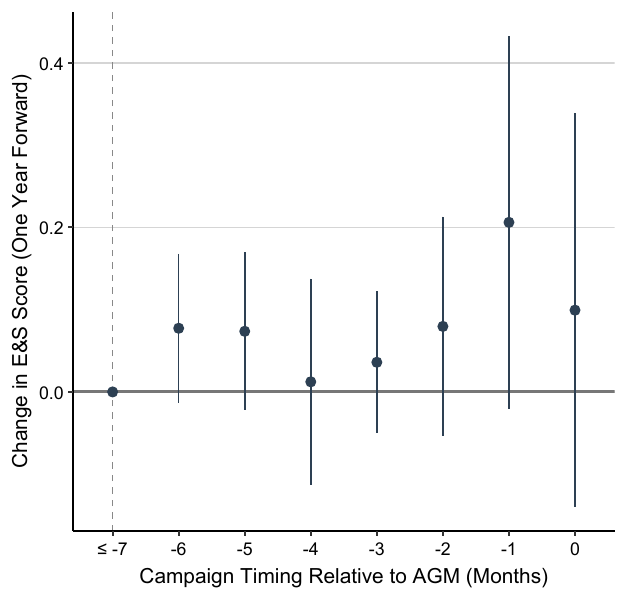}
\begin{minipage}{1\textwidth}
{\footnotesize
Note: This figure plots event-study coefficients for the one-year change in a firm's aggregate E\&S score following an NGO campaign, using Equation~\eqref{eq:eventstudy} without the maturity split. Coefficients compare campaign timing with the omitted bin of campaigns seven or more months before the AGM within the estimation window. The regression includes campaign-by-event-time interactions, event-time indicators, a campaign indicator, \auditrev{firm-by-AGM-date and month-by-year fixed effects}. Error bars denote 95\% confidence intervals based on standard errors clustered by firm and calendar month. Data combine ISS, MSCI ESG scores, Sigwatch, and Compustat. \par}
\end{minipage}
\end{figure}

\begin{figure}[!htbp]
	\centering
	\caption{NGO campaign timing and pro-social changes}
	\label{fig:campaign_other_dep}
	\begin{subfigure}[t]{0.49\textwidth} %
		\centering
		\includegraphics[width=\textwidth]{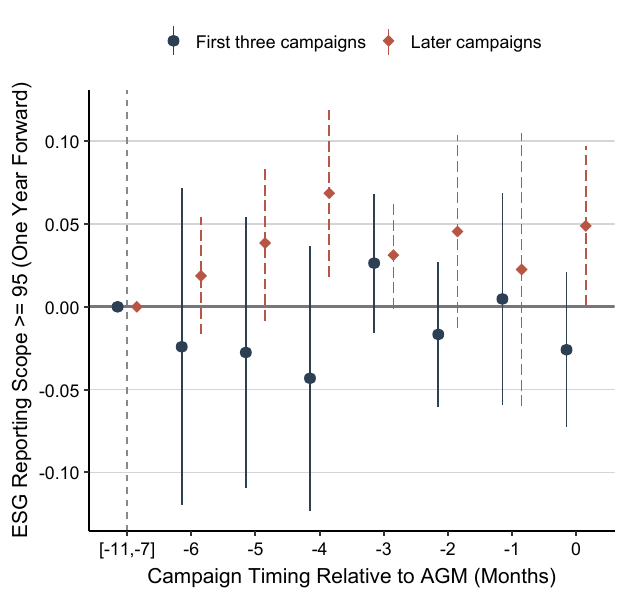}
		\subcaption{ESG Reporting Scope}
		
	\end{subfigure}
	\begin{subfigure}[t]{0.49\textwidth}
		\centering
		\includegraphics[width=\textwidth]{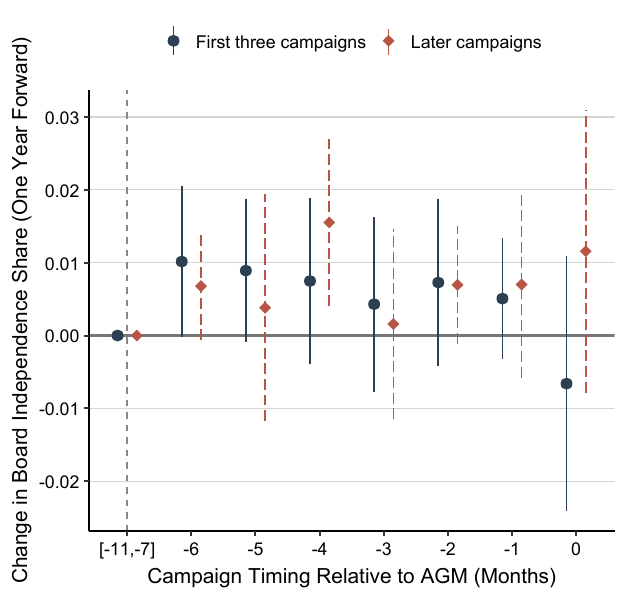}
		\subcaption{Board Independence}
	\end{subfigure}
	\begin{minipage}{\textwidth} %
		{\footnotesize Notes: Panel (a) uses an indicator equal to one if the firm's ESG reporting in the following year covers at least 95\% of its activities, as reported by LSEG. Panel (b) uses the one-year change in the share of independent board members, constructed from BoardEx. Each panel splits campaigns into the NGO's first three (blue circles) and later campaigns (red diamonds). The omitted bin pools campaigns seven or more months before the AGM. Both regressions include campaign-by-event-time interactions, event-time indicators, a campaign indicator, \auditrev{firm-by-AGM-date and month-by-year-by-industry fixed effects}. Error bars denote 95\% confidence intervals based on standard errors clustered by firm and calendar month. Data combine ISS, LSEG, BoardEx, Sigwatch, and Compustat. \par}
	\end{minipage}
\end{figure}

\begingroup\color{black}
Potentially, one could attribute the conduct results to the maturity split, or dismiss E\&S scores as ratings management. Without the split, the E\&S-score estimates are generally positive but imprecise, with all 95\% confidence intervals including zero (Figure~\ref{fig:campaign_esg_score_all}). The evidence nevertheless extends beyond scores to the two outcomes of Section~\ref{s:ES}: the scope of ESG reporting and board independence (Figure~\ref{fig:campaign_other_dep}). The responses to established NGOs' campaigns suggest changes in what firms disclose and who governs them, as well as in how they are rated.
\endgroup

\begin{figure}
\color{black}%
\caption{NGO past campaigns and the timing of current NGO campaigns (Alternative Fixed Effects)} \label{ngo_campaign_timing_robustness_fe}
    \centering
\begin{subfigure}[t]{0.66\textwidth}
 \centering
 \includegraphics[width= \linewidth]{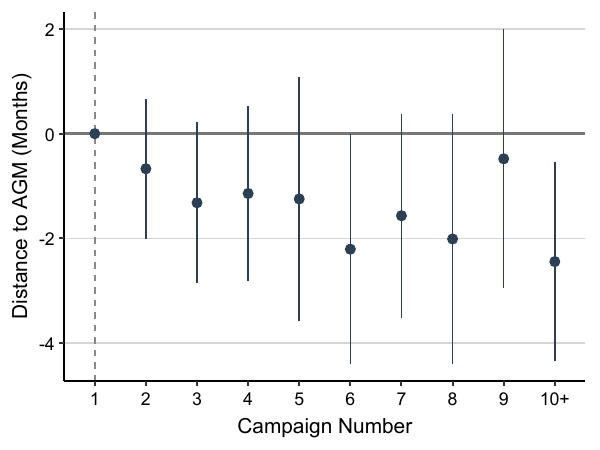}
\end{subfigure}
\begin{minipage}{1\textwidth}%
{\footnotesize Note: This figure plots coefficients from a regression of campaign distance to the AGM, measured in months, on indicators for the focal NGO's campaign order, following Equation~\eqref{eq:timing}. Relative to Figure~\ref{fig:timing_exp_gt}, Panel (a), the baseline NGO-by-firm-by-year effects are replaced by separate NGO and firm-by-year effects, retaining topic-by-year effects. This allows identification to use variation across NGO--firm--year cells. The first campaign is omitted, and the tenth and later campaigns are pooled. Error bars denote 95\% confidence intervals based on standard errors clustered by NGO. Data combine ISS and Sigwatch. \par}
\end{minipage}
\end{figure}

\begin{figure}
\color{black}%
\caption{Campaign timing for NGOs whose first campaign is after the AGM} \label{ngo_campaign_timing_before_after}
    \centering
\begin{subfigure}[t]{0.66\textwidth}
 \centering
 \includegraphics[width= \linewidth]{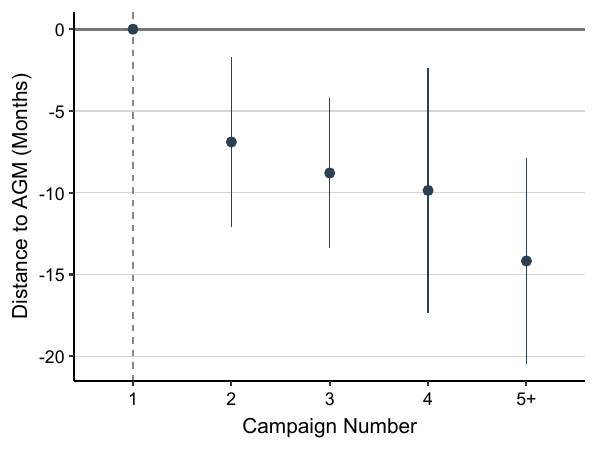}
\end{subfigure}

\begin{minipage}{1\textwidth}%
{\footnotesize Note: This figure plots coefficients from a regression of campaign distance to the AGM on indicators for the focal NGO's campaign order, following Equation~\eqref{eq:timing}. The sample is restricted to NGOs whose first recorded campaign occurred 1 to 30 days after a target firm's AGM. The specification includes NGO-by-firm-by-year and topic-by-year fixed effects. The first campaign is omitted, and the fifth and later campaigns are pooled. Data combine ISS and Sigwatch. \par}
\end{minipage}
\end{figure}

\begingroup\color{black}%
\clearpage
\subsection{The Timing Shift under Alternative Specifications}
\begingroup\color{black}
The timing shift could, in principle, be built into the specification or reflect a single lesson rather than a \auditrev{life cycle}. Replacing the baseline NGO-by-firm-by-year effects with separate NGO and firm-by-year effects, while retaining topic-by-year effects, leaves a generally negative but less precise pattern (Figure~\ref{ngo_campaign_timing_robustness_fe}). NGOs whose first recorded campaign came in the 30 days after a target's AGM also move toward earlier campaigns as they accumulate experience (Figure~\ref{ngo_campaign_timing_before_after}). Although this restricted regression retains only 21 observations, the pattern is consistent with a gradual shift rather than a one-time realization that votes have already been cast.
\endgroup
\endgroup%

\clearpage
\subsection{Alternative Reputation Proxies}
\begin{figure}[t]
\color{black}%
     \centering
          \caption{NGO reputation and campaign timing \label{fig:timing_exp_us}} 
     \begin{subfigure}{0.49 \linewidth}
    \centering\includegraphics[width=\textwidth]{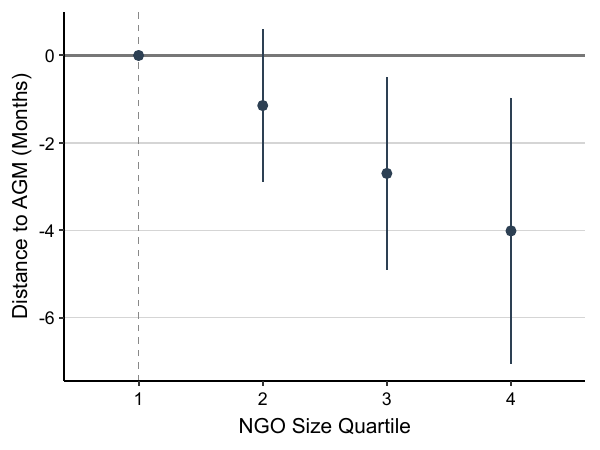}
    \subcaption{NGOs with larger assets tend to campaign before the target's AGM date}
  \end{subfigure} \hfill
  \begin{subfigure}{0.49 \linewidth}
    \centering\includegraphics[width=\textwidth]{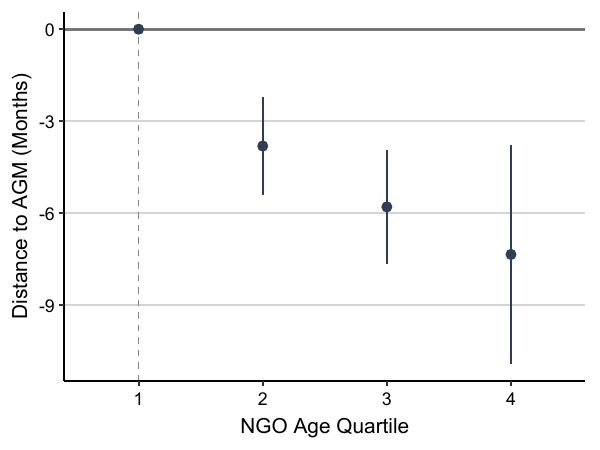}
    \subcaption{Older NGOs tend to campaign before the target's AGM date}
  \end{subfigure}\hfill
   \begin{minipage}{1 \textwidth} %
 {\footnotesize Note: This figure plots coefficients from regressions of campaign distance to the AGM, measured in months, on quartiles of NGO reputation proxies, following the approach in Equation~\eqref{eq:timing}. The sample consists of \auditrev{U.S.} NGOs with IRS Form 990 / Urban Institute financial data. Panel (a) uses lagged total assets, and Panel (b) uses NGO age, ordered by negative formation year. Both regressions include NGO-by-firm and topic-by-year fixed effects. Error bars denote 95\% confidence intervals based on standard errors clustered by NGO. Data combine Sigwatch, IRS Form 990 / Urban Institute, and ISS. \par}
 \end{minipage}
 \end{figure}

Finally, the campaign count could bundle reputation with whatever else grows over an NGO's life, and, because it starts with the Sigwatch sample in 2010, it understates the prior activity of NGOs already campaigning earlier; our long panel keeps the truncation limited. For the \auditrev{U.S.} NGOs whose Form~990 filings let us look, the \auditrev{life cycle} survives substituting the proxy entirely: older and larger NGOs campaign earlier, away from the AGM (Figure~\ref{fig:timing_exp_us}), and the point estimates for reputation and proposal success are concave for search visibility, Sigwatch's NGO power measure, assets, revenue, and expenditures (Table~\ref{ngo_size_success}). The \auditrev{life cycle} is a property of the organization's standing, not of how we count its campaigns.

\begin{table}[!htb]
\color{black}
   \caption{\textbf{NGOs' strategies and support for related shareholder proposals}} \label{ngo_size_success}
   \begin{center}
   \resizebox{ 0.95\textwidth}{!}{
\begin{tabular}{lccccc}
\toprule
Dependent variable: & \multicolumn{5}{c}{Proposal success (0/1)} \\
Model: & (1) & (2) & (3) & (4) & (5) \\
NGO characteristic & Search (std.) & NGO power (raw) & Assets decile & Revenue decile & Expenses decile \\
Distance to AGM (std.) & -0.005 & -0.008 & -0.091 & -0.113 & -0.097 \\
 & (0.030) & (0.030) & (0.066) & (0.075) & (0.066) \\
Same topic $\times$ Distance to AGM & -0.033 & -0.028 & -0.029 & 0.004 & -0.013 \\
 & (0.038) & (0.030) & (0.054) & (0.058) & (0.042) \\
Campaign on same topic & -0.028 & -0.638*** & -0.395 & -0.573** & -0.399* \\
 & (0.049) & (0.164) & (0.245) & (0.265) & (0.220) \\
Same topic $\times$ NGO characteristic & 0.027 & 0.709*** & 0.144* & 0.191** & 0.148** \\
 & (0.034) & (0.186) & (0.084) & (0.085) & (0.073) \\
Same topic $\times$ Characteristic squared & -0.015 & -0.184*** & -0.011 & -0.014** & -0.012* \\
 & (0.018) & (0.052) & (0.007) & (0.007) & (0.006) \\
\midrule
Firm $\times$ NGO $\times$ Year FE & $\checkmark$ & $\checkmark$ & $\checkmark$ & $\checkmark$ & $\checkmark$ \\
Topic $\times$ Month-Year FE & $\checkmark$ & $\checkmark$ & $\checkmark$ & $\checkmark$ & $\checkmark$ \\
\midrule
Observations & 2,110 & 2,361 & 889 & 785 & 889 \\
R$^2$ & 0.63231 & 0.64027 & 0.60022 & 0.58850 & 0.60021 \\
\bottomrule
\end{tabular}}
   \end{center}
   \begin{tablenotes}[flushleft,para]
\footnotesize\color{black}
\item Note: This table reports coefficients from Equation~\eqref{eq:main_reg}, replacing past campaigns with standardized Google search popularity (Column~1), NGO power in raw units (Column~2), or NGO size measured by total assets, revenue, and expenses (Columns~3--5). Each proxy enters through its interaction with same-topic campaigning and the corresponding squared interaction. The financial proxies are within-year deciles lagged one year, available for \auditrev{U.S.} NGOs with Form 990 data. The dependent variable equals one for a withdrawn or passed proposal and zero otherwise. The sample includes campaigns 240 to 30 days before the AGM. All regressions include Firm $\times$ NGO $\times$ Year and Topic $\times$ Meeting Month-Year fixed effects, which absorb the characteristic main effects. Firm-clustered standard errors are reported in parentheses. Data combine ISS, Google Trends, Sigwatch, and IRS Form 990 / Urban Institute. * p $<$ 0.10; ** p $<$ 0.05; *** p $<$ 0.01.
\end{tablenotes}
   \end{table}

\clearpage

\section{The Fashion Transparency Index}\label{apndx:additional_selection}
\setcounter{table}{0}\renewcommand{\thetable}{F\arabic{table}}
\setcounter{equation}{0}\renewcommand{\theequation}{F\arabic{equation}}
\setcounter{figure}{0}\renewcommand{\thefigure}{F\arabic{figure}}

This appendix collects the exhibits for the case study of Section~\ref{s:case}: the balance check behind the exogeneity of publication timing, the market response around publication, and the shareholder outcomes that mirror the main results.

\begin{table}[!htb]
\color{black}
   \caption{\textbf{Exogeneity of the index publication relative to firms' AGMs}} \label{fti_balance_check}
\begin{center}
\begin{tabular}{lccc}
\toprule
Dependent variable: & \multicolumn{3}{c}{Report before AGM (0/1)} \\
Model: & (1) & (2) & (3) \\
Market capitalization (decile) & -0.028 &  &  \\
 & (0.025) &  &  \\
Europe &  & -0.157 &  \\
 &  & (0.125) &  \\
North America &  & 0.003 &  \\
 &  & (0.003) &  \\
FTI score &  &  & 0.005 \\
 &  &  & (0.005) \\
\midrule
Year FE & $\checkmark$ & $\checkmark$ & $\checkmark$ \\
\midrule
Observations & 91 & 92 & 92 \\
R$^2$ & 0.08113 & 0.10792 & 0.10244 \\
\bottomrule
\end{tabular}
\end{center}
\begin{tablenotes}[flushleft,para]
\footnotesize\color{black}
\item Note: This table reports OLS regressions of an indicator for FTI publication before a firm's AGM on the firm's market-capitalization decile (Column~1), headquarters region (Column~2), or initial FTI score (Column~3). The sample contains first reports with observed publication and AGM dates. In Column~2, regions other than Europe and North America form the reference category. All regressions include year fixed effects. Standard errors are clustered by country and reported in parentheses. Data combine ISS (AGM dates), LSEG (financials), and Fashion Revolution (FTI scores). * p $<$ 0.10; ** p $<$ 0.05; *** p $<$ 0.01.
\end{tablenotes}
\end{table}

 \begin{figure}[!htb]
\color{black}%
     \centering
          \caption{Stock returns around inclusion \label{fig:fti_cer}} 
     \begin{subfigure}{0.49 \linewidth}
    \centering\includegraphics[width=\textwidth]{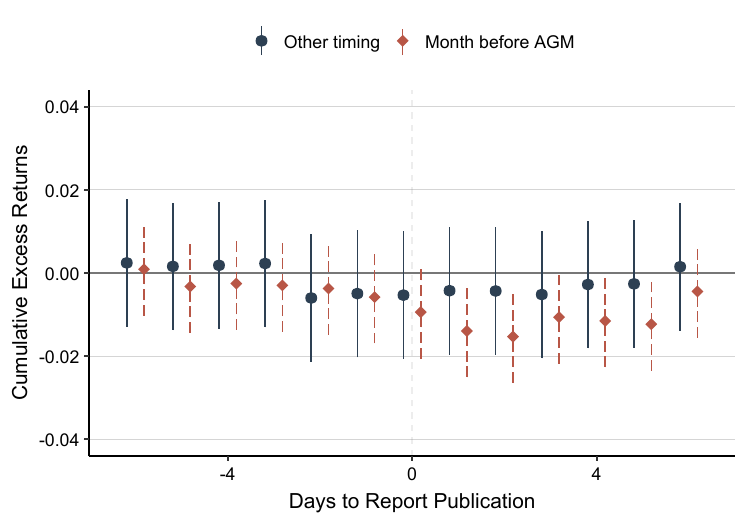}
    \subcaption{Stock returns depending on timing}
  \end{subfigure} \hfill
  \begin{subfigure}{0.49 \linewidth}
    \centering\includegraphics[width=\textwidth]{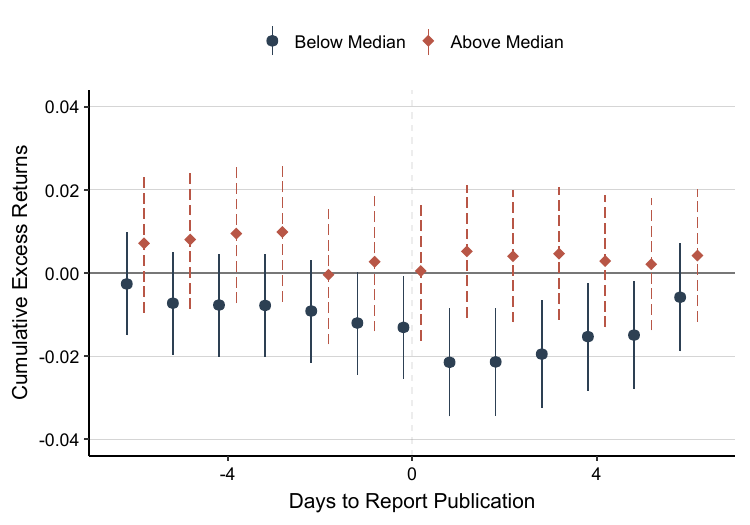}
    \subcaption{Stock returns by FTI initial score}
  \end{subfigure}\hfill
   \begin{minipage}{1 \textwidth} %

 {\footnotesize Note: This figure plots cumulative excess returns around publication of the first \textit{Fashion Transparency Index} (FTI) covering the firm. Excess returns are the difference between daily stock returns and the corresponding national stock-index return. Panel (a) compares publication in the month before the AGM with other timing; Panel (b) splits event-day observations into two within-year groups by the firm's initial FTI score. Daily excess returns are winsorized at the 1st and 99th percentiles within event day and cumulated over the window of six trading days on either side of publication. Error bars denote 95\% confidence intervals. Data combine LSEG (stock and index returns) and Fashion Revolution (FTI scores). \par} %
 \end{minipage}
 \end{figure}

\begin{figure}[!htb]
    \centering
        \caption{Index publication and support for E\&S shareholder proposals \label{fig:fti_proposals}}
    \begin{subfigure}{0.49\textwidth}
\includegraphics[width=\linewidth]{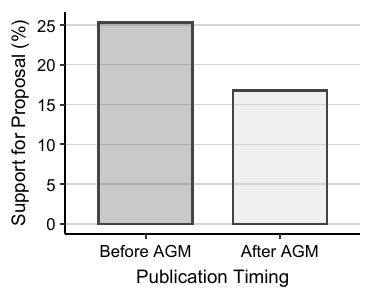}
\subcaption{Voting support}
\end{subfigure}\hfill
\begin{subfigure}{0.49\textwidth}
\includegraphics[width=\linewidth]{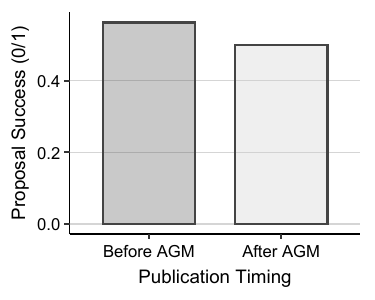}
\subcaption{Firm-year proposal success}
\end{subfigure}
    \begin{minipage}{1 \textwidth} %
 {\footnotesize Note: This figure compares shareholder-proposal outcomes at firms included in the \textit{Fashion Transparency Index} (FTI), according to whether publication occurs before or after the firm's AGM. Panel (a) reports average voting support for E\&S proposals, using Diligent voting outcomes; the before/after means are 25.32\% and 16.76\%. Panel (b) reports the fraction of ISS firm-years with at least one withdrawn or passed E\&S proposal; the corresponding means are 56.25\% and 50.00\%. Data combine Diligent, ISS, and Fashion Revolution. \par} %
 \end{minipage}
\end{figure}

 \begin{figure}[!ht]
     \centering
          \caption{Publication timing and firm outcomes} \label{fig:fti_regs}
     \begin{subfigure}{0.49 \linewidth}
    \centering\includegraphics[width=\textwidth]{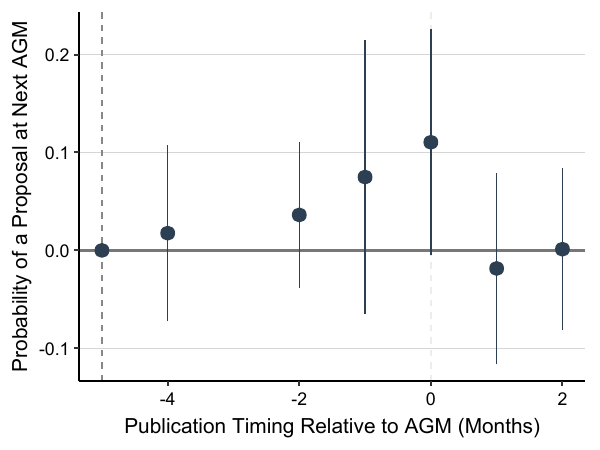}
    \subcaption{Publication timing and related shareholder proposals at the next AGM}
  \end{subfigure} \hfill
  \begin{subfigure}{0.49 \linewidth}
    \centering\includegraphics[width=\textwidth]{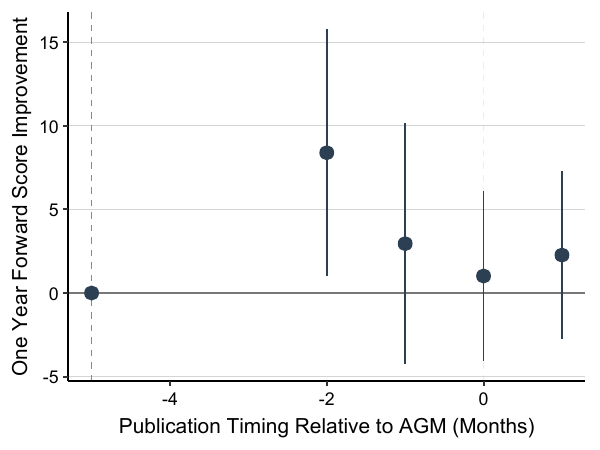}
    \subcaption{Publication timing and FTI score improvement}
  \end{subfigure}\hfill
   \begin{minipage}{1 \textwidth} %
 {\footnotesize Note: This figure presents event-study coefficients evaluating the relationship between \textit{Fashion Transparency Index} (FTI) publication timing and corporate outcomes. The coefficients are estimated from:
\begin{equation*}
Y_{n,y}=\sum_{\tau=-4}^{2}\beta_\tau\,\mathbf{1}_{\{D_{n,y}=\tau\}}\,\text{Treated}_{n,y}
+\mathbf{X}_{n,y}\gamma+\alpha_n+\alpha_y+u_{n,y}.
\end{equation*}
Here $Y_{n,y}$ is an indicator for a related proposal at the next AGM in Panel (a), or the change in FTI score over the following year in Panel (b). $\text{Treated}_{n,y}$ indicates FTI coverage. The timing index $D_{n,y}$ pools values below $-5$ at $-5$, the omitted category. The vector $\mathbf{X}_{n,y}$ includes timing indicators, the coverage indicator, and market-capitalization deciles. Both regressions include firm ($\alpha_n$) and year ($\alpha_y$) fixed effects. Error bars denote 95\% confidence intervals based on standard errors clustered by firm. Data combine Fashion Revolution, Diligent, and firm financial data. \par}
 \end{minipage}
 \end{figure}

\clearpage

\clearpage
\begingroup\color{black}
\section{Solo and Collaborative Campaigns}\label{app:coalitions}
\setcounter{table}{0}\renewcommand{\thetable}{G\arabic{table}}
\setcounter{equation}{0}\renewcommand{\theequation}{G\arabic{equation}}
\setcounter{figure}{0}\renewcommand{\thefigure}{G\arabic{figure}}

The preceding analyses focus on single-NGO campaigns. This appendix extends the analysis to collaborative campaigns involving multiple NGOs and examines whether they generate similar patterns of visibility and influence. Collaborative campaigns are coded as involving up to five NGOs, although some may include additional participants. Overall, the results suggest limited individual benefits from collaboration, although the combined experience of participating NGOs increases campaign success.

\subsection{Visibility around Solo and Collaborative Campaigns}

Figure~\ref{fig:coalition_attention} compares attention responses for NGOs participating in both solo and collaborative campaigns. We jointly estimate four trajectories, distinguishing solo from collaborative campaigns and, within each group, the first three campaigns from subsequent ones. Each group has its own calendar-month and NGO-by-campaign-date-by-year fixed effects. Standard errors are clustered at the NGO level across campaigns and groups.

For solo campaigns, the results remain similar to those in Figure~\ref{fig:gtrends_campaigns}, despite the smaller sample: AGM campaigns among an NGO's first three campaigns generate substantial visibility gains that disappear as NGOs gain experience. In contrast, collaborative campaigns generate no significant visibility gains, regardless of experience. Unreported results similarly show no visibility gains for non-AGM campaigns.

\begin{figure}[!htbp]
\centering
\caption{NGO visibility around solo and collaborative campaigns}\label{fig:coalition_attention}
\begin{subfigure}{0.49\textwidth}
\includegraphics[width=\linewidth]{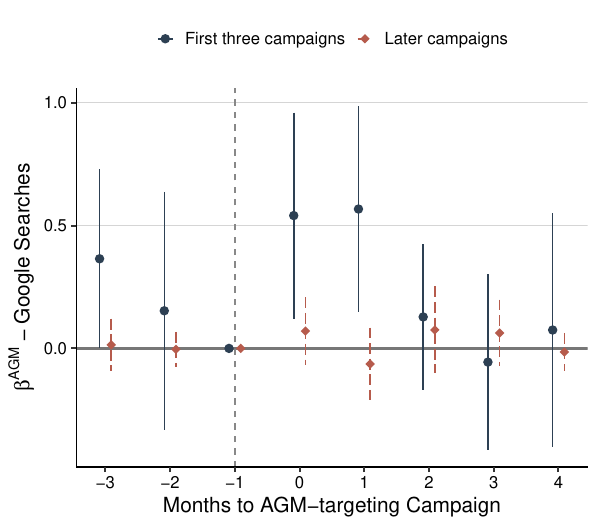}
\subcaption{Solo campaigns}
\end{subfigure}\hfill
\begin{subfigure}{0.49\textwidth}
\includegraphics[width=\linewidth]{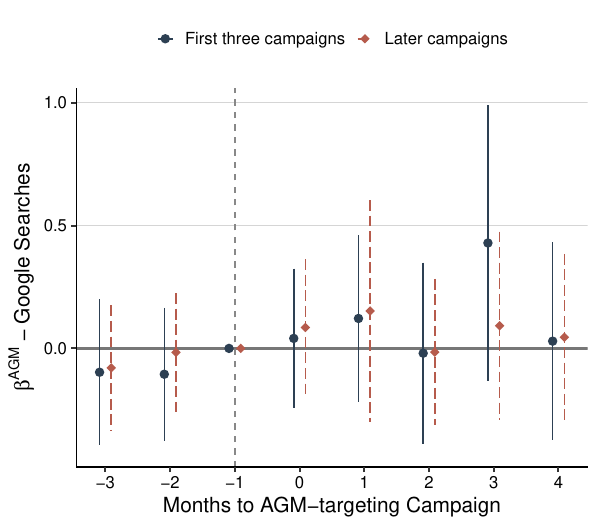}
\subcaption{Collaborative campaigns}
\end{subfigure}
\begin{minipage}{\textwidth}\footnotesize
\color{black}
Note: The dependent variable is monthly NGO search activity (standardized). The figure reports the AGM interaction relative to the all-campaign benchmark in each stratum, with month $-1$ omitted and an event window from $-3$ to $4$. Blue circles denote the first three campaigns and red diamonds later campaigns. Joint estimation allows stratum-specific calendar-month and NGO-by-campaign-date-by-year fixed effects. Error bars show 95\% confidence intervals based on NGO-clustered standard errors. The attention sample includes NGOs observed in both solo and collaborative campaigns. Data: Google Trends, Sigwatch, and ISS.
\end{minipage}
\end{figure}

\subsection{Experience and Proposal Success}

Table~\ref{tab:coalition_success} distinguishes between prior solo and collaborative campaign experience. Columns (1) and (2) focus on collaborative campaigns, aggregating members' prior experience using either the sum or the maximum across participants, respectively. Experience measures are standardized, and the regressions control for coalition size, campaign distance, and prior campaigns involving the same set of NGOs, along with same-topic interactions and the corresponding level effects. \auditrev{All specifications include topic-by-meeting-month-year fixed effects; columns (1) and (2) include firm-by-year fixed effects, while column (3) includes firm-by-NGO-by-year fixed effects. Standard errors are clustered at the firm level.}

\begingroup\color{black}%
\begingroup\color{black}
Prior solo AGM-campaign experience predicts proposal success in both coalition specifications at the 5\% level. Prior solo non-AGM experience is also positive and significant at the 10\% level when members' experience is aggregated by its maximum (Column~2), while the collaborative-history interactions are imprecise. These findings suggest that coalitions benefit from their members' individual experience, with limited evidence of additional benefits from prior collaboration.
\endgroup
\endgroup%

\begingroup\color{black}%
\begingroup\color{black}
Column (3) returns to solo campaigns and asks whether prior solo or collaborative experience predicts proposal success, regardless of AGM timing. Since only eight NGOs have prior collaborative experience, both histories enter as indicators. The solo-experience interaction is positive and significant at the 10\% level, while the collaborative-experience interaction is not. The former point estimate is nearly four times the latter, although this is not a formal test of their difference. The results thus provide stronger evidence for benefits from individual experience than from past collaboration.
\endgroup
\endgroup%

\begin{table}[htbp]
\color{black}\centering\caption{Campaign success and prior solo and collaborative experience}\label{tab:coalition_success}\begin{adjustbox}{max width=\textwidth}\begin{tabular}{lccc}\toprule
Dependent variable: & Proposal success (0/1) & Proposal success (0/1) & Proposal success (0/1) \\
Model: & (1) & (2) & (3) \\
Current campaign & Collaboration & Collaboration & Solo \\
Experience measure & Sum & Maximum & Any prior \\
\midrule
Distance to AGM (std.) & -0.024 & -0.020 & -0.005 \\
 & (0.018) & (0.017) & (0.030) \\
Same topic $\times$ Distance to AGM & -0.001 & 0.000 & -0.031 \\
 & (0.029) & (0.029) & (0.034) \\
Campaign on same topic & -0.019 & -0.012 & -0.087 \\
 & (0.049) & (0.050) & (0.055) \\
Same topic $\times$ Past Solo AGM Campaigns & 0.083** & 0.063** &  \\
 & (0.033) & (0.025) &  \\
Same topic $\times$ Past Collaborative AGM Campaigns & 0.006 & 0.005 &  \\
 & (0.042) & (0.039) &  \\
Same topic $\times$ Past Solo Non-AGM Campaigns & 0.116 & 0.106* &  \\
 & (0.089) & (0.062) &  \\
Same topic $\times$ Past Collaborative Non-AGM Campaigns & -0.070 & -0.035 &  \\
 & (0.069) & (0.052) &  \\
Same topic $\times$ Past Exact-Coalition Campaigns & -0.022 & -0.023 &  \\
 & (0.033) & (0.037) &  \\
Same topic $\times$ Past Solo Campaigns (0/1) &  &  & 0.056* \\
 &  &  & (0.030) \\
Same topic $\times$ Past Collaborative Campaigns (0/1) &  &  & 0.015 \\
 &  &  & (0.029) \\
\midrule
Coalition size controls & Yes & Yes & No \\
Firm $\times$ NGO $\times$ Year FE & & & $\checkmark$ \\
Firm $\times$ Year FE & $\checkmark$ & $\checkmark$ & \\
Topic $\times$ Month-Year FE & $\checkmark$ & $\checkmark$ & $\checkmark$ \\
\midrule
Observations & 1,119 & 1,119 & 2,361 \\
R$^2$ & 0.603 & 0.604 & 0.639 \\
\bottomrule\end{tabular}\end{adjustbox}\par\medskip\begin{tablenotes}[flushleft,para]
\footnotesize\color{black}
\item Note: This table relates proposal success to prior solo and collaborative campaign experience. The dependent variable equals one for a withdrawn or passed proposal and zero otherwise. Histories count AGM-matched campaigns through the previous year; AGM timing means the meeting date or preceding day.\\
\item \textbf{Columns (1)--(2).} Collaborative campaigns, with one observation per campaign--proposal pair. Experience is aggregated by its sum or maximum across distinct NGOs in the five participant fields and standardized. Exact-coalition history counts prior campaigns by the same member set. Coalition size, histories, and standardized distance enter with same-topic interactions and level effects where identified. Both columns include Firm $\times$ Year and Topic $\times$ Meeting Month-Year fixed effects.\\
\item \textbf{Column (3).} Solo campaigns, with indicators for any prior solo or collaborative experience, regardless of AGM timing. The regression includes Firm $\times$ NGO $\times$ Year and Topic $\times$ Meeting Month-Year fixed effects, which absorb annual indicator levels. Distance and its same-topic interaction enter all three columns. Standard errors are clustered by firm and reported in parentheses. * p $<$ 0.10; ** p $<$ 0.05; *** p $<$ 0.01.
\end{tablenotes}\end{table}
\clearpage
\endgroup

\bibliographystylesecvthree{ecca-mod}
\bibliographysecvthree{main_RR1_v3_references}

\end{document}